\documentclass[%
 reprint,
 amsmath,amssymb,
prl,
]{revtex4-2}

\usepackage{graphicx}
\usepackage{pgfplots}
\usepackage[caption=false]{subfig}
\usepackage[colorinlistoftodos]{todonotes}
\usepackage{dcolumn}
\usepackage{bm}
\usepackage{upgreek}

\usepackage[maxfloats=256]{morefloats}

\begin{document}

\author{Divyam Neer Verma$^{1}$}
\author{KV Chinmaya$^{1,3}$}
\author{Jan R. Heck$^{4,5,+}$}
\author{G Mohan Rao$^{1}$}
\author{Sonia Contera$^{6,2}$}
\author{Moumita Ghosh$^{1,2,3*}$}
\author{Siddharth Ghosh$^{7,1,2,3}$}
\email{corresponding email - moumita@centerfornanodevices.com and sg915@cam.ac.uk}
\altaffiliation{
$^+$Current affiliation: Research Center for Advanced Science and Technology, University of Tokyo, Tokyo, Japan.}

\affiliation{$^1$International Center for Nanodevices, INCeNSE-TBI, Indian Institute of Science Campus, Bengaluru, IN and High Tech Campus Eindhoven, NL.}
\affiliation{$^2$Maxapiens, Cambridge, UK.}
\affiliation{$^3$Open Academic Research, Cambridge, UK and Kolkata, India.}
\affiliation{$^4$Cavendish Laboratory, University of Cambridge, Cambridge, UK.}
\affiliation{$^5$Centre for Misfolding Diseases, Yusuf Hamied Department of Chemistry, University of Cambridge, Cambridge, UK.}
\affiliation{$^6$Clarendon Laboratory, University of Oxford, Oxford, UK.}
\affiliation{$^7$Department of Applied Mathematics and Theoretical Physics, University of Cambridge, Cambridge, UK.}

\title{Single-molecule motion control}

\begin{abstract}
Achieving dynamic manipulation and control of single molecules at high spatio-temporal resolution is pivotal for advancing atomic-scale computing and nanorobotics. 
However, this endeavour is critically challenged by complex nature of atomic and molecular interactions, high-dimensional characteristics of nanoscale systems, and scarcity of experimental data. 
Here, we present a toy model for controlling single-molecule diffusion by harnessing electrostatic forces arising from elementary surface charges within a lattice structure, mimicking embedded charges on a surface.
We investigate the interplay between quantum mechanics and electrostatic interactions in single molecule diffusion processes using a combination of state-dependent diffusion equations and Green's functions.
We find that surface charge density critically influences diffusion coefficients, exhibiting linear scaling akin to Coulombic forces. 
We achieve accurate predictions of experimental diffusion constants and extending the observed range to values reaching up to 6000 $\upmu\text{m}^2\text{ms}^{-1}$ and 80000 $\upmu\text{m}^2\text{ms}^{-1}$. 
The molecular trajectories predicted by our model bear resemblance to planetary motion, particularly in their gravity-assisted acceleration-like behaviour.
It holds transformative implications for nanorobotics, motion control at the nanoscale, and computing applications, particularly in the areas of molecular and quantum computing where the trapping of atoms and molecules is essential.
Beyond the state-of-the-art optical lattice and scanning tunnelling microscopy for atomic/molecular manipulation, our findings give unambiguous advantage of precise control over single-molecule dynamics through quantum manipulation at the angstrom scale.
\end{abstract}

\maketitle

\maxdeadcycles=1000

Fast precision control for manipulating molecular dynamics \cite{ruggeri2017single}, with implications for enhancing molecular control in nanorobotics \cite{koumura1999light, zhou2023toward}, single-atom/molecule trap computing \cite{nelson2007imaging, bluvstein2024logical}, and other emerging fields is critically missing.
Single-molecule diffusion, governed by the stochastic motion of individual molecules \cite{xu1997direct}, is therefore a phenomenon of interest in computing and various fields, including materials science \cite{zurner2007visualizing}, biophysics \cite{hedgeland2009measurement, reed2022real} and molecular engineering \cite{li2023electrically}. 
Reaching the capacity to achieve control over trajectories of single molecule diffusion will transform the field of computing and nanorobotics and the development of advanced functional materials \cite{kocherzhenko2011single,leake2013physics}. 
Here, we investigate the interplay between single-molecule diffusion and electrostatic forces arising from surface charges within a lattice structured surface. 
By elucidating the mechanisms through which these forces influence molecular trajectories, we aim to identify novel approaches for controlling single-molecule motion with quantum mechanical precision.
The physics of single molecule traps has been extensively studied \cite{cohen2005control, fields2011electrokinetic,  ruggeri2017single, ruggeri2017lattice, ruggeri2018entropic}. In particular, molecular diffusion, due to the speeds achieved by molecular translational and rotational diffusion, has been explored for its potential in e.g. unconventional computing \cite{clerc2005additive, sagues2007spatiotemporal, faez2014optical, skaug2018nanofluidic,  syed2023physics, pisarchik2023coherence} and quantum metrology \cite{skaug2018nanofluidic, ghosh2020single, bovskovic2022nanopore, ronceray2023liquid} at room temperature. Previously, we measured the electrokinetics of 1D molecular mass transport inside nanoconfined water, using two-focus fluorescence correlation spectroscopy, our measurements indicated an anomalous diffusion constant; theoretically, we interpreted these confinement-induced molecular interactions as quantum shot-noise \cite{ghosh2020single, ghosh2021feynman}.

\begin{figure*}
    \centering
    \includegraphics[width = 1\textwidth]{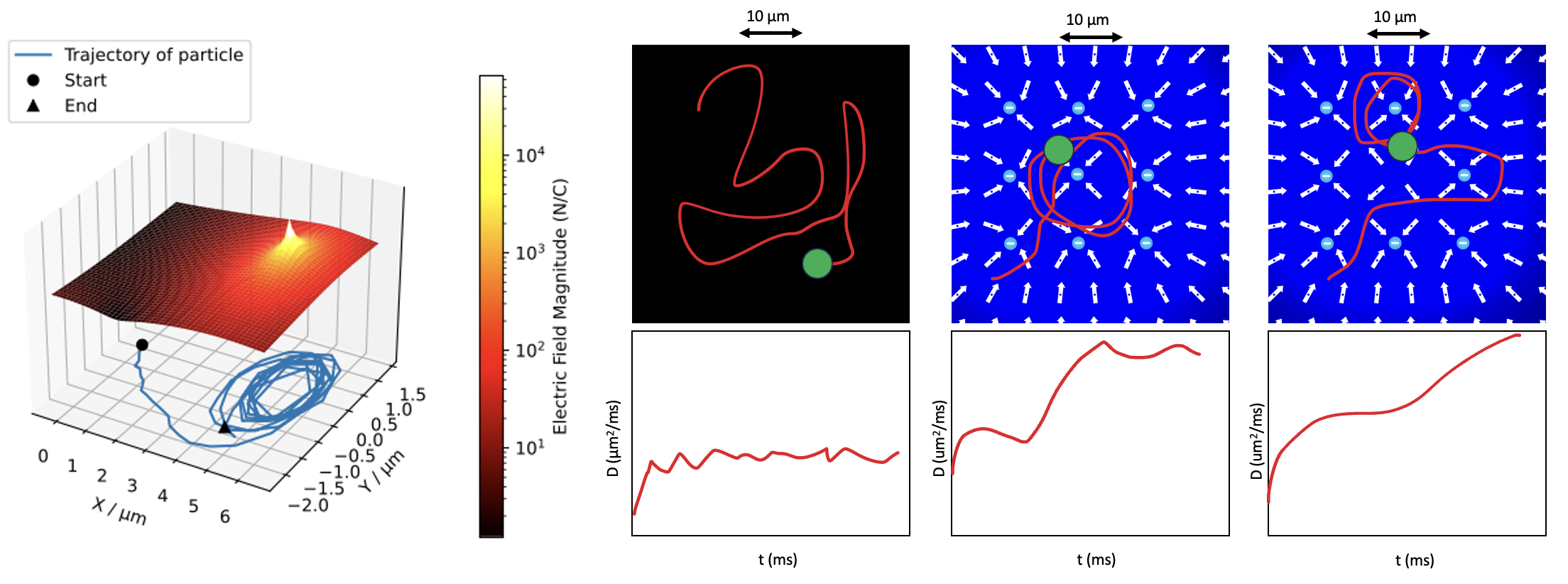}
    \caption{Single-electron controlled motion of single molecule undergoing free diffusion and electrostatic interactions in a lattice of surface charges without and with defects.}
    \label{Figure:1}
\end{figure*}

To investigate the origin of this complex behaviour, here, we address the fundamental question of what occurs when a single molecule comes into proximity with stationary charges \cite{millikan1913elementary, franklin1997millikan, einstein1905molekularkinetischen, from1906kinetic, bakken1999calculation}.
This requires consideration of long-range interactions. 
Beyond the domain of thermal forces, we consider in our analysis the subtle perturbations initiated by non-thermal forces \cite{ghosh2021feynman}, which include the gradual release of stress in slowly relaxing systems, energy injection into steady-state systems, and the dynamics of chemical reactions in active matter.
Previous reports explored the impact of the charge on molecular diffusion \cite{burgos2009directed, golestanian2009anomalous}. 
Their work focused on diverse examples using colloids \cite{agudo2018enhanced}; however, single molecules have remained unexplored.
Our analysis acknowledges results from previous works including,  Krishnan's electrostatic trap for single molecule electrometry \cite{ruggeri2017single, bespalova2019single}, Cohen and Moerner's earlier work on crafting an ABEL or Anti-Brownian Electrokinetic trap \cite{CohenPNAS2006} rooted in Enderlein's proposal of single molecule tracking using a rotating laser \cite{enderlein2000positional, enderlein2000tracking}.  
Our work explores quantum mechanical contributions to single-molecule motion, which include molecular trajectories similar to orbital and ballistic motions with an extraordinary diffusion behaviour of individual single molecules influenced by elemental surface charges within a lattice containing defects that cannot be explained by traditional theories.
We utilise a system of partial differential equations (PDEs) to model the behaviour of single molecules.
Our model accounts for factors such as diffusion coefficient, electrostatic force, friction coefficient, electrostatic potential, and single-molecule density. 
We analysed the dynamics of single molecules, including their energy balance and entropy production, and explore potential mechanisms for controlling and manipulating their motion at the nanoscale. 
We also discuss various extensions to the model, such as incorporating external fields, molecular interactions, and nonlinear effects, to provide a more comprehensive understanding of these complex systems.
To systematically explore this phenomenon, we considered several distinct scenarios, each characterised by a different random distribution of single-electron surface charges. 
Finally, we demonstrate that, following our analysis, it is possible to control the motion of individual molecules with single-electron precision.

\section{Theory}
\textit{Control of Single-Molecule Motion.---}
Let us first look at the interactions governing the motion of individual molecules in the presence of surface charges. 
To do this, we turn to Green's functions that describe the evolution of states under the influence of external potentials or forces.
The evolution of the system is described by the Schr\"odinger equation:

\begin{equation}
    i\hbar \frac{\partial\Psi(x, t)}{\partial t} = -\frac{\hbar^2}{2m} \nabla^2 \Psi(x, t) + V(x)\Psi(x, t)
\end{equation}
\(\Psi(x, t)\) represents the quantum state of the molecule at position \(x\) and time \(t\), \(\hbar\) is the reduced Planck constant, \(m\) is the mass of the molecule, and \(V(x)\) encompasses the potential energy landscape, including the Coulomb potential originating from the elementary surface charges.
To introduce Green's functions into our framework, we transform the Schr\"odinger equation into integral form:

\begin{equation}
    \Psi(x, t) = \int G(x, x', t, t')\Psi(x', t')dx'
\end{equation}

The Green's function \(G(x, x', t, t')\) encapsulates the quantum mechanical evolution of the system, revealing how the quantum state at position \(x\) and time \(t\) responds to the potential at position \(x'\) and time \(t'\).
The Green's function for quantum dynamics in response to Coulomb potentials --- the Green's function itself obeys a distinctive equation, often termed the time-dependent Schr\"odinger equation with a source term:

\begin{equation}
    i\hbar \frac{\partial G(x, x', t, t')}{\partial t} = -\frac{\hbar^2}{2m} \nabla^2 G(x, x', t, t') + \delta(x - x')\delta(t - t')
\end{equation}
In this equation, \(\delta(x - x')\) and \(\delta(t - t')\) denote Dirac delta functions, signifying the presence of the molecule at position \(x'\) and time \(t'\).
Solving this equation for \(G(x, x', t, t')\) unravels the quantum mechanical evolution of the system as influenced by Coulomb forces arising from surface charges. 

By incorporating Green's functions into our theoretical foundation, we gain a deeper understanding of the quantum mechanical underpinnings that govern single-molecule trajectories in the presence of surface charges. 
This insight enriches our comprehension of non-equilibrium phenomena at the nanoscale and underscores the pivotal role of quantum control in shaping the dynamics of individual molecules.
This theoretical framework integrates the concept of state-dependent diffusion, shedding light on the unusual behaviour of single molecules as they navigate within a lattice structure housing surface charges. 
State-dependent diffusion, a fundamental aspect of our study, captures the essential idea that the motion of a molecule is contingent upon its specific state and the surrounding environment \cite{lau2007state}.
Incorporating this concept into our theoretical foundation, we consider how the quantum state of the molecule, described by the wavefunction \(\Psi(x, t)\), evolves in response to both external potentials and state-dependent factors. 
The presence of surface charges introduces state-dependent forces that influence the molecule's trajectory. 
These forces are intricately interwoven with the molecule's quantum state, leading to non-trivial alterations in its motion.
Our utilisation of Green's functions allows us to rigorously account for these state-dependent effects. By solving the Schr\"odinger equation with a source term that incorporates state-dependent factors, we obtain a comprehensive understanding of how the quantum state of the molecule responds to the Coulomb potentials generated by surface charges.
This approach provides a robust foundation for comprehending the quantum control exerted over individual molecules within complex environments, marking a significant step forward in our exploration of non-equilibrium phenomena at the nanoscale.
This framework for single-electron controlled motion of single molecules combines the Schr\"odinger equation, quantum control techniques, electron-molecule interactions, and the quantum Langevin equation to describe how quantum manipulation can influence the diffusion properties of individual molecules. 
This mathematical formulation can be used experimentally for the precise engineering of molecular diffusion at the quantum level.

\begin{figure*}[]
    \centering
    \includegraphics[width = 0.95\textwidth]{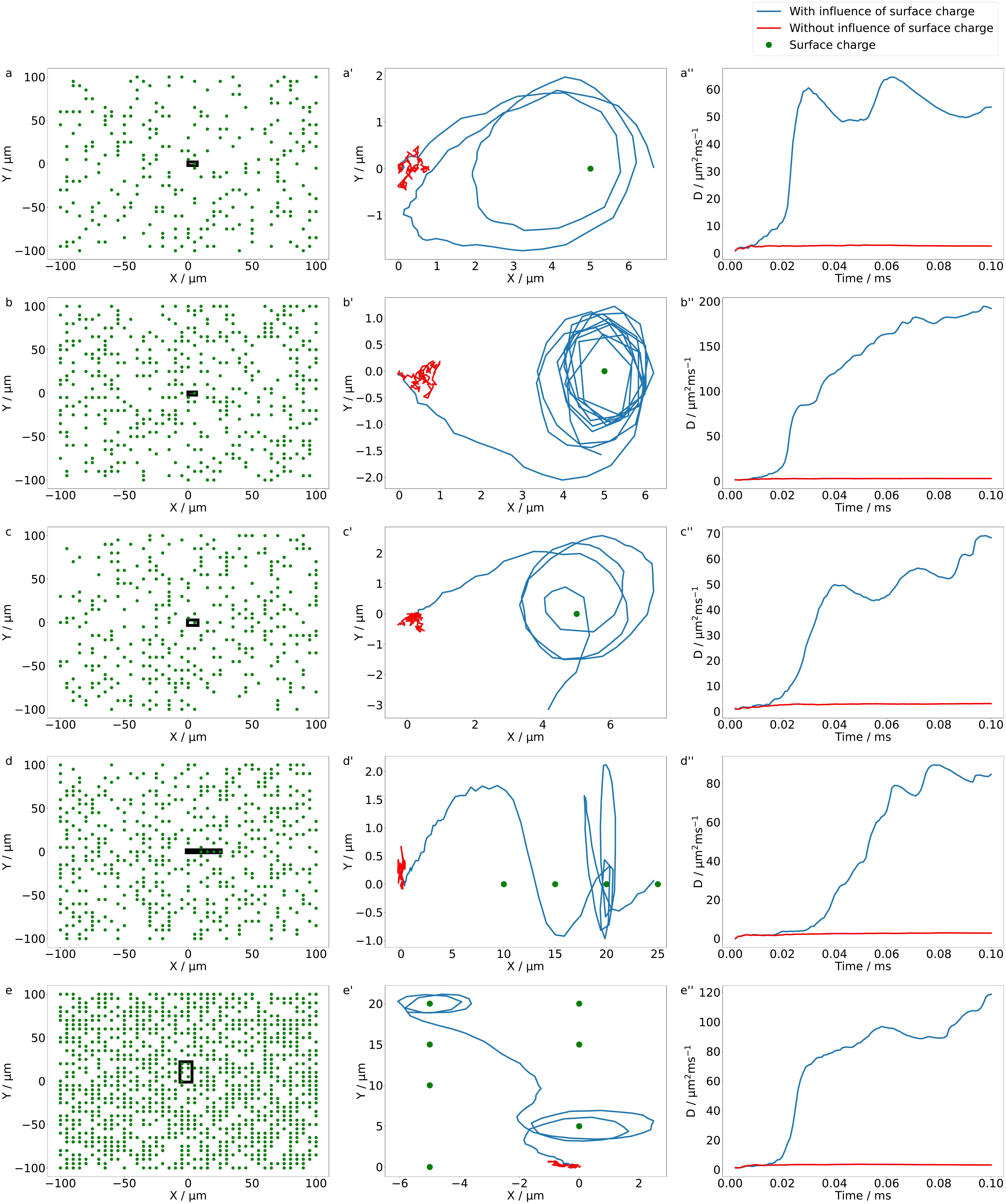}
    \caption{
    Single-molecule trajectories in the presence of single-electron surface charges in contrast to absence of them. 
    The circular trajectory occurred within specific black insets.
    (a) The first distribution consisted of 300 charges and (a') had corresponding trajectory plots and (a") diffusion coefficient plots. 
    (b) The second distribution consisted of 500 charges and (b') had corresponding trajectory plots and (b") diffusion coefficient plots. 
    (c) The third distribution also consisted of 500 charges and (c') had corresponding trajectory plots and (c") diffusion coefficient plots. 
    (d) The fourth distribution consisted of 600 charges and (d') had corresponding trajectory plots and (d") diffusion coefficient plots. 
    (e) The fifth distribution consisted of 1500 charges and (e') had corresponding trajectory plots and (e") diffusion coefficient plots.
    }
    \label{Figure:2}
\end{figure*}

\begin{figure*}[]
    \centering
    \includegraphics[width = 0.95\textwidth]{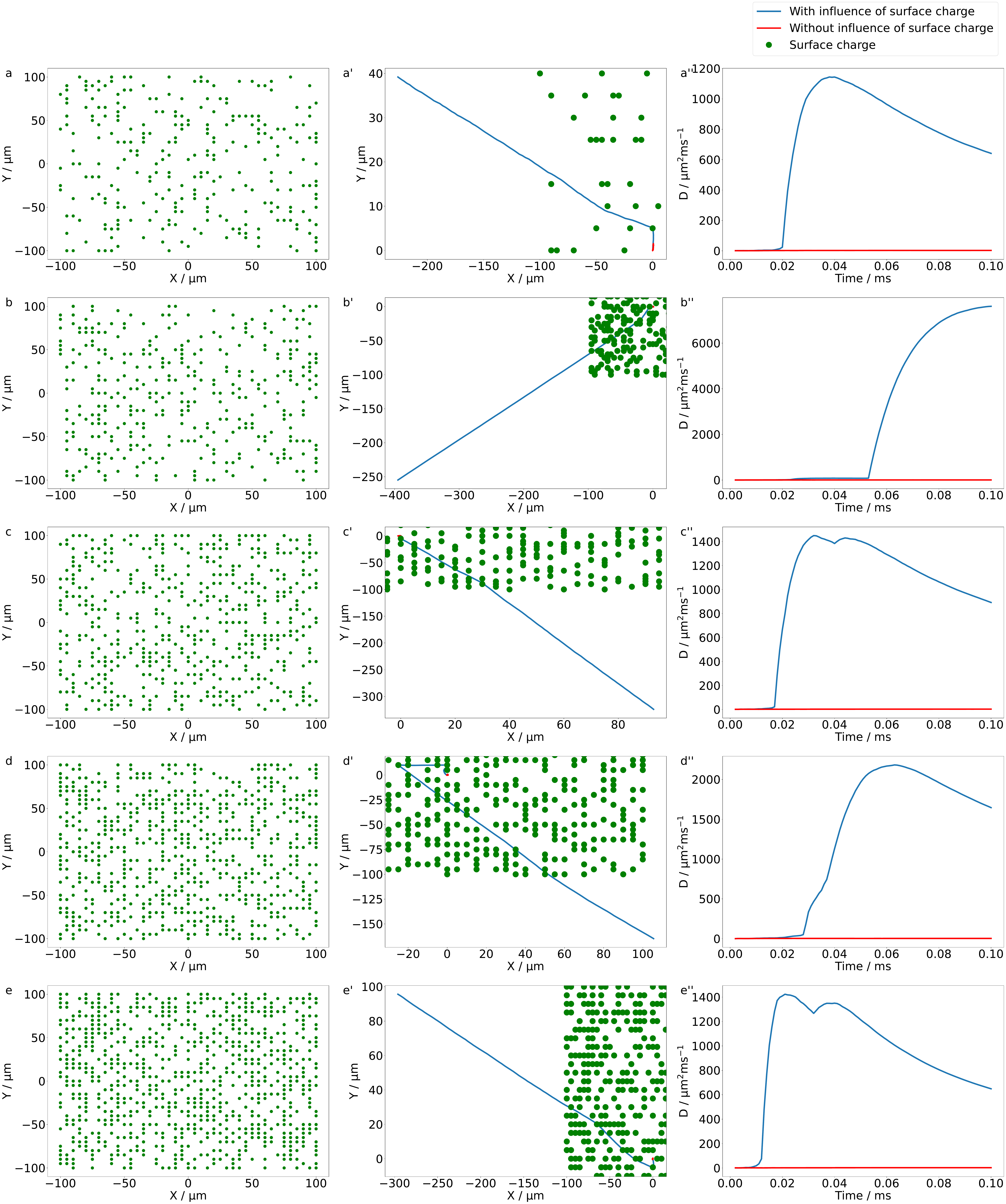}
    \caption{Single molecule trajectories in the presence and absence of surface charge influence.
    (a), First random surface charge distribution consisting of 300 charges.
    (a'), Trajectory plots corresponding to the first distribution.
    (a"), Diffusion coefficient plots corresponding to the first distribution.
    (b), Second random surface charge distribution consisting of 400 charges.
    (b'), Trajectory plots corresponding to the second distribution.
    (b"), Diffusion coefficient plots corresponding to the second distribution.
    (c), Third random surface charge distribution consisting of 600 charges.
    (c'), Trajectory plots corresponding to the third distribution.
    (c"), Diffusion coefficient plots corresponding to the third distribution.
    (d), Fourth random surface charge distribution consisting of 800 charges.
    (d'), Trajectory plots corresponding to the fourth distribution.
    (d"), Diffusion coefficient plots corresponding to the fourth distribution.
    (e), Fifth random surface charge distribution consisting of 1000 charges.
    (e'), Trajectory plots corresponding to the fifth distribution.
    (e"), Diffusion coefficient plots corresponding to the fifth distribution.}
    \label{Figure:3}
\end{figure*}

\begin{figure*}[]
    \centering
    \includegraphics[width = 0.95\textwidth]{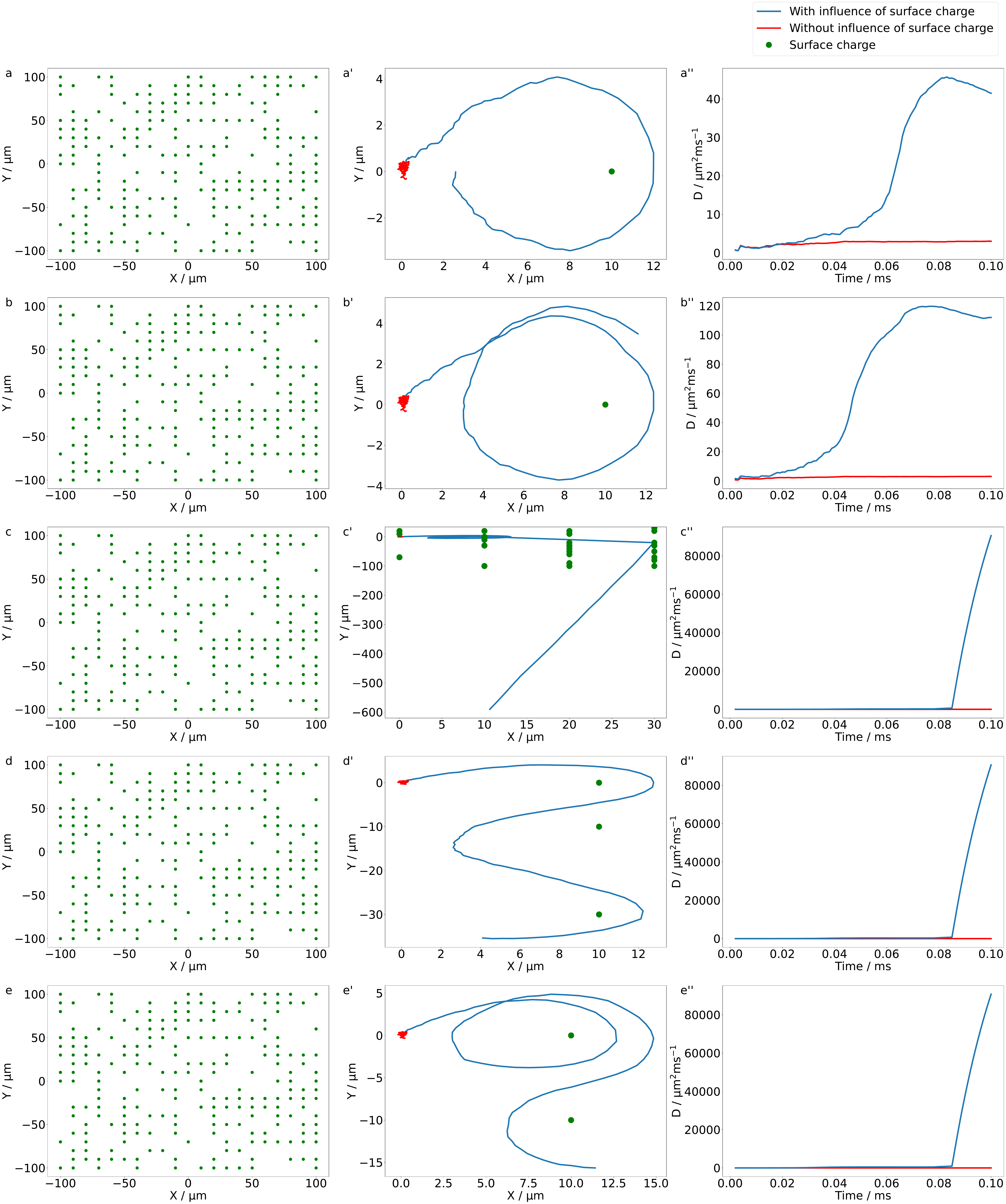}
    \caption{\textbf{Increased electrostatic field ---} single particle trajectories in the presence and absence of surface charge influence with varying surface charge magnitude and constant distribution of 300 charges.
    (a) The magnitude of each surface charge is $-1.6 \times 10^{-19}$ C.
    (a'), Trajectory plots corresponding to the first magnitude.
    (a"), Diffusion coefficient plots corresponding to the first magnitude.
    (b), The Magnitude of each surface charge is $-3.2 \times 10^{-19}$ C.
    (b'), Trajectory plots corresponding to the second magnitude.
    (b"), Diffusion coefficient plots corresponding to the second magnitude.
    (c), The Magnitude of each surface charge is $-4.8 \times 10^{-19}$ C.
    (c'), Trajectory plots corresponding to the third magnitude.
    (c"), Diffusion coefficient plots corresponding to the third magnitude.
    (d), The Magnitude of each surface charge is $-6.4 \times 10^{-19}$ C.
    (d'), Trajectory plots corresponding to the fourth magnitude.
    (d"), Diffusion coefficient plots corresponding to the fourth magnitude.
    (e), The Magnitude of each surface charge is $-8 \times 10^{-19}$ C.
    (e'), Trajectory plots corresponding to the fifth magnitude.
    (e"), Diffusion coefficient plots corresponding to the fifth magnitude.}
    \label{Figure:4}
\end{figure*}

\begin{figure*}[]
    \centering
    \includegraphics[width = 1\textwidth]{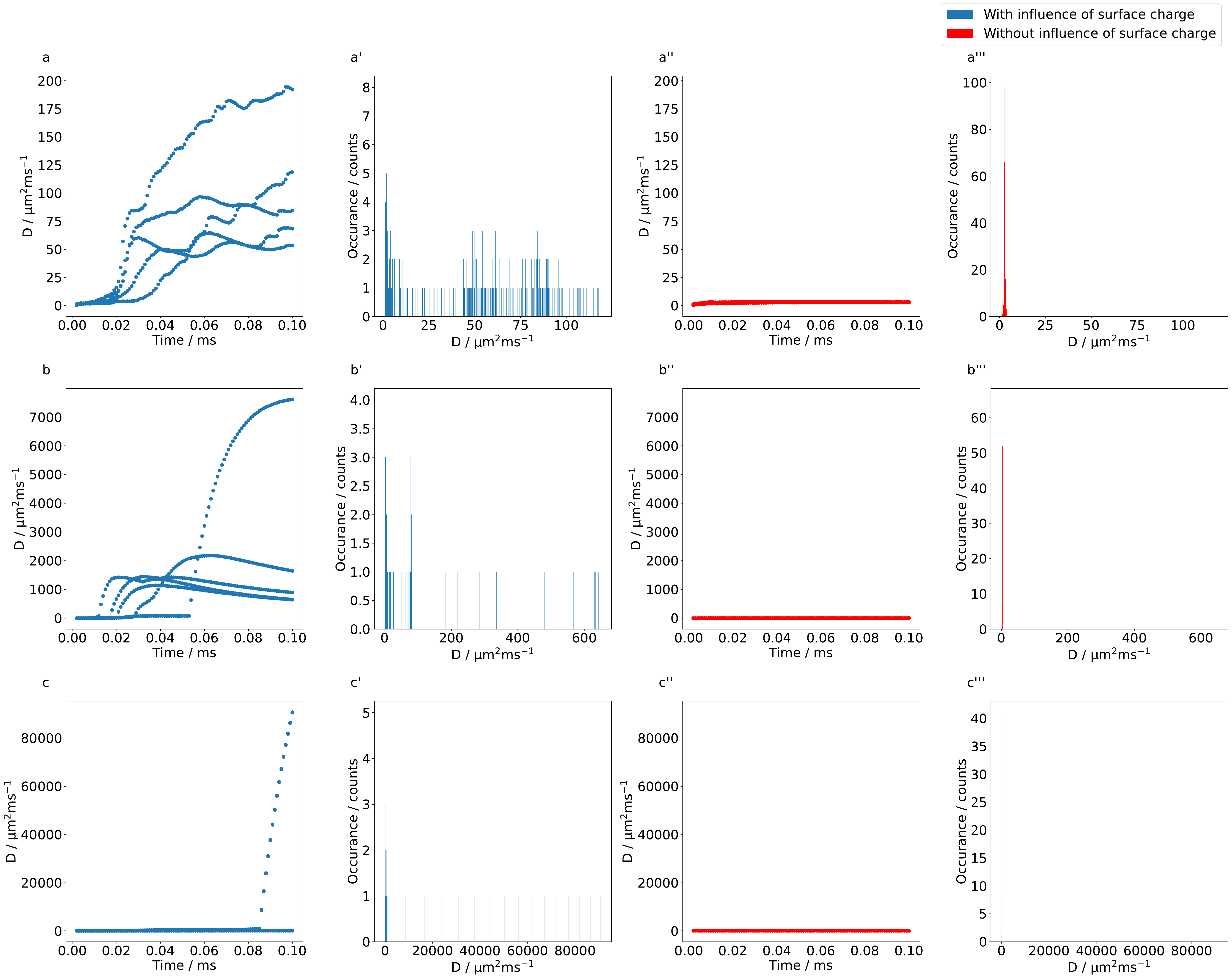}
    \caption{\textbf{Quantification of diffusion coefficients with and without influence of surface charges.} The width of each bar in the frequency distribution is 1 $\upmu\text{m}^2\text{ms}^{-1}$. 
    Behaviour of  circular trajectories --
    (a), All diffusion coefficients with influence of surface charge.  
    (a'), Frequency distribution of diffusion coefficients.
    (a''), All diffusion coefficients without influence of surface charge. 
    (a'''), Frequency distribution of diffusion coefficients.
    Behaviour of ballistic trajectories -- 
    (b), All diffusion coefficients with influence of surface charge.  
    (b'), Frequency distribution of diffusion coefficients.
    (b''), All diffusion coefficients without influence of surface charge. 
    (b'''), Frequency distribution of diffusion coefficients. 
    Behaviour of trajectories with increasing surface charge and constant non-surface charge influenced diffusion pattern and surface charge distribution
    (c), All diffusion coefficients with influence of surface charge.  
    (c'), Frequency distribution of diffusion coefficients.
    (c''), Diffusion coefficients without influence of surface charge. 
    (c'''), Frequency distribution of diffusion coefficients.}
    \label{Figure:5}
\end{figure*}

\begin{figure*}[]
    \centering
    \includegraphics[width = 1\textwidth]{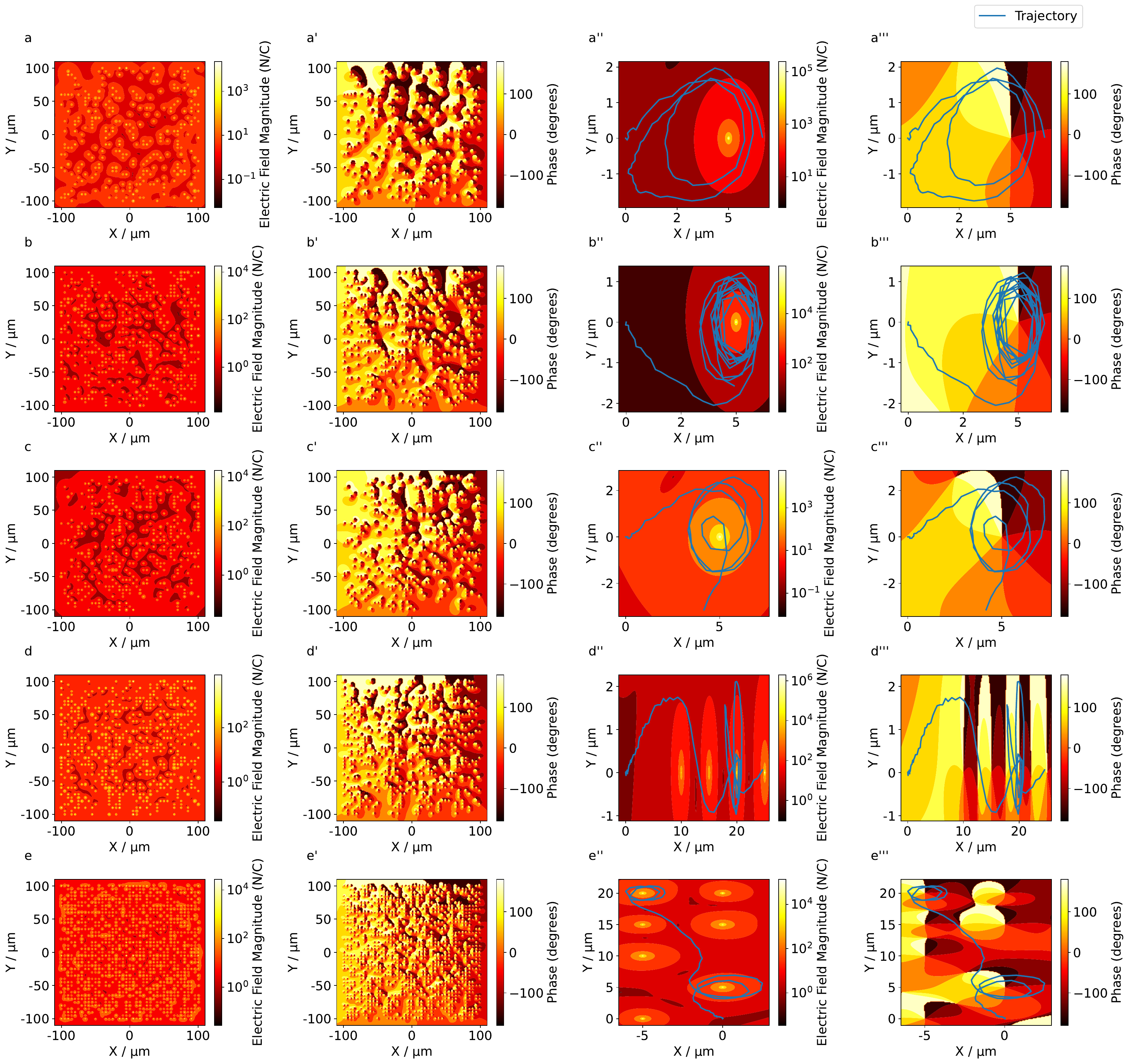}
    \caption{\textbf{Distribution of electric field across surface charge distributions that result in a circular trajectory.} Horizontally, first and second images represent the magnitude and phase of the electric field across the area of surface charge distribution respectively. Third and fourth images represent the magnitude and phase of the electric field across the area of trajectory respectively. (a) Random surface charge distribution consisting of 300 charges. (b) Random surface charge distribution consisting of 500 charges. (c) Random surface charge distribution consisting of 500 charges. (d) Random surface charge distribution consisting of 600 charges. (e) Random surface charge distribution consisting of 1500 charges.}
    \label{Figure:6}
\end{figure*}

\textit{Energy balance and non-equilibriumness.---}
To account for the time-dependent nature of the system, we introduce time derivatives into the analysis using PDEs. 
We focus on the diffusion of single molecules under the influence of electrostatic forces in a 2D plane. 
We use the following variables and parameters:
$\rho(x, y, t)$ as single-molecule density at position $(x, y)$ and time  $t$, 
$D(x, y, t)$ as diffusion coefficient, which may depend on position and time, $F_{es}(x, y, t)$ as electrostatic force acting on the single molecule. 
We start by writing the continuity equation, which describes the conservation of single molecules in the system.
\begin{equation}
    \frac{\partial\rho}{\partial t} + \nabla\cdot\left(-D\nabla\rho + \rho\mathbf{V}\right) = 0
\end{equation}
Here, $\mathbf{V} = \frac{F_{es}}{\zeta}$,  where  $\zeta$ is the friction coefficient for the single molecule, and $\nabla\cdot$ represents the divergence operator. 
This equation accounts for the change in single-molecule density over time due to diffusion and advection in the presence of electrostatic forces.
To describe the evolution of the electrostatic forces, we can use the Poisson equation for electrostatics:
\begin{equation}
    \nabla^2\phi = -\frac{\rho}{\epsilon_0}
\end{equation}
Here, $\phi(x, y, t)$ is the electrostatic potential, and  $\nabla^2$ is the Laplacian operator. The electrostatic force $F_{es}$ can be calculated as $F_{es} = -\nabla\phi$.
Together, these two equations form a system of coupled PDEs that describe the time evolution of the single-molecule density and the electrostatic forces in the system. 
The energy balance and entropy production are derived from these equations by considering the energy and entropy associated with the diffusing molecules and the electrostatic interactions.
The energy balance can be obtained by considering the time derivative of the total energy in the system, which includes the kinetic energy of the molecules, the electrostatic potential energy, and the energy dissipated as heat:
\begin{equation}
   \frac{\partial E}{\partial t} = \frac{\partial E_{conv}}{\partial t} + \frac{\partial E_{diss}}{\partial t} 
\end{equation}
Here, $E = \int(K_{mol} + U_{es})\rho dA$ is integrating over the 2D domain, $A$.
$E_{conv}$ represents the energy convected by the molecules as they move through the system under the influence of electrostatic forces. 
$E_{diss}$ represents the energy dissipated as heat due to friction and other dissipative processes in the system. 
To incorporate entropy production, we can use the following relation:
\begin{equation}
    \sigma = \frac{\partial S}{\partial t} - \int\left(\frac{1}{T}\right)\nabla\cdot\left(\frac{\dot{Q}}{T}\right) dA
\end{equation}
Here,  $S$ is the entropy of the system and $\dot{Q}$ is the heat dissipation rate.
Solving these PDEs and calculating the energy balance and entropy production provides insights into the time-dependent dynamics of the single molecules under the influence of electrostatic forces, including the circular and ballistic trajectories. 
Analytical solutions for this given system of PDEs are challenging to obtain due to the complexity and nonlinearity of the equations, especially when considering extensions such as external fields, molecular interactions, or nonlinear effects. 

\section{Results}
\textit{Single-Molecules at the interface of Single-Electrons.---}
We have numerically simulated single-molecule diffusion influenced by surface charges, comparing it to the reference case of diffusion without surface charges, depicted in blue and red respectively. Our simulations utilise a model particle resembling the behaviour of a charged Atto-488 molecule in a medium that constitutes a mixture of acetone and water in the ratio 1:1, the equivalent dynamic viscosity being $6 \times {10}^{-4}$ Pa s. The mass and charge of the particle are taken as $1.3350734 \times 10^{-23}$ kg and $-1.6 \times 10^{-19}$ C respectively. The surface charges have fixed positions and the initial diffusion coefficient of the particle is assumed to be 0.5907458 $\upmu\text{m}^2\text{ms}^{-1}$. 
This initial diffusion coefficient is used to calculate the range of displacement of the particle, and free diffusion is simulated by adding a random displacement within this specified range in both $x$ and $y$ directions. 
The time interval used is $10^{-6}$ seconds and the code is run for 100 time-steps. The calculations for the range of displacement are done using the Stokes-Einstein equation and the mean squared displacement formula. 
The limits of the range are set by the magnitudes of the displacements with time steps one order of magnitude lesser and one order of magnitude greater than the timestep.
Two separate trajectories are generated for a single run, one showing the influence of the surrounding surface charges and the second without their influence. The dynamics observed in these simulations show how surface charges can dramatically impact molecular diffusion. The simulated system is not in equilibrium, since the surface charges are stationary despite the action of electrostatic forces, implying an external force application keeps them stationary.

To initiate our simulations, particles consistently start from the origin (0 $\upmu$m, 0 $\upmu$m). Two different particle classes are created for the particle being observed and the surface charges.
This serves as our baseline scenario for further exploration. We introduce a key element of variability by considering different random arrangements of surface charges. These charges are randomly placed as defect lattices, ensuring they are spaced more than 10 micrometers apart. The lattice is represented by a grid of dimensions 200 $\upmu$m $\times$ 200 $\upmu$m. 300 surface charges are assigned random positions in the grid, with each run of the code having a different orientation of the surface charges. The specific configuration of surface charges plays a pivotal role in determining the subsequent particle trajectories. Four separate position arrays are generated, two which store the ${x}$ and ${y}$ positions of the particle with the influence of surrounding electrostatic particles, and two which store the positions without the influence. 
The surface charges are assumed to be stationary throughout the run. The free diffusion displacement is a random float value in the range for both directions. This displacement is added to both sets of arrays for every interval.
For every interval, the electrostatic forces on the particle due to each surface charge are calculated based on the distance between them, and they're split into corresponding ${x}$ and ${y}$ components. The force components are arithmetically added, and this generates the net total force on the particle for the duration of the interval. Subsequently, the acceleration is calculated, and the displacement of the particle in both directions is obtained.
These displacements are only added to the first set of arrays to display the effect of surface charges.
The calculation of diffusion coefficient is done in a similar way as it has been done for the change-point detection algorithm and the values are stored in two arrays, one which considers the influence of surface charges and one which does not consider their influence. One major difference is that the initial diffusion coefficient for all simulation runs is 0.5907458 $\upmu\text{m}^2\text{ms}^{-1}$ to obtain a non-zero initial diffusion coefficient.
The position and diffusion arrays are then used to generate trajectory plots and diffusion coefficient versus time plots respectively. 

\textit{Presence and absence of surface charge influence.---}
The presence and absence of surface charge influence are examined through the analysis of single-molecule trajectories. 
The black rectangles in the diagram serve as visual representations of the area where the circular trajectory occurs. 
In the first instance shown in Figure 2 (a), a random surface charge distribution is illustrated, which consists of a total of 300 charges. 
The trajectory plots corresponding to this distribution are shown, providing a visual representation of the particle's movement under the influence of the surface charges. 
Additionally, diffusion coefficient plots are included (Figure 2 (a')), which demonstrate the relationship between the surface charge distribution and the molecule's $D$ increased from 2 $\upmu\text{m}^2/\text{ms}$ to 50 $\upmu\text{m}^2/\text{ms}$ and 60 $\upmu\text{m}^2/\text{ms}$. 
We observed two peaks and two valleys in $D$, which correlates with the two rotations observed in Figure 2 (a")
In the second random surface charge distribution in Figure 2 (b), this distribution consists of 500 charges.
Similar to the first distribution, trajectory plots corresponding to this distribution are provided, enabling a comparison of particle movement between the two distributions. 
$D$ is impacted significantly by the single charge with an incremental response as it continues to perform multiple rotations until 180 $\upmu\text{m}^2/\text{ms}$.
Continuing the same analysis, the third random surface charge distribution is introduced, which once again consists of 500 charges as shown in Figure 2 (c). Trajectory plots corresponding to this distribution show similar circular rotation (Figure 2 (c')) but with lower number of turns $D$ variations with lower number of peaks and valleys with maximum attainment of 68 $\upmu\text{m}^2/\text{ms}$ within the same time duration of 0.1 ms.
Increasing the random surface charge distribution comprising of a total of 600 charges (Figure 2(d)), trajectory plots starting loop around multiple charges, facilitating a comparison of the particle's movement in this scenario with the previous distributions (Figure 2(d')). $D$ variations show a similar response as before (Figure 2(d")) peaks and valleys with maximum attainment of 85 $\upmu\text{m}^2/\text{ms}$.
Lastly, we increased a much higher number of charges, specifically 1500 as shown in Figure 2 (e). 
Figure 2 (e') shows that the particle's movement differs significantly when subjected to a greater number of surface charges - it loops from one single charge to another single charge. 
Furthermore, $D$ reaches 118 $\upmu\text{m}^2/\text{ms}$ with a diverse changes with prominent peaks at 65 $\upmu\text{m}^2/\text{ms}$ and 90 $\upmu\text{m}^2/\text{ms}$,.
These reveal a diverse range of single-molecule trajectories. 
The circular trajectories in Figure \ref{Figure:2} for different charge distributions are always accompanied by multiple diffusion change-points with remarkable two-order-of-magnitude change in $D$.

We also observe ballistic single-molecule trajectories as shown in Figure \ref{Figure:3}. 
In these cases, the $D$ exhibits significant changes, sometimes even spanning three orders of magnitude.
The effect of the magnitude of surface charge on trajectories in a similar distribution is visualised in Figure \ref{Figure:4}. 
An increase in the magnitude of the surface charge also increases the range of diffusion coefficients, as well as leads to varying trajectories.
The arrangement of charges, as well as the resulting trajectories and associated changes in diffusion coefficients, highlight the rich dynamics at play in these systems. 
Figure \ref{Figure:5} presents an overview of diffusion coefficients obtained from six distinct surface charge arrangements, clearly demonstrating the substantial influence of these charges on particle diffusion. 
To better understand the statistical distribution of these coefficients, Figure \ref{Figure:5}(a) and Figure \ref{Figure:5}(c) plot all diffusion coefficients for circular and ballistic trajectories respectively and Figure \ref{Figure:5}(a')  and Figure \ref{Figure:5}(c') provide the corresponding frequency distribution, shedding light on the variability in diffusion behaviour attributed to different surface charge configurations.
In stark contrast, Figure \ref{Figure:5}(b) and Figure \ref{Figure:5}(d) showcase the diffusion coefficients for the same particles, but in this case, without the presence of surface charges. Complementing this, Figure \ref{Figure:5}(b') and Figure \ref{Figure:5}(d') offer the frequency distribution of these coefficients in the absence of surface charge effects.
This agreement validates the accuracy of our simulation model in capturing real-world behaviour. 
Conversely, the introduction of surface charges leads to a striking contrast, with the diffusion coefficients experiencing a three order-of-magnitude increase. 
Figure \ref{Figure:6} presents the magnitude and phase contour plots of the electric field generated by different surface charge distributions. 
The total number of elementary surface charges present in \ref{Figure:6}(a)-(e) are 300, 500, 500, 600 and 1500 respectively. Figures \ref{Figure:6}(a)-(e) display the electric field magnitude due to surface charge, and areas with high magnitude denote close proximity to a surface charge. Figures \ref{Figure:6}(a')-(e') show the phase distribution of the net electric field. Figures \ref{Figure:6}(a")-(e") show us the electric field magnitude in the area traversed by the molecule, and it can be observed that the orbits occur near a surface charge. Figures \ref{Figure:6}(a"')-(e"') show us the electric field phase in the traversed area. The trajectories of the molecule being tracked are shown along with the contour plots, and it can be observed that the orbits tend to occur in close proximity to a surface charge. It can also be observed that multiple orbits can occur in a single trajectory, as shown in Figure \ref{Figure:6}(e").

\section{Discussion}
The motivation of our investigation lies in the interplay between single molecules and electrostatic forces arising from surface charges forming a lattice structure. 
Our work has shown how these forces play a pivotal role in shaping the trajectories of diffusing molecules.
The distinction made between circular and ballistic trajectories highlights the varying influence of electrostatic forces on single molecular motion. 
The ability to manipulate the diffusion coefficients through precise control of elementary charges at the quantum level opens up exciting possibilities for tailoring single-molecule motion. 
The connection established between theoretical concepts such as state-dependent diffusion, effective temperatures, and the quasi-fluctuation-dissipation theorem adds further conditions to the study. 
We emphasise the importance of understanding and characterising non-equilibrium systems, where traditional descriptions of temperature and diffusion may not suffice.
In the case of circular trajectories, the diffusing molecule closely emulates the behaviour of celestial bodies orbiting a central star. 
Surface charges act as the central force (similar to gravitational pull) sustaining the molecule in a circular path. 
This equilibrium (resembling the gravitational balance in planetary dynamics) is a prominent feature of these circular trajectories.
In contrast, ballistic trajectories bear a striking resemblance to the phenomenon of gravity assist \cite{dyson2023gravitational}. 
Molecules undergoing ballistic trajectories gain momentum through interactions with elementary surface charges, resulting in high-speed motion reminiscent of gravity-assisted acceleration \cite{hadid2024bepicolombo}. 
This parallel underscores the non-equilibrium nature of the system, akin to systems involving self-propelled particles in active matter.
Our numerical analysis unambiguously establishes a direct correlation between elementary surface charge density and the magnitude of diffusion coefficients. 
Elevated surface charge densities engender stronger electrostatic forces, thereby inducing more pronounced trajectory alterations and magnifying diffusion coefficients. 
This relationship is particularly noteworthy when juxtaposed with the typical diffusion coefficient of a molecule, which hovers around 2.5 $\upmu\text{m}^2\text{ms}^{-1}$.
Of paramount significance is our revelation of quantum mechanical control over single-molecule motion. 
Precise manipulation of elementary charges at the quantum level amplifies the influence of surface charges on molecular trajectories, providing a potent tool for tailoring single-molecule motion with unparalleled precision.
In the context of our earlier theoretical exploration, particularly within the realms of state-dependent diffusion \cite{lau2007state} and effective temperatures \cite{golestanian2015enhanced}, our findings offer profound insights. 
The alterations in molecular trajectories wrought by surface charges underscore the non-equilibrium nature of the system. 
The injected energy through electrostatic interactions leads to deviations from equilibrium-based descriptions of temperature and diffusion. 
Instead, we observe the emergence of effective temperatures that accurately capture these non-equilibrium deviations, aligning seamlessly with the quasi-fluctuation-dissipation theorem.
Our study adeptly bridges the chasm between theoretical foundations and experimental observations, highlighting the profound impact of electrostatic forces on single-molecule trajectories. 
These insights deepen our comprehension of molecular motion, with diffusion coefficients ranging up to 200 $\upmu\text{m}^2\text{ms}^{-1}$ for circular trajectories and up to 6000 $\upmu\text{m}^2\text{ms}^{-1}$ for ballistic trajectories, presenting a stark departure from the typical experimental diffusion coefficient of around 0.4 $\upmu\text{m}^2\text{ms}^{-1}$ for these molecules.  
In certain cases, diffusion coefficient goes as high as 80000 $\upmu\text{m}^2\text{ms}^{-1}$ -- for a single molecule, this is remarkably high and could have significant implications in various scientific fields since it can be achieved and controlled effectively according to our findings. 
To put this into context, the diffusion coefficient for most molecules in aqueous solutions is typically on the order of 10-100 $\upmu\text{m}^2\text{ms}^{-1}$, and even for fast-diffusing molecules, it rarely exceeds 1000 $\upmu\text{m}^2\text{ms}^{-1}$.
Atomically smooth surfaces, such as those found in high-quality graphene and hBN, can potentially facilitate enhanced molecular diffusion by reducing surface roughness and providing a low-friction environment for molecules to move. 
These materials also possess unique electronic properties that may enable efficient manipulation of electrostatic forces to control molecular motion.
The realisation of ultra-fast molecular diffusion, characterised by exceptionally high diffusion coefficients, has the potential to revolutionise various aspects of semiconductor technology and 2D materials research. 
In the semiconductor industry, achieving large diffusion coefficients could enable faster, more precise fabrication of nanoscale structures, ultimately pushing the boundaries of device scaling and performance. 
By harnessing the fundamental mechanisms behind ultra-fast diffusion, we can develop novel materials or processing techniques that optimise molecular transport for advanced semiconductor applications.
Similarly, in 2D materials, understanding and controlling molecular diffusion will generate new possibilities for controlling and enhancing material properties. 
This can lead to breakthroughs in the performance and functionality of devices incorporating 2D materials, such as electronics, energy storage, and sensing applications.

Authors' contribution: 
MG and SG developed the concept. 
MG's earlier works on atomic-scale electromechanics conceptualised the electrostatic interactions in this work. 
SG's earlier works on single-molecule's coupled with MG's work developed this interdisciplinary approach.
SC's earlier work on atomic force microscopy and surface probe microscopy influenced this research.
DNV performed the numerical computation. 
KVC participated in the research discussions.
JRH and SG have worked on the change-point algorithm and the earlier experiment in Cambridge, which has led to this work.
MG, DNV, SC, and SG have written the paper.
SG has worked on the optical setup.    
GMR, MG, and SG brought funding and organised the facility for this research at IISc Bangalore and led the research. 
SC contributed with critical inputs on single molecule motions. 
SG has developed the theory and supervised research.
All authors contributed to the writing of the paper.
    
Acknowledgement: 
International Center for Nanodevices (ICeNd) acknowledges the funding from Honeywell for this research. 
DNV and KVC are supported by ICeNd. 
JRH is supported by the EPSRC Sensor CDT, University of Cambridge.
SG is grateful to the German Research Foundation and Isaac Newton Trust for funding and The Royal Society for endorsement.
ICeNd is thankful to the Indian Nano User Program --- Indian Nanoelectronics Users Program (INUP-i2i) funding from the Ministry of Electronics and Information Technology, the Union Government of the Republic of India awarded by IISc Bangalore. 
SG, MG, and SC are thankful to Brian Corbett for facilitating their intellectual collaborations through Maxapiens.

\bibliography{Ref}

\clearpage
\newpage

\section{SUPPORTING INFORMATION}

\section{Simulation of Diffusion of Single Molecules with Surface Charge}
We simulated single-molecule diffusion influenced by surface charges, comparing it to the reference case of diffusion without surface charges, depicted in blue and red respectively. Our simulations utilise a model particle resembling the behaviour of a charged Atto-488 molecule in a medium which constitutes a mixture of acetone and water in the ratio 1:1, the equivalent dynamic viscosity being $6 \times {10}^{-4}$ Pa s. The mass and charge of the particle is taken as $1.3350734 \times 10^{-23}$ kg and $-1.6 \times 10^{-19}$ C respectively. The surface charges have fixed positions and the initial diffusion coefficient of the particle is assumed to be 0.5907458 $\upmu\text{m}^2\text{ms}^{-1}$. 
This initial diffusion coefficient is used to calculate the range of displacement of the particle, and free diffusion is simulated by adding a random displacement within this specified range in both $x$ and $y$ directions. 
The time interval used is $10^{-6}$ seconds and the code is run for 100 time-steps. The calculations for the range of displacement is done using the the Stokes-Einstein equation and the mean squared displacement formula. 
The limits of the range are set by the magnitudes of the displacements with time-steps one order of magnitude lesser and one order of magnitude greater than the time-step.
\begin{verbatim}
time_step = 1e-6
ood1 = (2 * D * (time_step / 10)) ** 0.5
ood2 = (2 * D * (time_step * 10)) ** 0.5
\end{verbatim}
Two separate trajectories are generated for a single run, one showing the influence of the surrounding surface charges and the second without their influence. The dynamics observed in these simulations show how surface charges can dramatically impact molecular diffusion.

To initiate our simulations, particles consistently start from the origin (0 $\upmu$m, 0 $\upmu$m). Two different particle classes are created for the particle being observed and the surface charges.
\begin{verbatim}
class Particle:
    def __init__(self, mass, charge, position1, 
    position2, velocity):
        self.mass = mass
        self.charge = charge
        self.position1 = position1
        self.position2 = position2
        self.velocity = velocity
        self.acceleration = np.zeros(2)

class SCharge:
    def __init__(self, charge, position):
        self.charge = charge
        self.position = position
\end{verbatim}
This serves as our baseline scenario for further exploration. We introduce a key element of variability by considering different random arrangements of surface charges. These charges are randomly placed as defect lattice, ensuring they are spaced more than 10 micrometers apart. The lattice is represented by a grid of dimensions 200 $\upmu$m $\times$ 200 $\upmu$m. 300 surface charges are assigned random positions in the grid, with each run of the code having a different orientation of the surface charges. The specific configuration of surface charges plays a pivotal role in determining the subsequent particle trajectories. Four separate position arrays are generated, two which store the ${x}$ and ${y}$ positions of the particle with the influence of surrounding electrostatic particles, and two which store the positions without the influence. 
\begin{verbatim}
x1 = [0] * (steps + 1)
y1 = [0] * (steps + 1)
x2 = [0] * (steps + 1)
y2 = [0] * (steps + 1)
\end{verbatim}
The surface charges are assumed to be stationary throughout the run. The free diffusion displacement is a random float value in the range for both directions. This displacement is added to both sets of arrays for every interval.
\begin{verbatim}
diffx1 = np.random.uniform(ood1, ood2)
diffx2 = np.random.uniform(-ood2, -ood1)
diffx = random.choice([diffx1, diffx2])
diffy1 = np.random.uniform(ood1, ood2)
diffy2 = np.random.uniform(-ood2, -ood1)
diffy = random.choice([diffy1, diffy2])
diff = np.random.uniform(-1, 1, (2,))
diff[0] = diffx
diff[1] = diffy
particle.position1 += diff
particle.position2 += diff
\end{verbatim}
For every interval, the electrostatic forces on the particle due to each surface charge are calculated based on the distance between them, and they're split into corresponding ${x}$ and ${y}$ components. The force components are arithmetically added, and this generates the net total force on the particle for the duration of the interval. Subsequently, the acceleration is calculated, and the displacement of the particle in both directions is obtained using Newton's laws of motion. These displacements are only added to the first set of arrays to display the effect of surface charges.
\begin{verbatim}
r = other_particle.position - particle.position1
distance = np.linalg.norm(r)
force_magnitude = (k * particle.charge * 
    other_particle.charge) / distance**2
force_direction = r / distance
particle.acceleration += (force_magnitude 
/ particle.mass) * force_direction
particle.velocity += particle.acceleration * 
    time_step
particle.position1 += particle.velocity * 
    time_step
\end{verbatim}
The calculation of diffusion coefficient is done in a similar way as it has been done for the change-point detection algorithm and the values are stored in two arrays, one which considers the influence of surface charges and one which does not consider their influence. One major difference is that the initial diffusion coefficient for all simulation runs is 0.5907458 $\upmu\text{m}^2\text{ms}^{-1}$ to obtain a non-zero initial diffusion coefficient.
\begin{verbatim}
dxr[count10][0] = x1[count1] - x1[count1 - 1]
dyr[count10][0] = y1[count1] - y1[count1 - 1]
dxr[count10][1] = x2[count1] - x2[count1 - 1]
dyr[count10][1] = y2[count1] - y2[count1 - 1]

for count in range(1, steps + 1):
    dx0 = 0
    dy0 = 0
    dx1 = 0
    dy1 = 0
    for count1 in range(count + 1):
        dx0 += dxr[count1][0]
        dy0 += dyr[count1][0]
        dx1 += dxr[count1][1]
        dy1 += dyr[count1][1]
    vhatxr[count][0] = dx0 / (count * time_step)
    vhatyr[count][0] = dy0 / (count * time_step)
    vhatxr[count][1] = dx1 / (count * time_step)
    vhatyr[count][1] = dy1 / (count * time_step)
    varx1[count][0] = (dxr[count][0] - 
        vhatxr[count][0] * time_step) ** 2
    vary1[count][0] = (dyr[count][0] - 
        vhatyr[count][0] * time_step) ** 2
    varx1[count][1] = (dxr[count][1] - 
        vhatxr[count][1] * time_step) ** 2
    vary1[count][1] = (dyr[count][1] - 
        vhatyr[count][1] * time_step) ** 2

for count in range(1, steps + 1):
    for count1 in range(count + 1):
        varx2[count][0] += varx1[count1][0]
        vary2[count][0] += vary1[count1][0]
        varx2[count][1] += varx1[count1][1]
        vary2[count][1] += vary1[count1][1]
    Dhatxr[count][0] = varx2[count][0] / 
        (2 * count * time_step)
    Dhatyr[count][0] = vary2[count][0] / 
        (2 * count * time_step)
    Dhatxr[count][1] = varx2[count][1] / 
        (2 * count * time_step)
    Dhatyr[count][1] = vary2[count][1] / 
        (2 * count * time_step)
    Dhat[count][0] = np.sqrt(Dhatxr[count][0] ** 2
        + Dhatyr[count][0] ** 2)
    Dhat[count][1] = np.sqrt(Dhatxr[count][1] ** 2
        + Dhatyr[count][1] ** 2)
\end{verbatim}
The position and diffusion arrays are then used to generate trajectory plots and diffusion coefficient versus time plots respectively. 

\section{Dynamic Equations}
The electrostatic interaction between charged particles in the presence of Brownian motion can be described using the Smoluchowski equation, which combines electrostatic forces and Brownian motion. The Smoluchowski equation is a fundamental equation which describes the time evolution of the probability distribution function of particle positions.
\begin{equation}
    \frac{\partial P(\vec{r}, t)}{\partial t} = D\nabla(\frac{1}{k_{B}T}\nabla P(\vec{r}, t) + \frac{\vec{F}}{k_{B}T}P(\vec{r}, t))
\end{equation}
where $P(\vec{r},t)$ is the probability distribution function for the position of the particle at position $r$ at time $t$, $D$ is the diffusion coefficient, $k_{B}$ is Boltzmann's constant, $T$ is the temperature and $\vec{F}$ is the net force acting on the particle, which can be described using Coulomb's law for electrostatic interactions and Brownian motion.
\begin{equation}
    \vec{F} = q\vec{E} + \vec{F}_{Brownian}
\end{equation}
\begin{equation}
    F_{Brownian} = \sqrt{2Dk_{B}T}\xi(t)
\end{equation}
where $q$ is the charge of the particle, $\vec{E}$ is the net electric field and $\xi(t)$ is a stochastic force with zero mean which satisfies the fluctuation-dissipation theorem.

\begin{figure*}[]
    \centering
    \includegraphics[width = 1\textwidth]{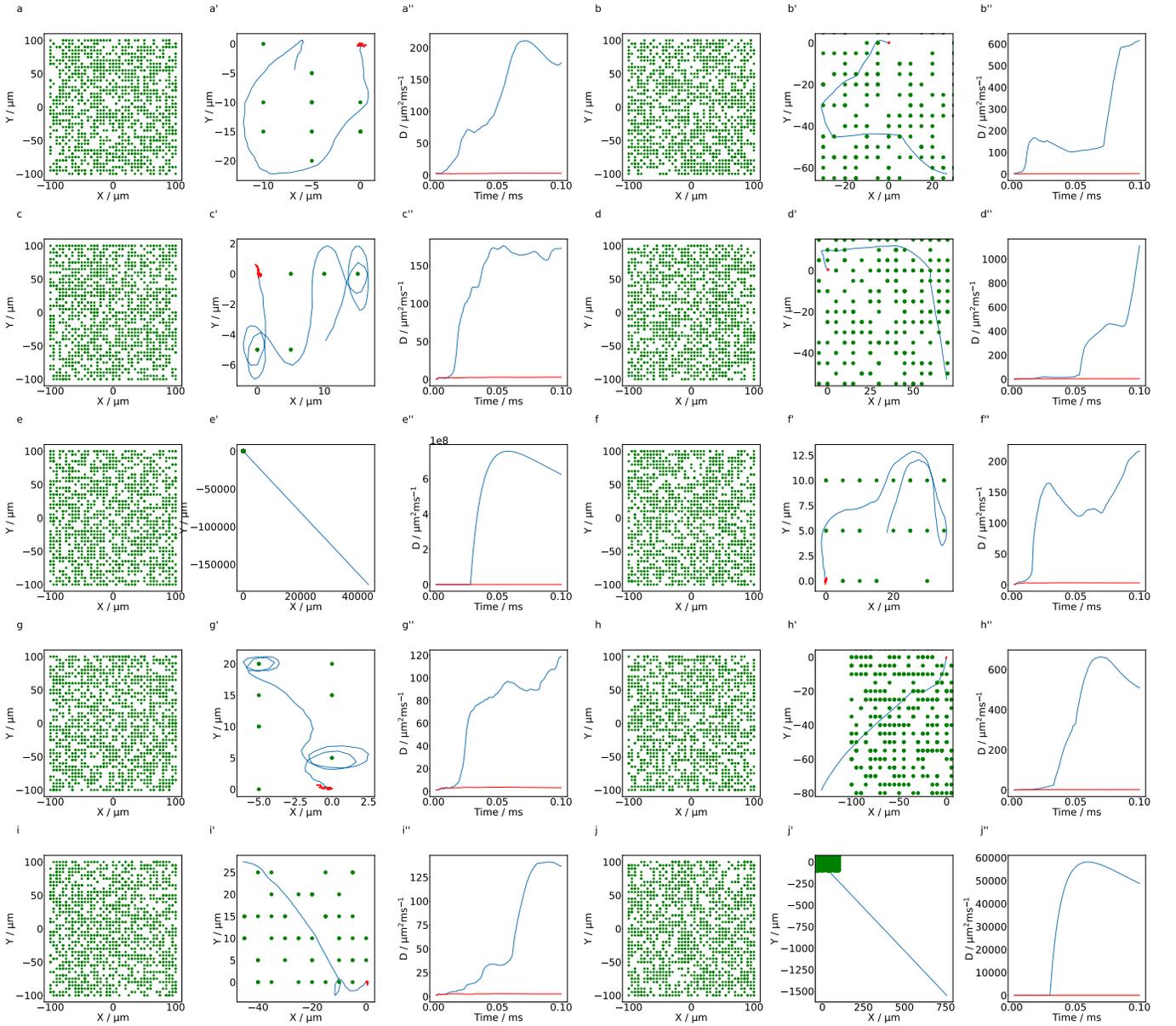}
    \caption{Simulated diffusion of a charged Atto-488-like single molecule under the influence of randomly positioned 1500 surface charges (blue), compared to diffusion without surface charges (red), with details shown in insets. The particles always start at the origin (0 $\upmu$m, 0 $\upmu$m) and the minimum distance between two surface charges is 5 $\upmu$m.
    (a - j), surface charge distribution.
    (a' - j'), diffusion trajectories.
    (a" - j"), their respective diffusion coefficient plots.}
    \label{Figure:1}
\end{figure*}

\begin{figure*}[]
    \centering
    \includegraphics[width = 1\textwidth]{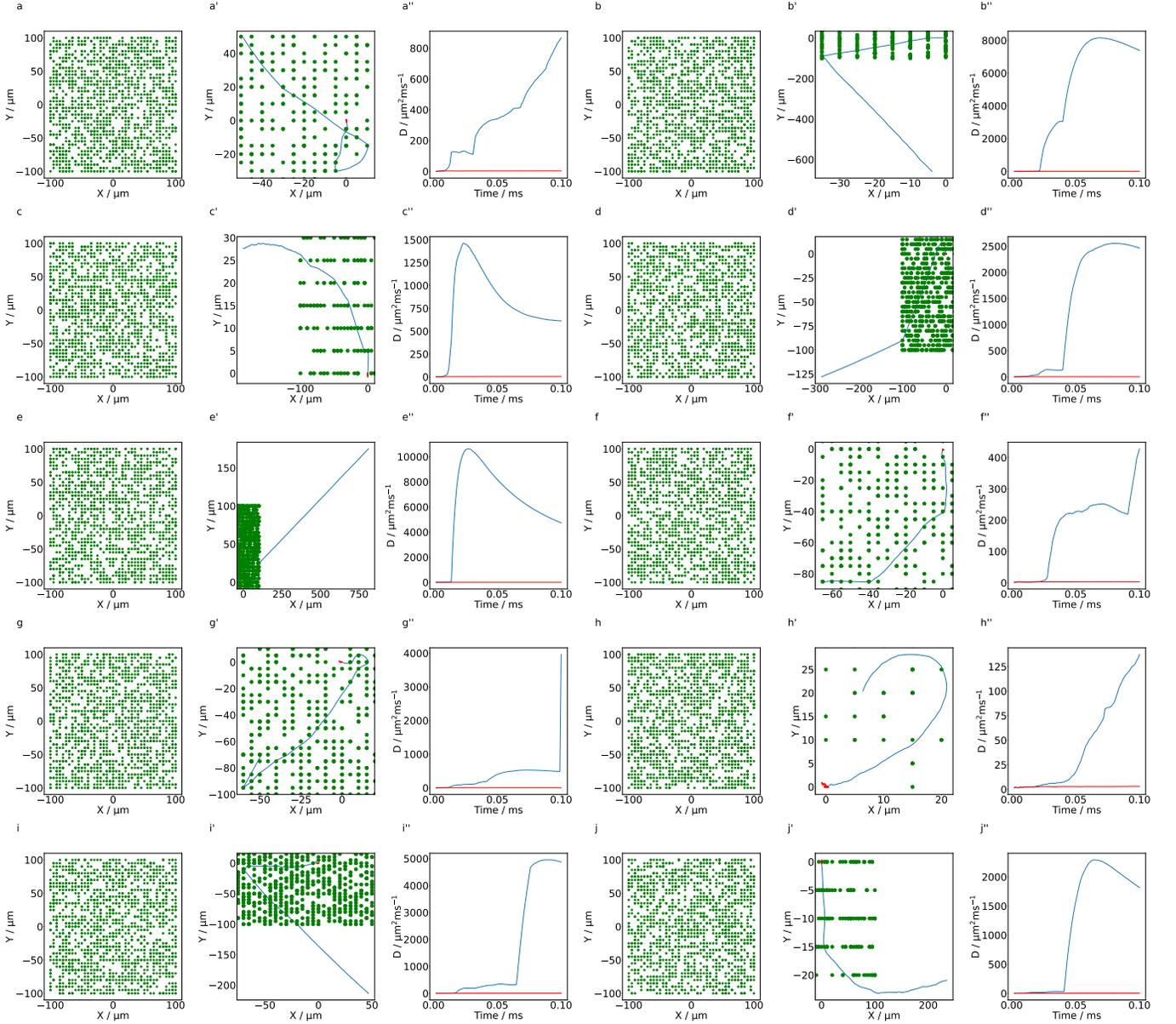}
    \caption{Simulated diffusion of a charged Atto-488-like single molecule under the influence of randomly positioned 1500 surface charges (blue), compared to diffusion without surface charges (red), with details shown in insets. The particles always start at the origin (0 $\upmu$m, 0 $\upmu$m) and the minimum distance between two surface charges is 5 $\upmu$m.
    (a - j), surface charge distribution.
    (a' - j'), diffusion trajectories.
    (a" - j"), their respective diffusion coefficient plots.}
    \label{Figure:2}
\end{figure*}

\begin{figure*}[]
    \centering
    \includegraphics[width = 1\textwidth]{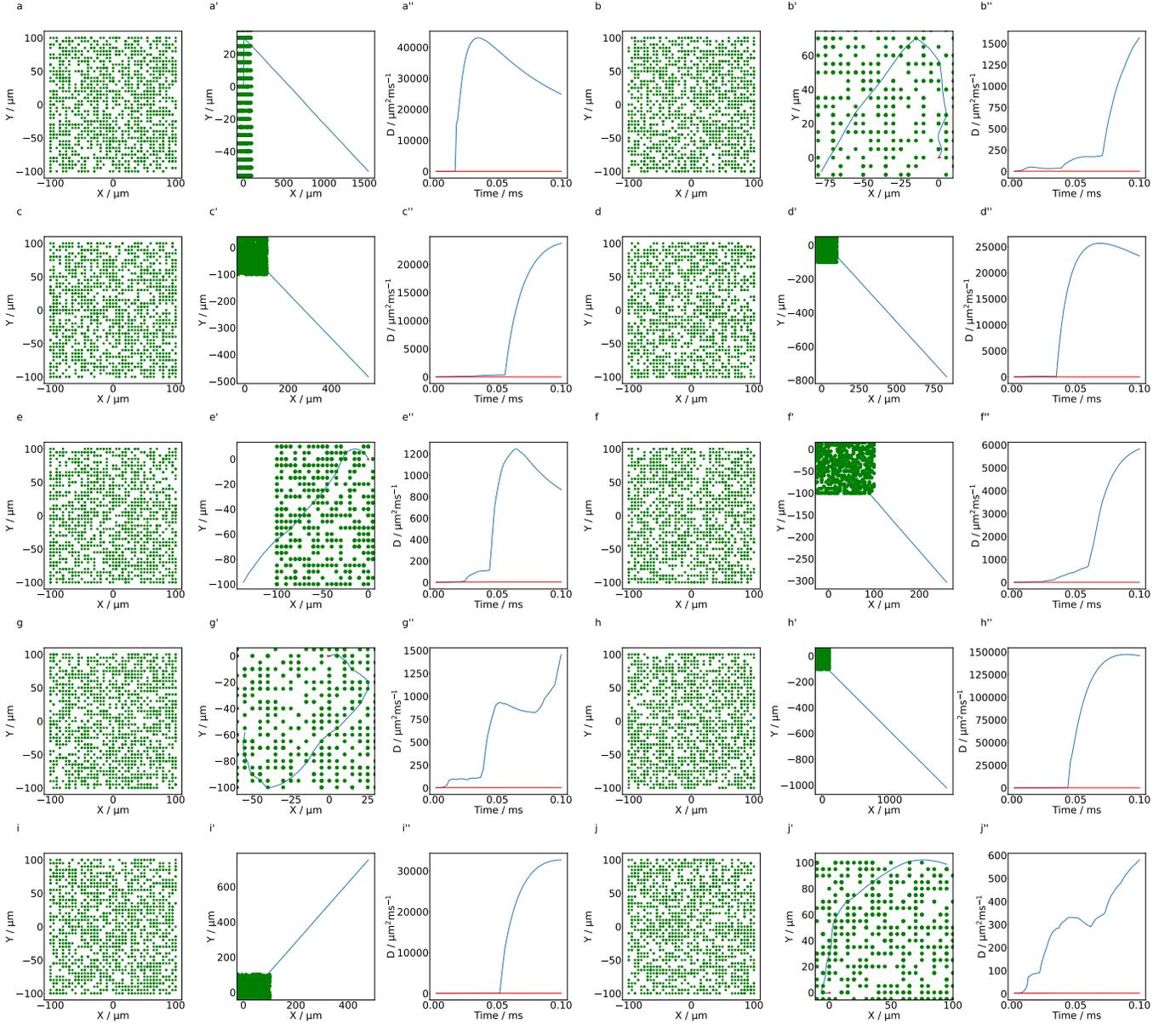}
    \caption{Simulated diffusion of a charged Atto-488-like single molecule under the influence of randomly positioned 1500 surface charges (blue), compared to diffusion without surface charges (red), with details shown in insets. The particles always start at the origin (0 $\upmu$m, 0 $\upmu$m) and the minimum distance between two surface charges is 5 $\upmu$m.
    (a - j), surface charge distribution.
    (a' - j'), diffusion trajectories.
    (a" - j"), their respective diffusion coefficient plots.}
    \label{Figure:S3}
\end{figure*}

\begin{figure*}[]
    \centering
    \includegraphics[width = 1\textwidth]{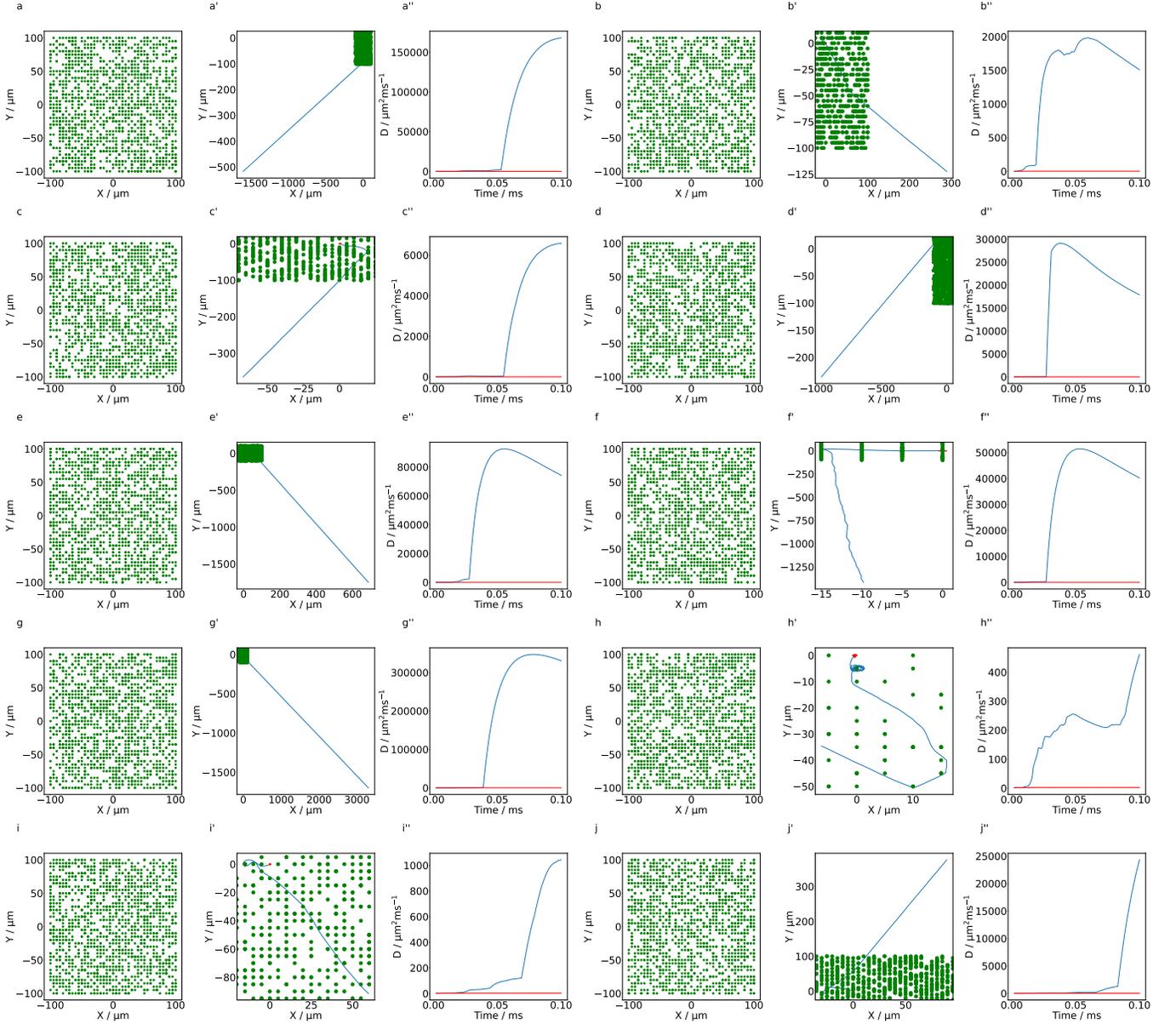}
    \caption{Simulated diffusion of a charged Atto-488-like single molecule under the influence of randomly positioned 1500 surface charges (blue), compared to diffusion without surface charges (red), with details shown in insets. The particles always start at the origin (0 $\upmu$m, 0 $\upmu$m) and the minimum distance between two surface charges is 5 $\upmu$m.
    (a - j), surface charge distribution.
    (a' - j'), diffusion trajectories.
    (a" - j"), their respective diffusion coefficient plots.}
    \label{Figure:S4}
\end{figure*}

\begin{figure*}[]
    \centering
    \includegraphics[width = 1\textwidth]{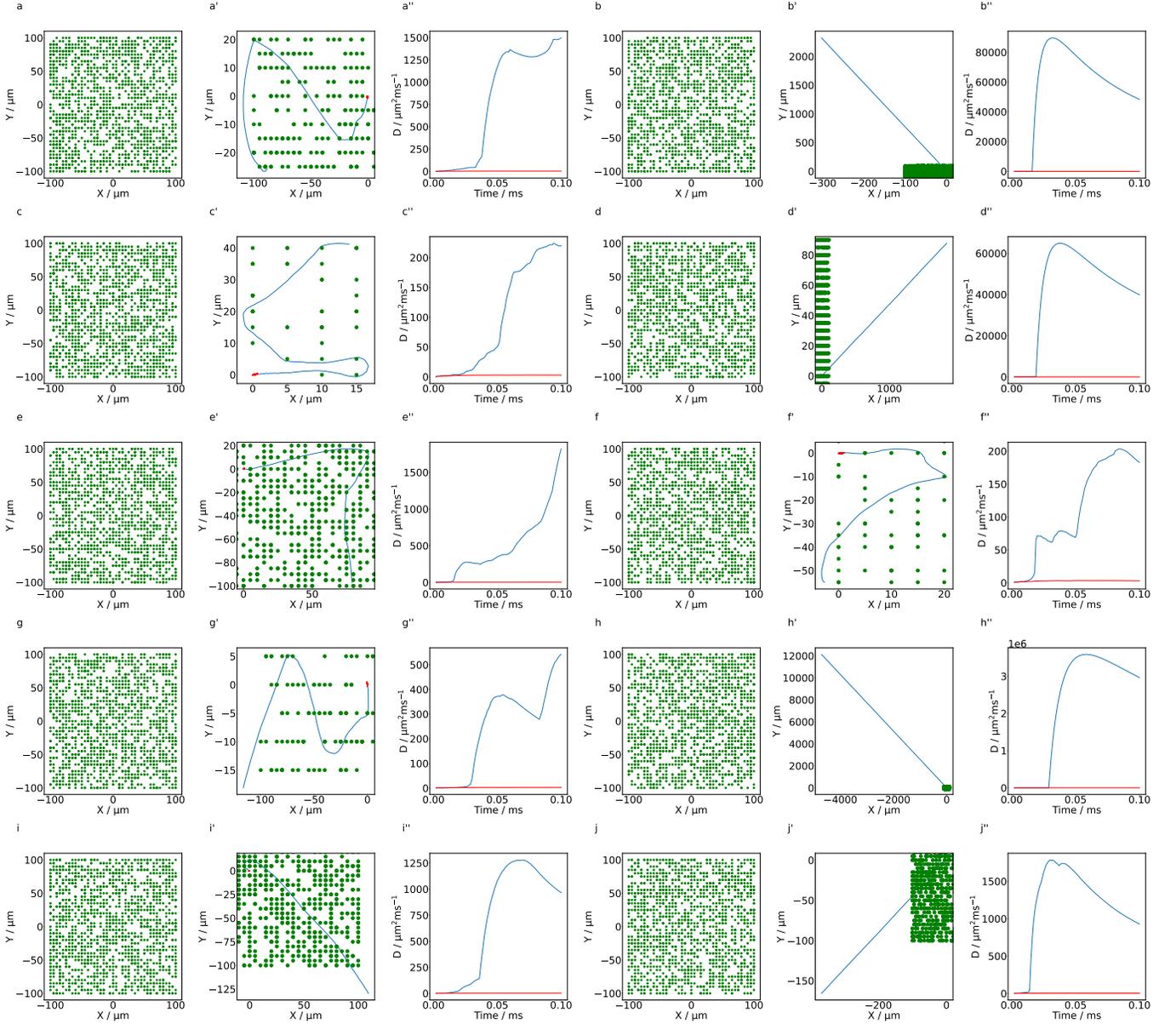}
    \caption{Simulated diffusion of a charged Atto-488-like single molecule under the influence of randomly positioned 1500 surface charges (blue), compared to diffusion without surface charges (red), with details shown in insets. The particles always start at the origin (0 $\upmu$m, 0 $\upmu$m) and the minimum distance between two surface charges is 5 $\upmu$m.
    (a - j), surface charge distribution.
    (a' - j'), diffusion trajectories.
    (a" - j"), their respective diffusion coefficient plots.}
    \label{Figure:S5}
\end{figure*}

\begin{figure*}[]
    \centering
    \includegraphics[width = 1\textwidth]{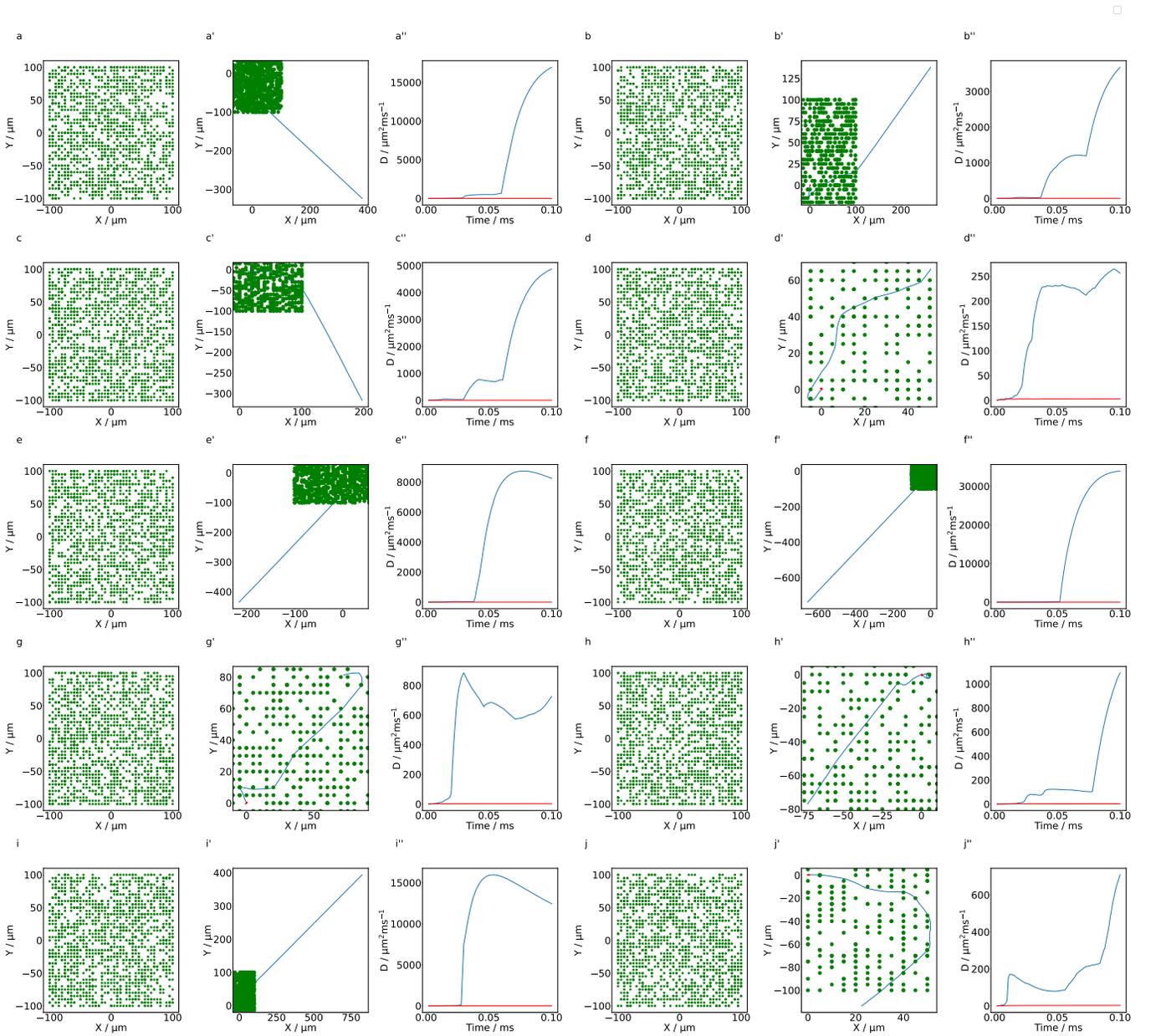}
    \caption{Simulated diffusion of a charged Atto-488-like single molecule under the influence of randomly positioned 1500 surface charges (blue), compared to diffusion without surface charges (red), with details shown in insets. The particles always start at the origin (0 $\upmu$m, 0 $\upmu$m) and the minimum distance between two surface charges is 5 $\upmu$m.
    (a - j), surface charge distribution.
    (a' - j'), diffusion trajectories.
    (a" - j"), their respective diffusion coefficient plots.}
    \label{Figure:S6}
\end{figure*}

\begin{figure*}[]
    \centering
    \includegraphics[width = 1\textwidth]{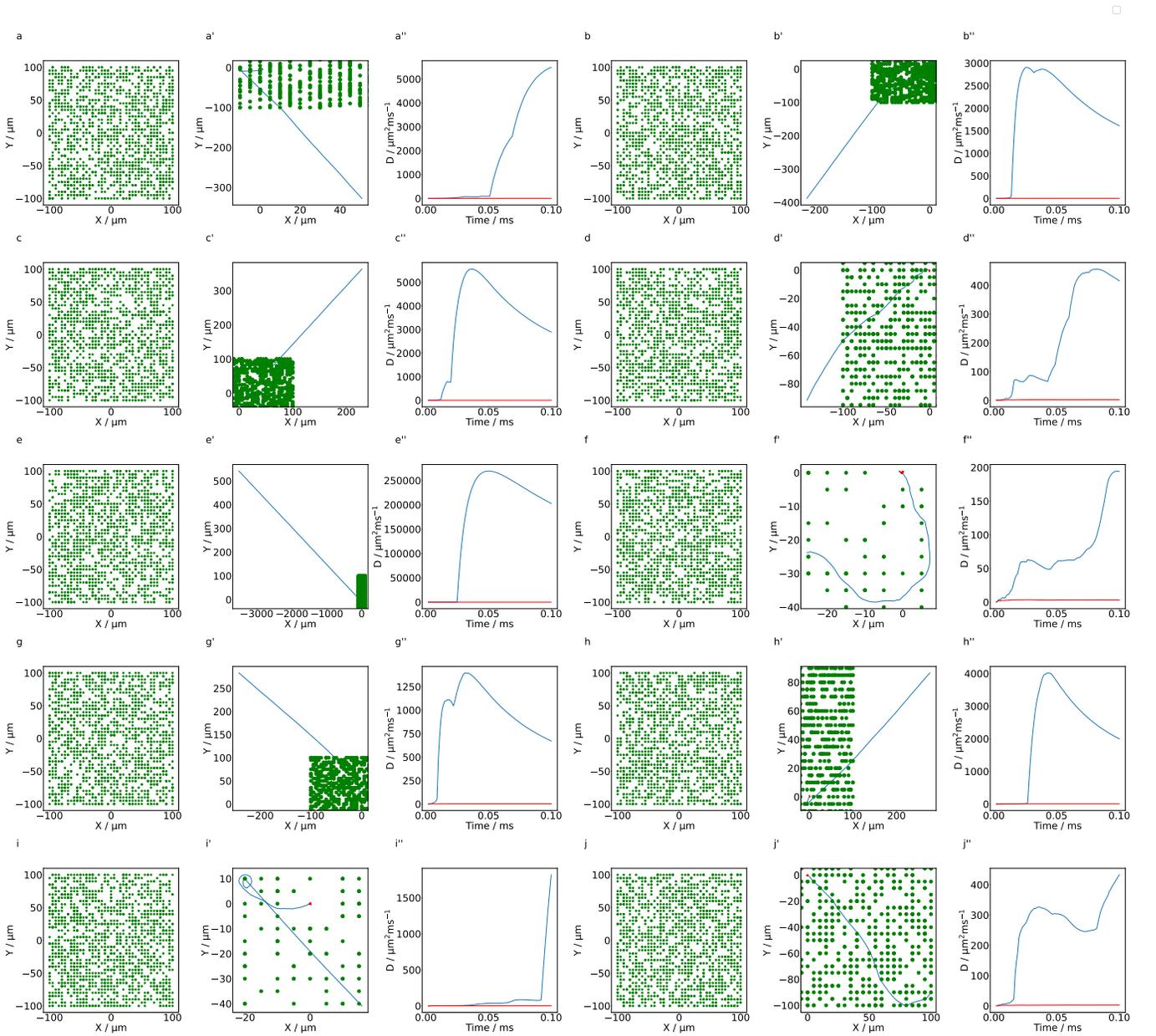}
    \caption{Simulated diffusion of a charged Atto-488-like single molecule under the influence of randomly positioned 1500 surface charges (blue), compared to diffusion without surface charges (red), with details shown in insets. The particles always start at the origin (0 $\upmu$m, 0 $\upmu$m) and the minimum distance between two surface charges is 5 $\upmu$m.
    (a - j), surface charge distribution.
    (a' - j'), diffusion trajectories.
    (a" - j"), their respective diffusion coefficient plots.}
    \label{Figure:7}
\end{figure*}

\begin{figure*}[]
    \centering
    \includegraphics[width = 1\textwidth]{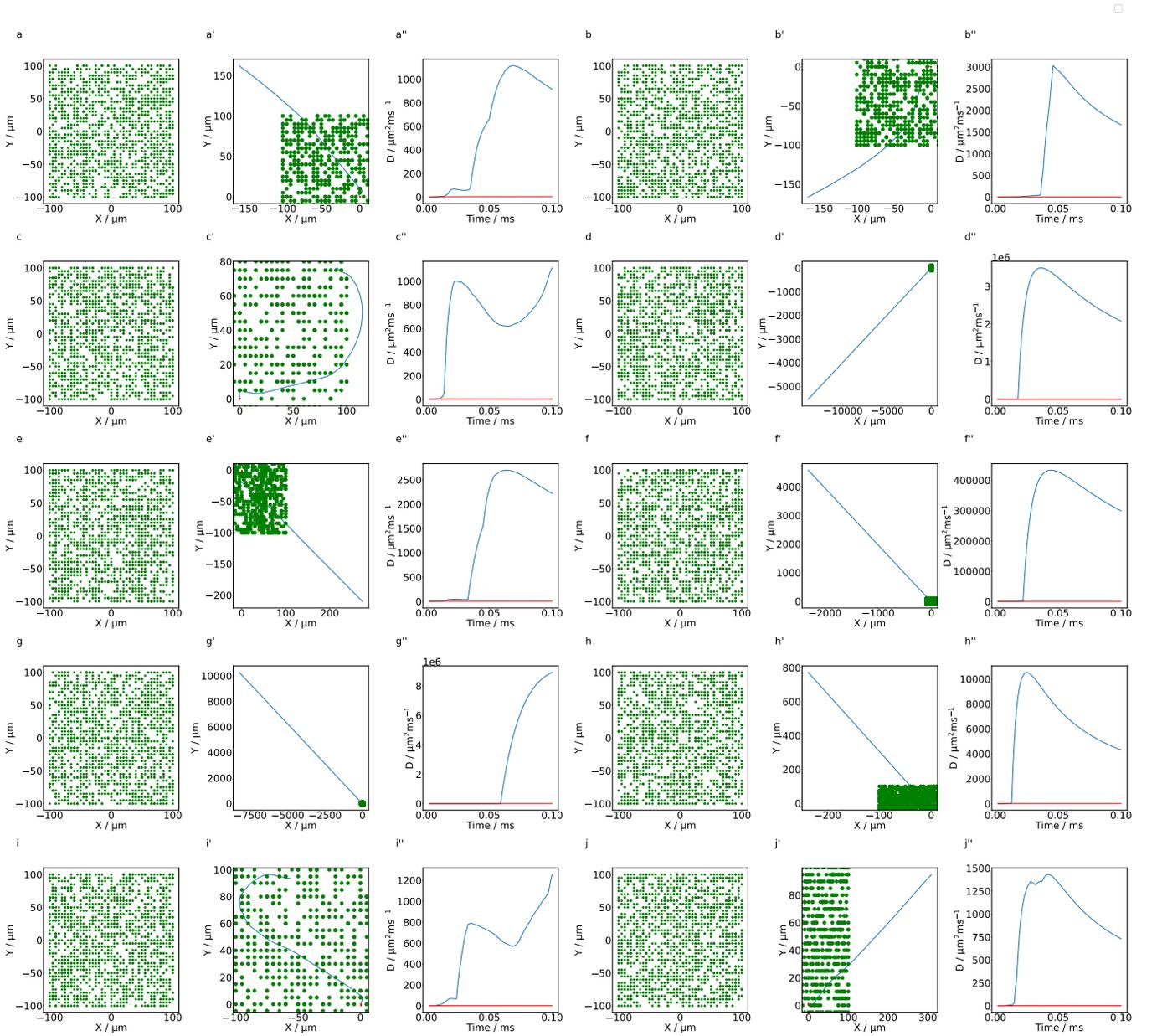}
    \caption{Simulated diffusion of a charged Atto-488-like single molecule under the influence of randomly positioned 1500 surface charges (blue), compared to diffusion without surface charges (red), with details shown in insets. The particles always start at the origin (0 $\upmu$m, 0 $\upmu$m) and the minimum distance between two surface charges is 5 $\upmu$m.
    (a - j), surface charge distribution.
    (a' - j'), diffusion trajectories.
    (a" - j"), their respective diffusion coefficient plots.}
    \label{Figure:8}
\end{figure*}

\begin{figure*}[]
    \centering
    \includegraphics[width = 1\textwidth]{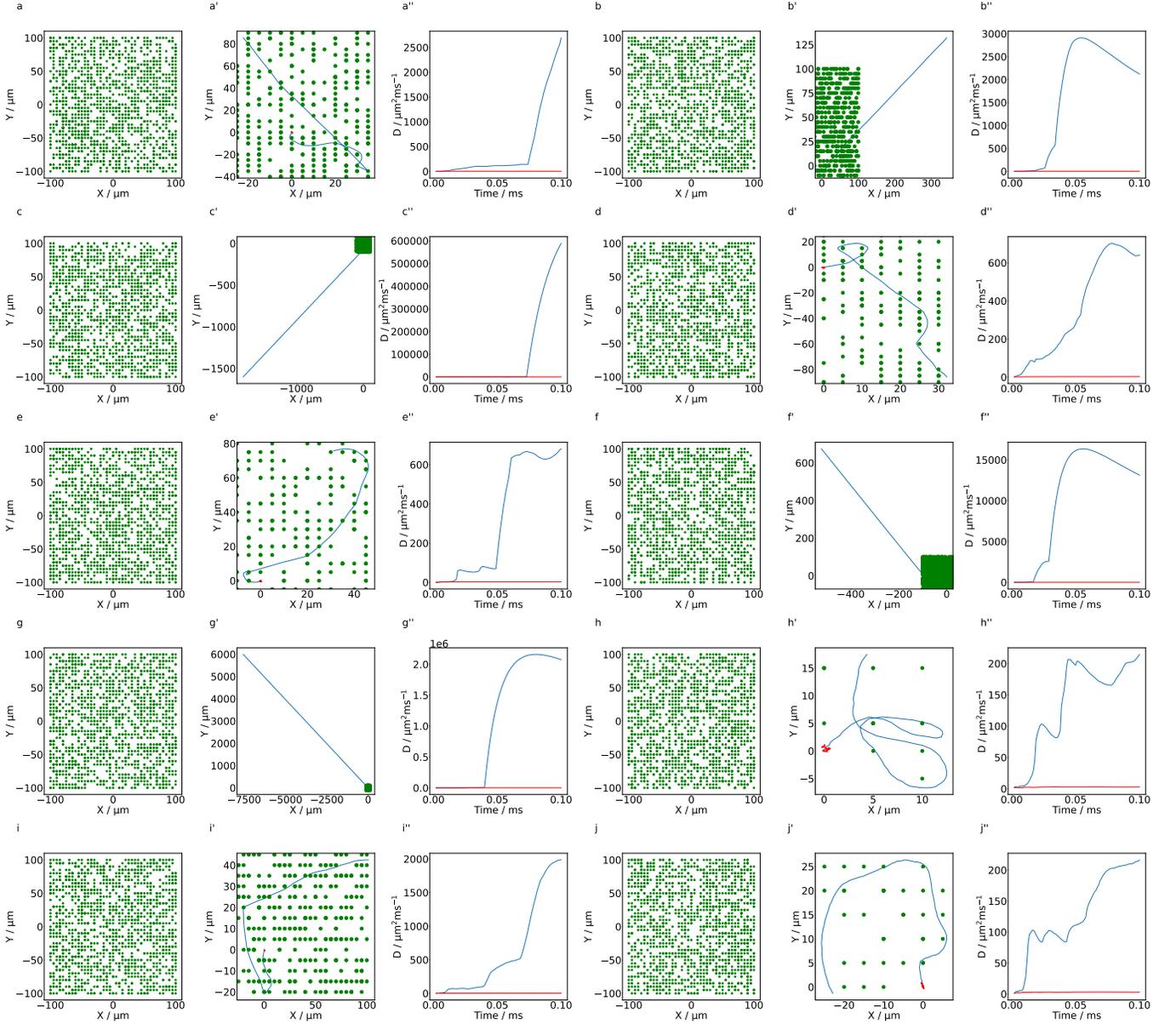}
    \caption{Simulated diffusion of a charged Atto-488-like single molecule under the influence of randomly positioned 1500 surface charges (blue), compared to diffusion without surface charges (red), with details shown in insets. The particles always start at the origin (0 $\upmu$m, 0 $\upmu$m) and the minimum distance between two surface charges is 5 $\upmu$m.
    (a - j), surface charge distribution.
    (a' - j'), diffusion trajectories.
    (a" - j"), their respective diffusion coefficient plots.}
    \label{Figure:9}
\end{figure*}

\begin{figure*}[]
    \centering
    \includegraphics[width = 1\textwidth]{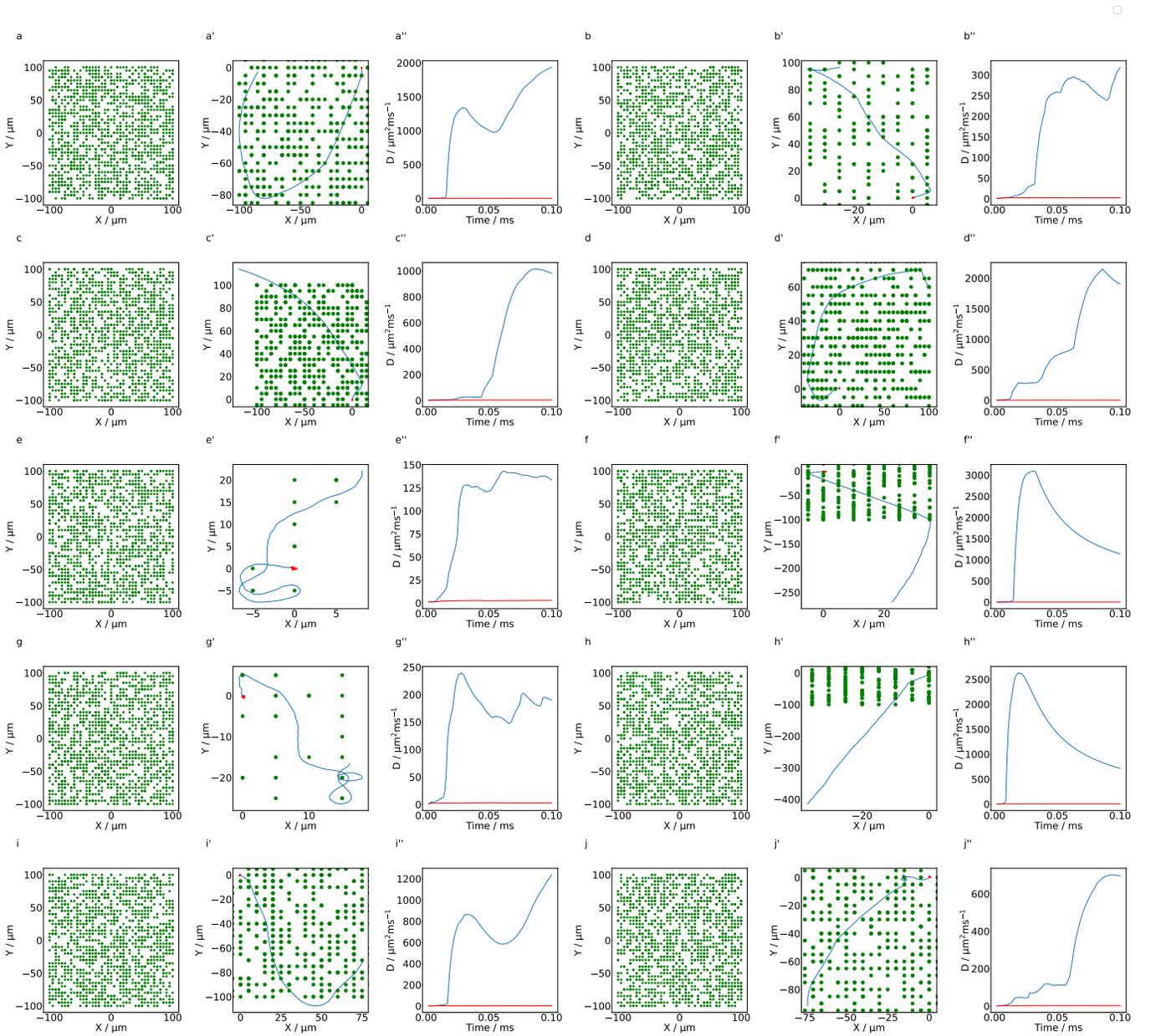}
    \caption{Simulated diffusion of a charged Atto-488-like single molecule under the influence of randomly positioned 1500 surface charges (blue), compared to diffusion without surface charges (red), with details shown in insets. The particles always start at the origin (0 $\upmu$m, 0 $\upmu$m) and the minimum distance between two surface charges is 5 $\upmu$m.
    (a - j), surface charge distribution.
    (a' - j'), diffusion trajectories.
    (a" - j"), their respective diffusion coefficient plots.}
    \label{Figure:10}
\end{figure*}

\begin{figure*}[]
    \centering
    \includegraphics[width = 1\textwidth]{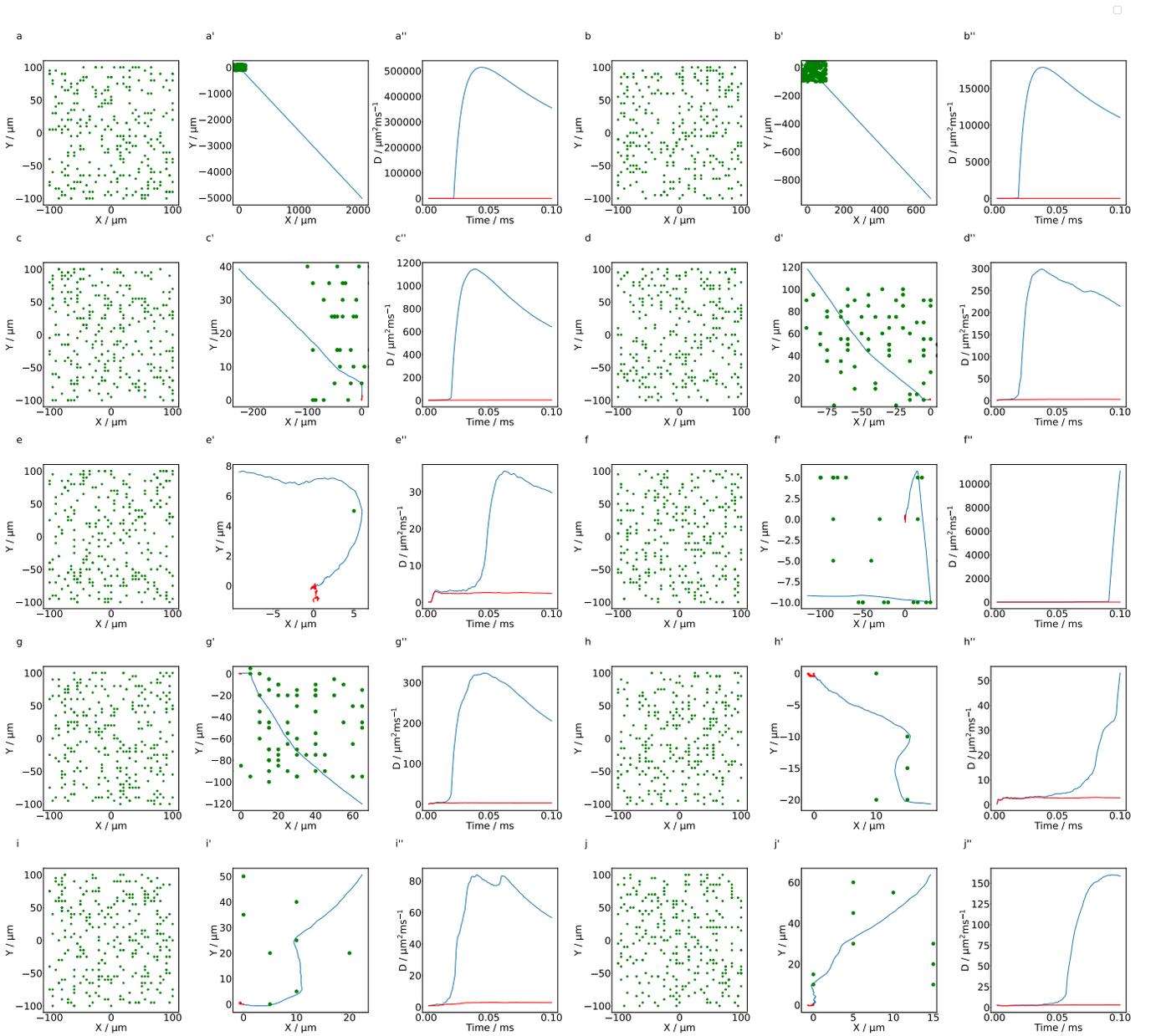}
    \caption{Simulated diffusion of a charged Atto-488-like single molecule under the influence of randomly positioned 300 surface charges (blue), compared to diffusion without surface charges (red), with details shown in insets. The particles always start at the origin (0 $\upmu$m, 0 $\upmu$m) and the minimum distance between two surface charges is 5 $\upmu$m.
    (a - j), surface charge distribution.
    (a' - j'), diffusion trajectories.
    (a" - j"), their respective diffusion coefficient plots.}
    \label{Figure:11}
\end{figure*}

\begin{figure*}[]
    \centering
    \includegraphics[width = 1\textwidth]{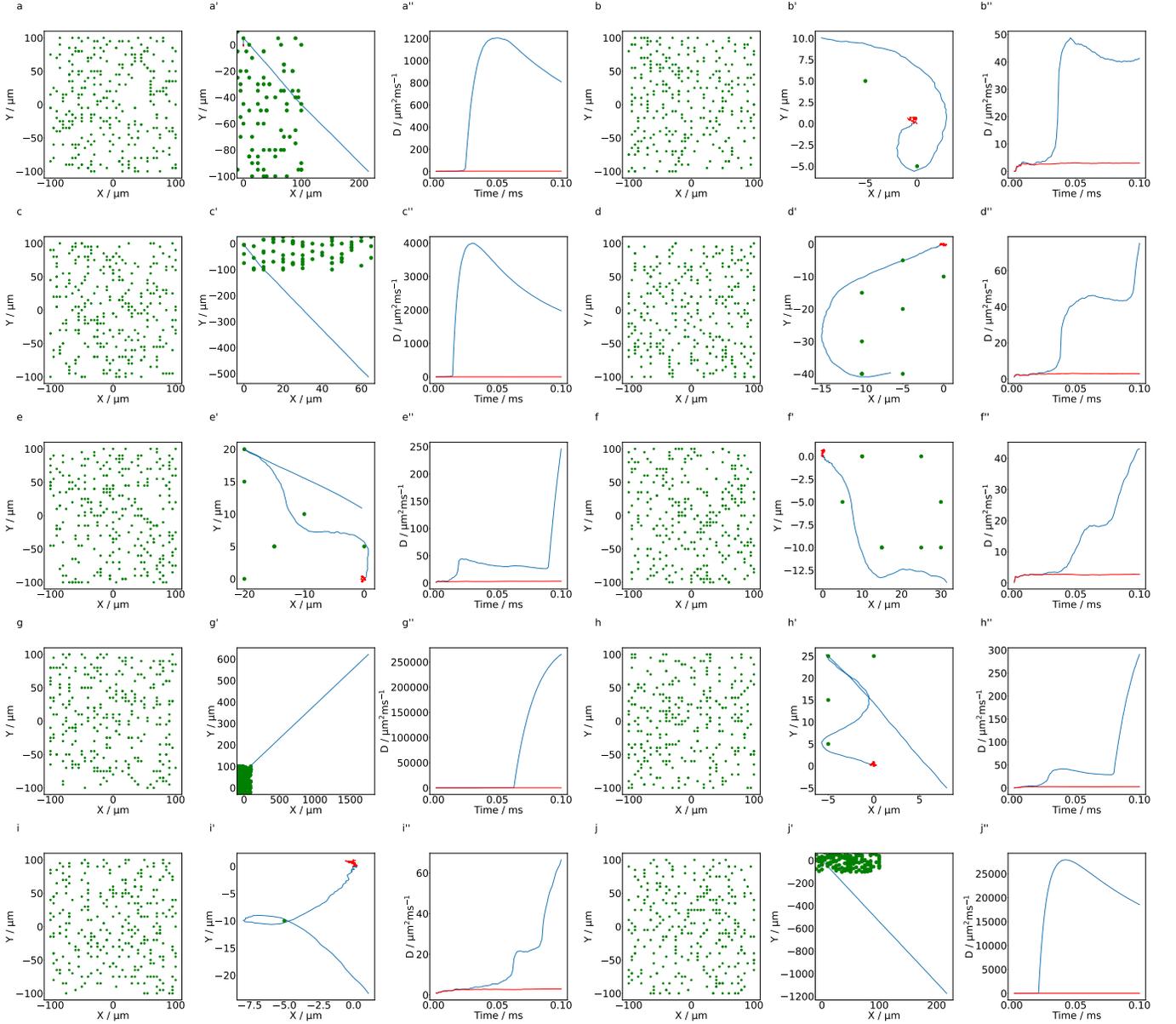}
    \caption{Simulated diffusion of a charged Atto-488-like single molecule under the influence of randomly positioned 300 surface charges (blue), compared to diffusion without surface charges (red), with details shown in insets. The particles always start at the origin (0 $\upmu$m, 0 $\upmu$m) and the minimum distance between two surface charges is 5 $\upmu$m.
    (a - j), surface charge distribution.
    (a' - j'), diffusion trajectories.
    (a" - j"), their respective diffusion coefficient plots.}
    \label{Figure:12}
\end{figure*}

\begin{figure*}[]
    \centering
    \includegraphics[width = 1\textwidth]{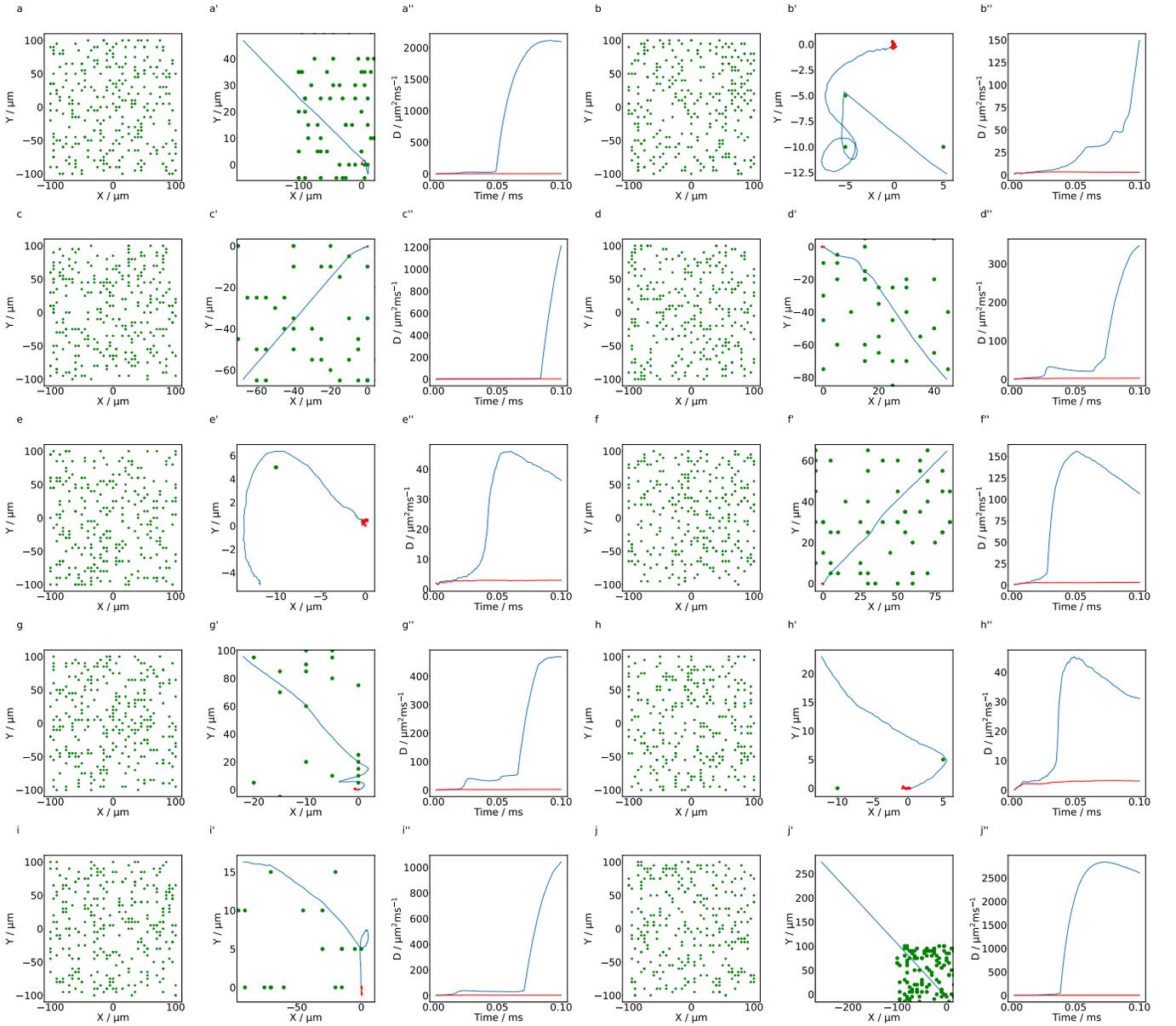}
    \caption{Simulated diffusion of a charged Atto-488-like single molecule under the influence of randomly positioned 300 surface charges (blue), compared to diffusion without surface charges (red), with details shown in insets. The particles always start at the origin (0 $\upmu$m, 0 $\upmu$m) and the minimum distance between two surface charges is 5 $\upmu$m.
    (a - j), surface charge distribution.
    (a' - j'), diffusion trajectories.
    (a" - j"), their respective diffusion coefficient plots.}
    \label{Figure:13}
\end{figure*}

\begin{figure*}[]
    \centering
    \includegraphics[width = 1\textwidth]{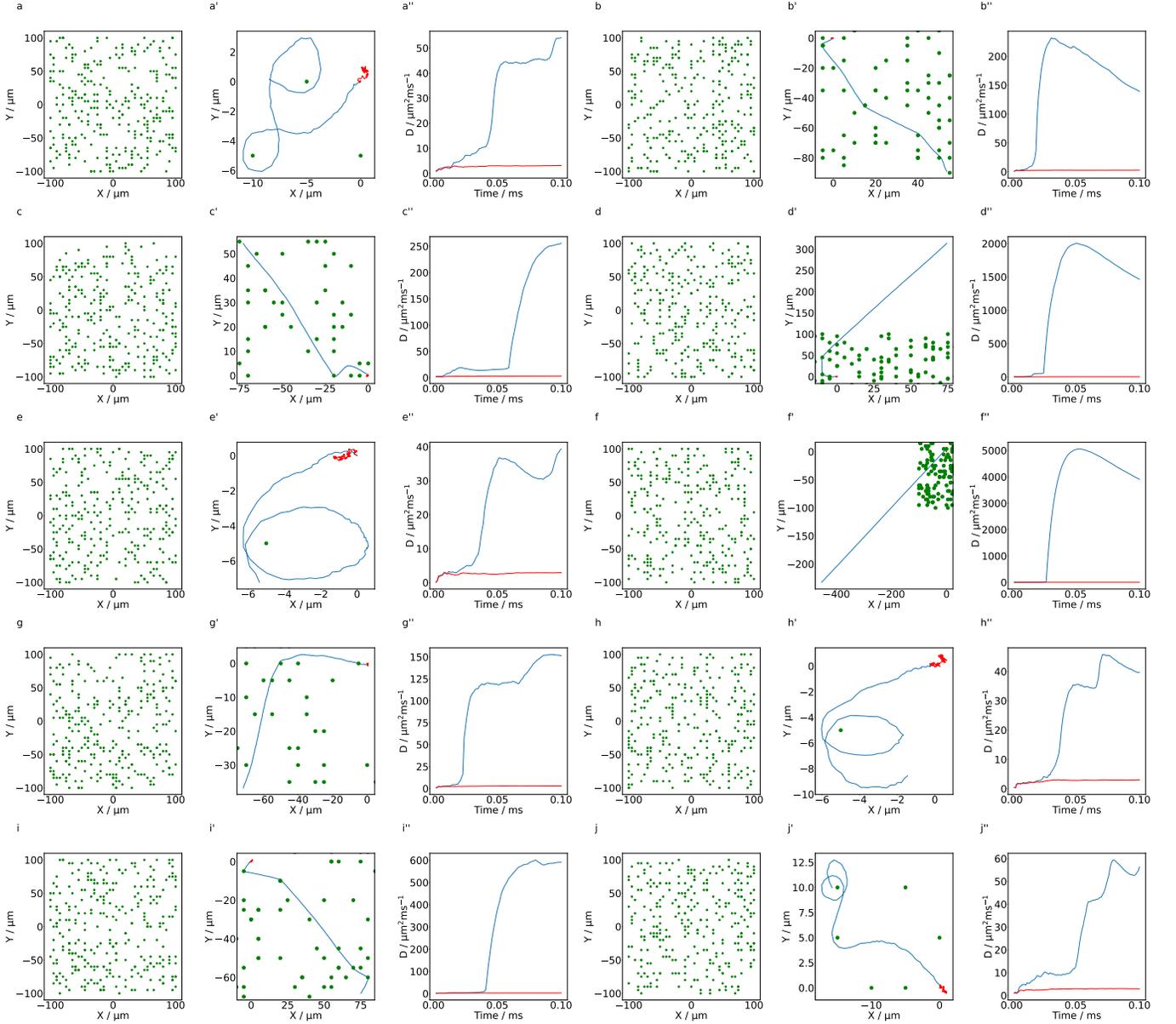}
    \caption{Simulated diffusion of a charged Atto-488-like single molecule under the influence of randomly positioned 300 surface charges (blue), compared to diffusion without surface charges (red), with details shown in insets. The particles always start at the origin (0 $\upmu$m, 0 $\upmu$m) and the minimum distance between two surface charges is 5 $\upmu$m.
    (a - j), surface charge distribution.
    (a' - j'), diffusion trajectories.
    (a" - j"), their respective diffusion coefficient plots.}
    \label{Figure:14}
\end{figure*}

\begin{figure*}[]
    \centering
    \includegraphics[width = 1\textwidth]{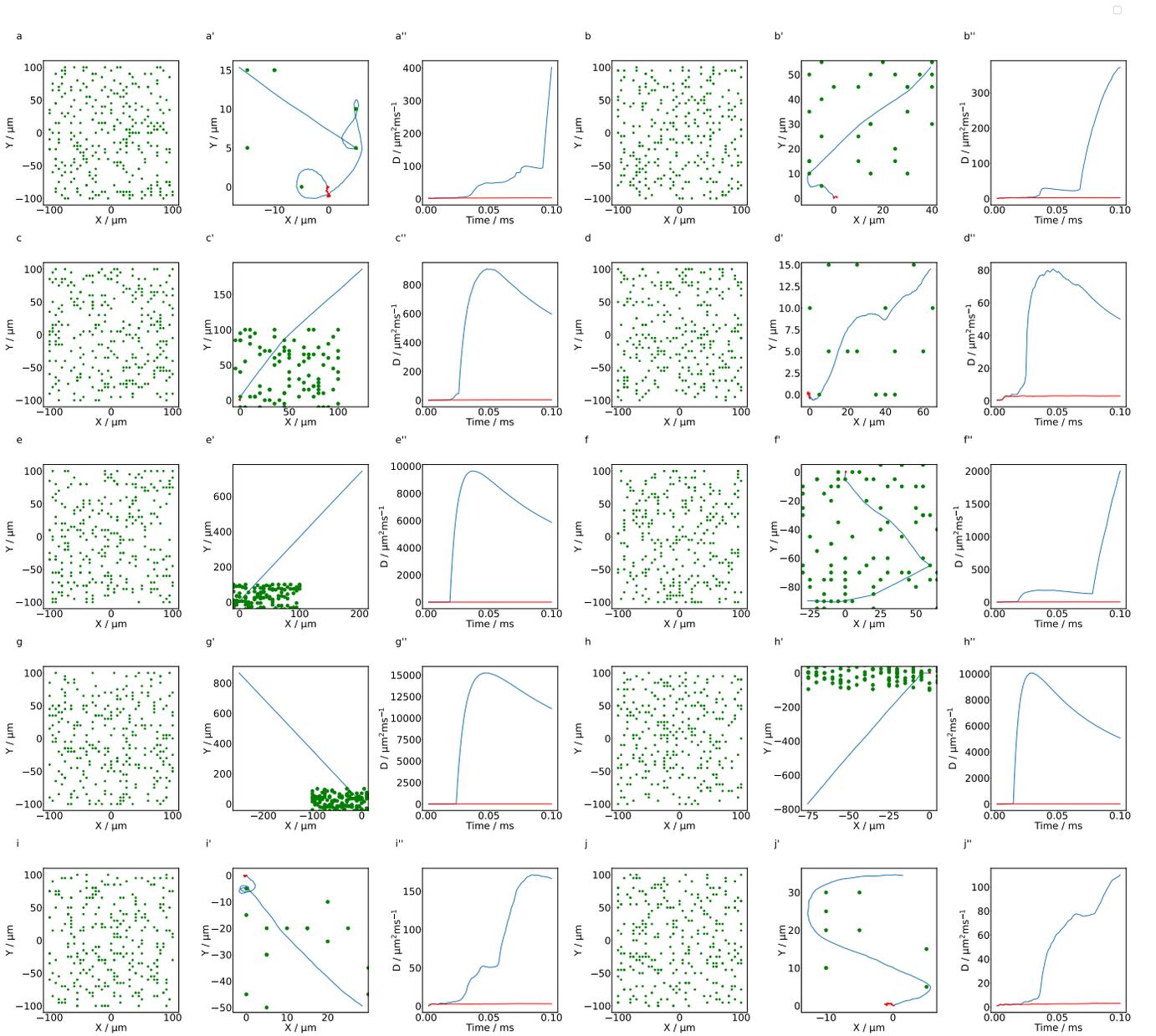}
    \caption{Simulated diffusion of a charged Atto-488-like single molecule under the influence of randomly positioned 300 surface charges (blue), compared to diffusion without surface charges (red), with details shown in insets. The particles always start at the origin (0 $\upmu$m, 0 $\upmu$m) and the minimum distance between two surface charges is 5 $\upmu$m.
    (a - j), surface charge distribution.
    (a' - j'), diffusion trajectories.
    (a" - j"), their respective diffusion coefficient plots.}
    \label{Figure:15}
\end{figure*}

\begin{figure*}[]
    \centering
    \includegraphics[width = 1\textwidth]{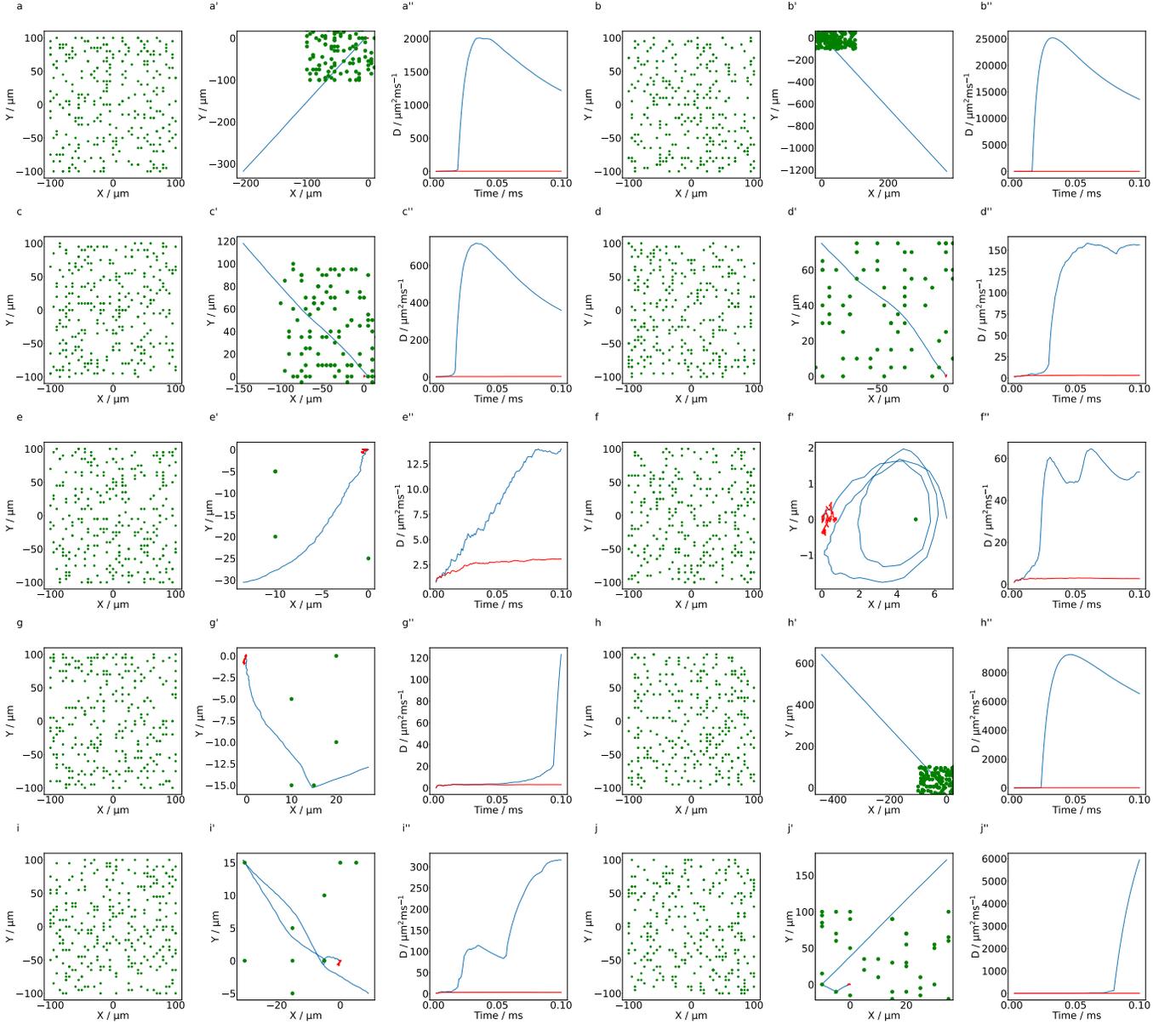}
    \caption{Simulated diffusion of a charged Atto-488-like single molecule under the influence of randomly positioned 300 surface charges (blue), compared to diffusion without surface charges (red), with details shown in insets. The particles always start at the origin (0 $\upmu$m, 0 $\upmu$m) and the minimum distance between two surface charges is 5 $\upmu$m.
    (a - j), surface charge distribution.
    (a' - j'), diffusion trajectories.
    (a" - j"), their respective diffusion coefficient plots.}
    \label{Figure:16}
\end{figure*}

\begin{figure*}[]
    \centering
    \includegraphics[width = 1\textwidth]{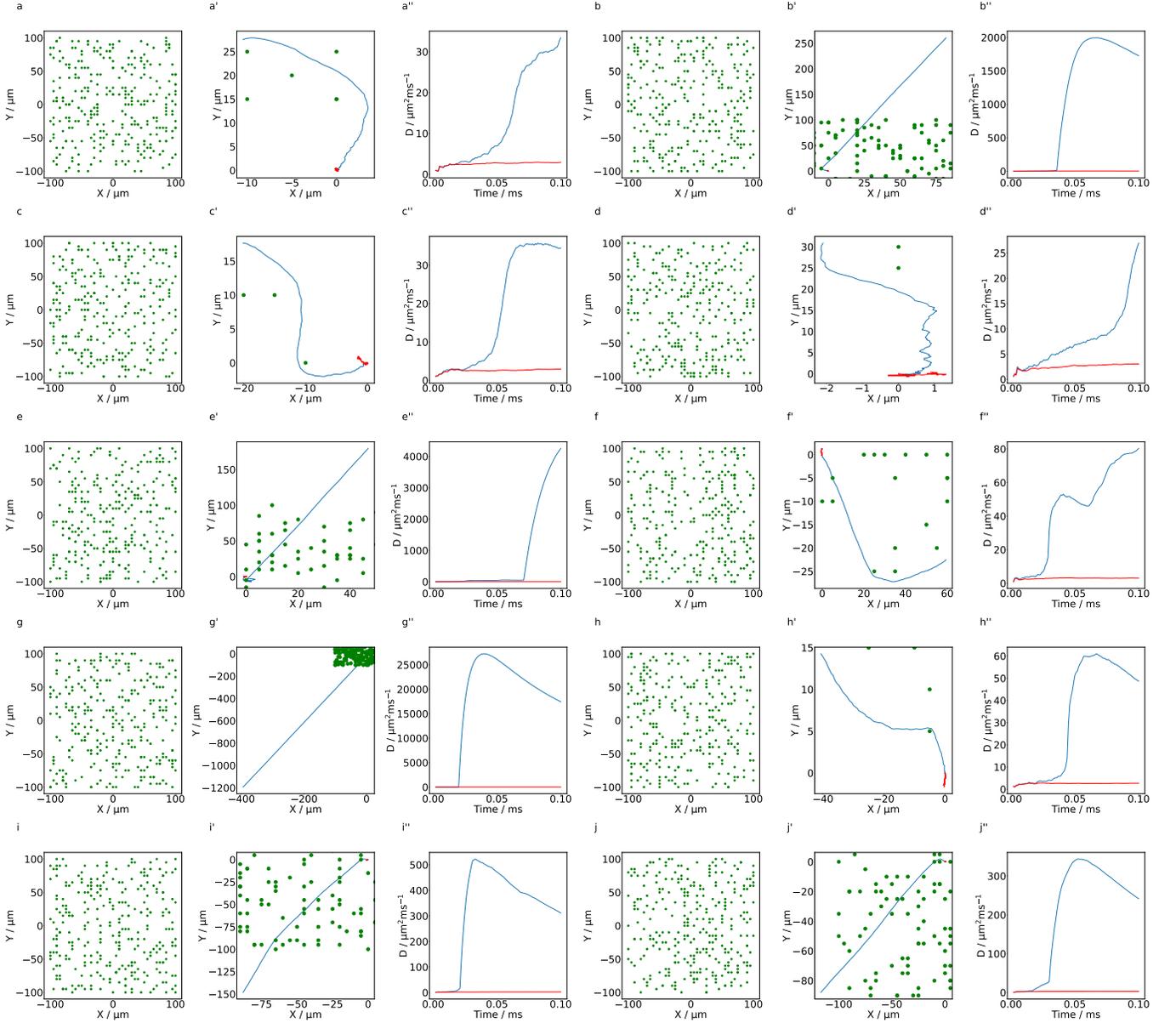}
    \caption{Simulated diffusion of a charged Atto-488-like single molecule under the influence of randomly positioned 300 surface charges (blue), compared to diffusion without surface charges (red), with details shown in insets. The particles always start at the origin (0 $\upmu$m, 0 $\upmu$m) and the minimum distance between two surface charges is 5 $\upmu$m.
    (a - j), surface charge distribution.
    (a' - j'), diffusion trajectories.
    (a" - j"), their respective diffusion coefficient plots.}
    \label{Figure:17}
\end{figure*}

\begin{figure*}[]
    \centering
    \includegraphics[width = 1\textwidth]{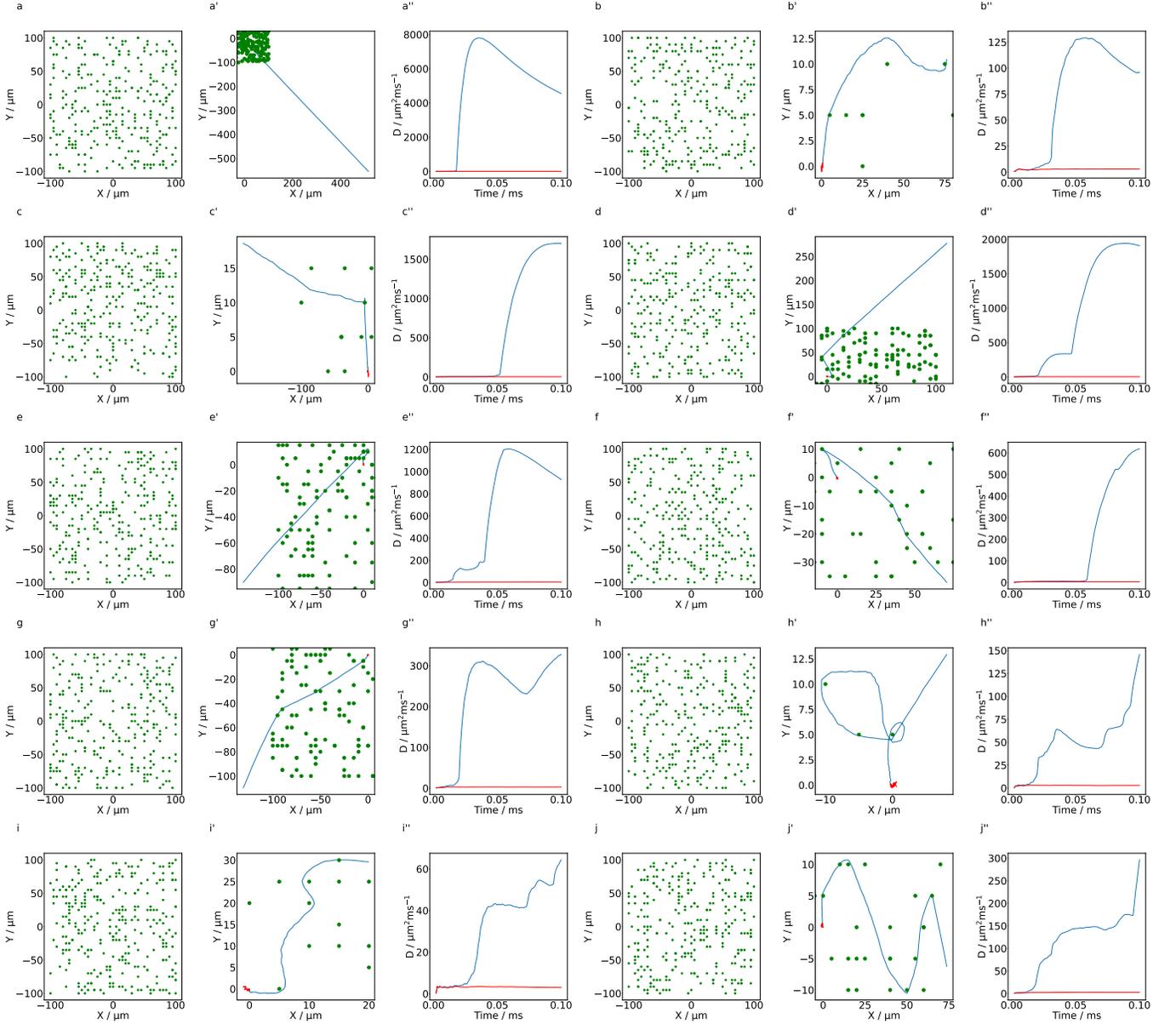}
    \caption{Simulated diffusion of a charged Atto-488-like single molecule under the influence of randomly positioned 300 surface charges (blue), compared to diffusion without surface charges (red), with details shown in insets. The particles always start at the origin (0 $\upmu$m, 0 $\upmu$m) and the minimum distance between two surface charges is 5 $\upmu$m.
    (a - j), surface charge distribution.
    (a' - j'), diffusion trajectories.
    (a" - j"), their respective diffusion coefficient plots.}
    \label{Figure:18}
\end{figure*}

\begin{figure*}[]
    \centering
    \includegraphics[width = 1\textwidth]{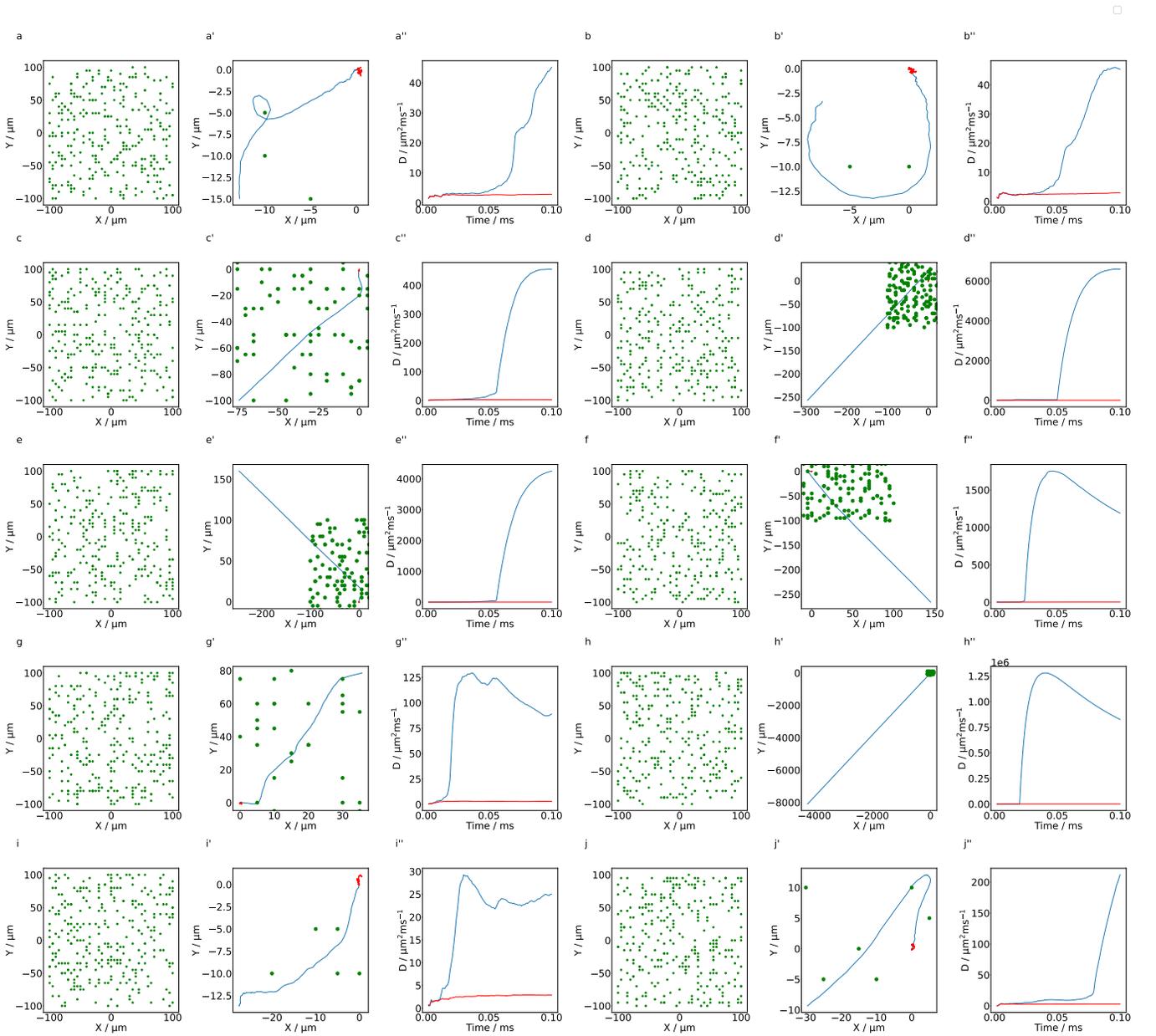}
    \caption{Simulated diffusion of a charged Atto-488-like single molecule under the influence of randomly positioned 300 surface charges (blue), compared to diffusion without surface charges (red), with details shown in insets. The particles always start at the origin (0 $\upmu$m, 0 $\upmu$m) and the minimum distance between two surface charges is 5 $\upmu$m.
    (a - j), surface charge distribution.
    (a' - j'), diffusion trajectories.
    (a" - j"), their respective diffusion coefficient plots.}
    \label{Figure:19}
\end{figure*}

\begin{figure*}[]
    \centering
    \includegraphics[width = 1\textwidth]{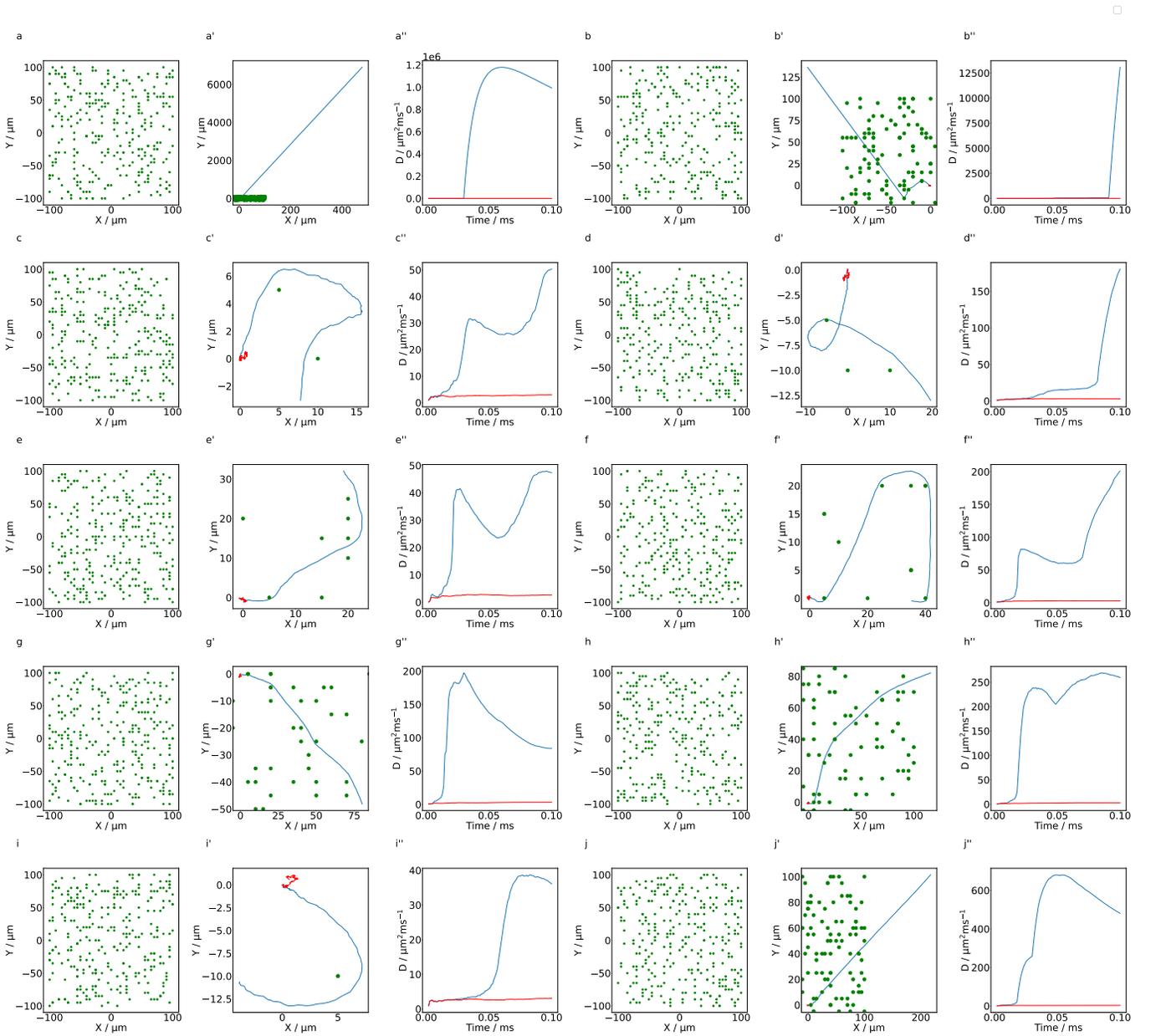}
    \caption{Simulated diffusion of a charged Atto-488-like single molecule under the influence of randomly positioned 300 surface charges (blue), compared to diffusion without surface charges (red), with details shown in insets. The particles always start at the origin (0 $\upmu$m, 0 $\upmu$m) and the minimum distance between two surface charges is 5 $\upmu$m.
    (a - j), surface charge distribution.
    (a' - j'), diffusion trajectories.
    (a" - j"), their respective diffusion coefficient plots.}
    \label{Figure:20}
\end{figure*}

\begin{figure*}[]
    \centering
    \includegraphics[width = 1\textwidth]{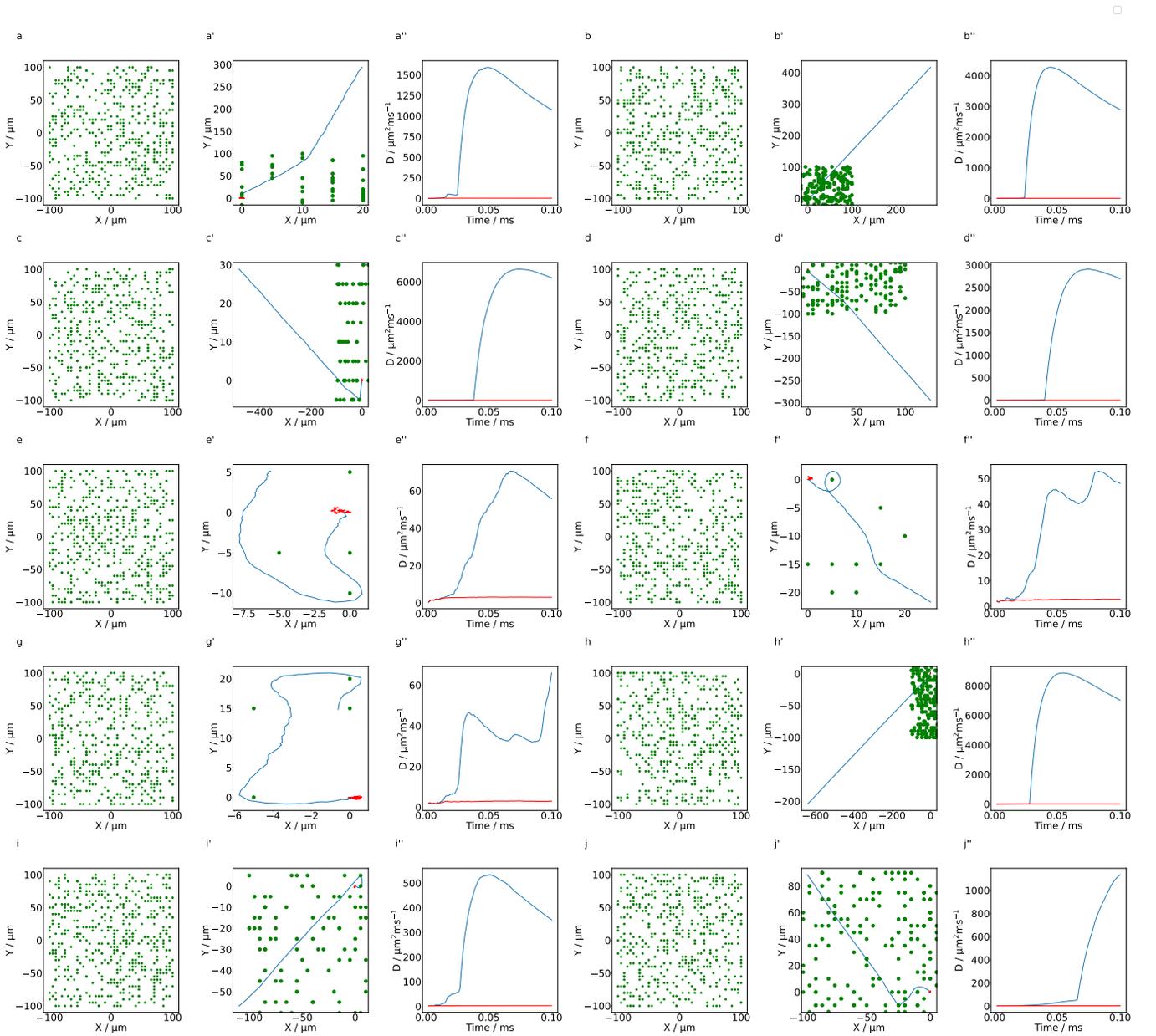}
    \caption{Simulated diffusion of a charged Atto-488-like single molecule under the influence of randomly positioned 500 surface charges (blue), compared to diffusion without surface charges (red), with details shown in insets. The particles always start at the origin (0 $\upmu$m, 0 $\upmu$m) and the minimum distance between two surface charges is 5 $\upmu$m.
    (a - j), surface charge distribution.
    (a' - j'), diffusion trajectories.
    (a" - j"), their respective diffusion coefficient plots.}
    \label{Figure:21}
\end{figure*}

\begin{figure*}[]
    \centering
    \includegraphics[width = 1\textwidth]{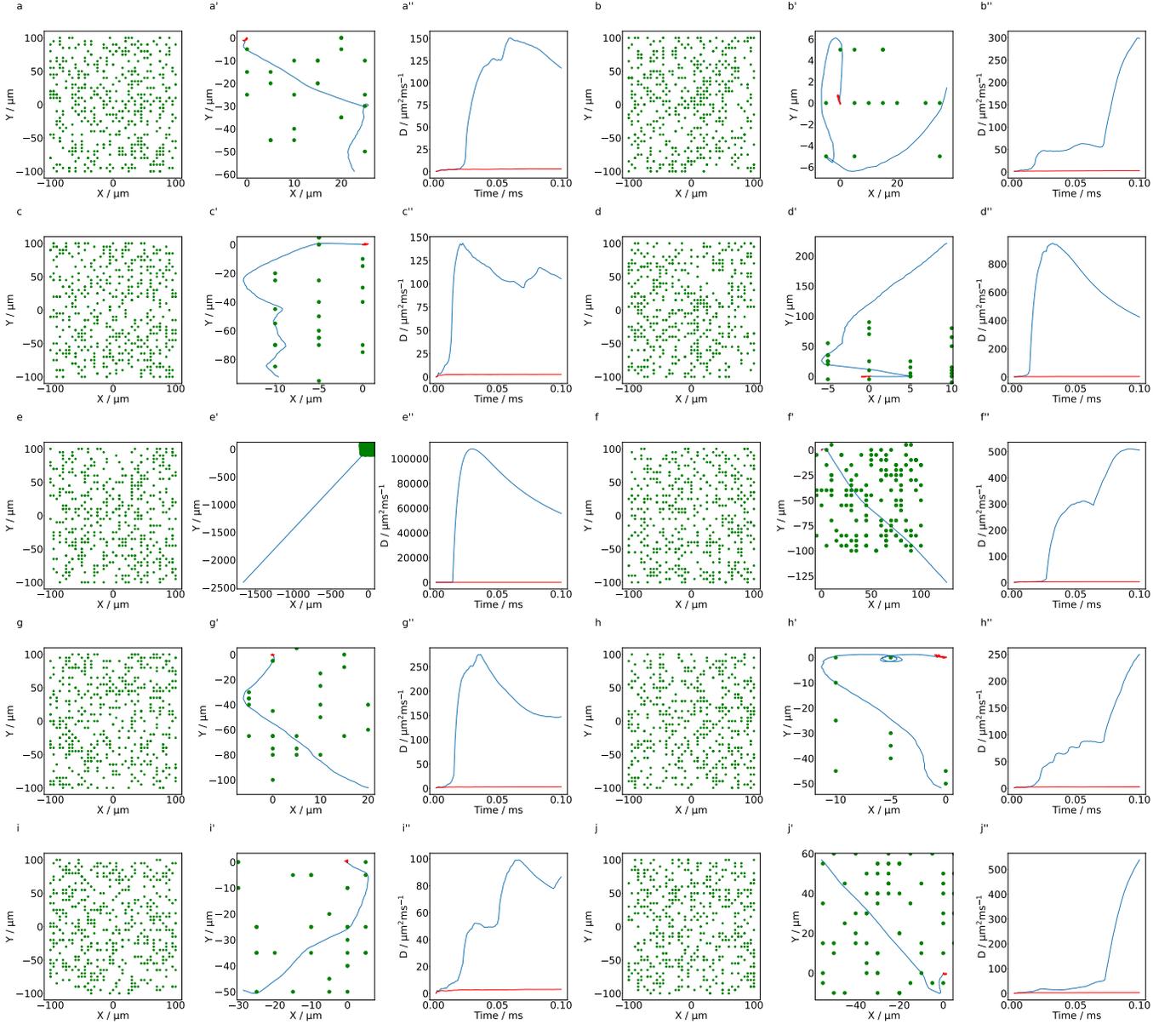}
    \caption{Simulated diffusion of a charged Atto-488-like single molecule under the influence of randomly positioned 500 surface charges (blue), compared to diffusion without surface charges (red), with details shown in insets. The particles always start at the origin (0 $\upmu$m, 0 $\upmu$m) and the minimum distance between two surface charges is 5 $\upmu$m.
    (a - j), surface charge distribution.
    (a' - j'), diffusion trajectories.
    (a" - j"), their respective diffusion coefficient plots.}
    \label{Figure:22}
\end{figure*}

\begin{figure*}[]
    \centering
    \includegraphics[width = 1\textwidth]{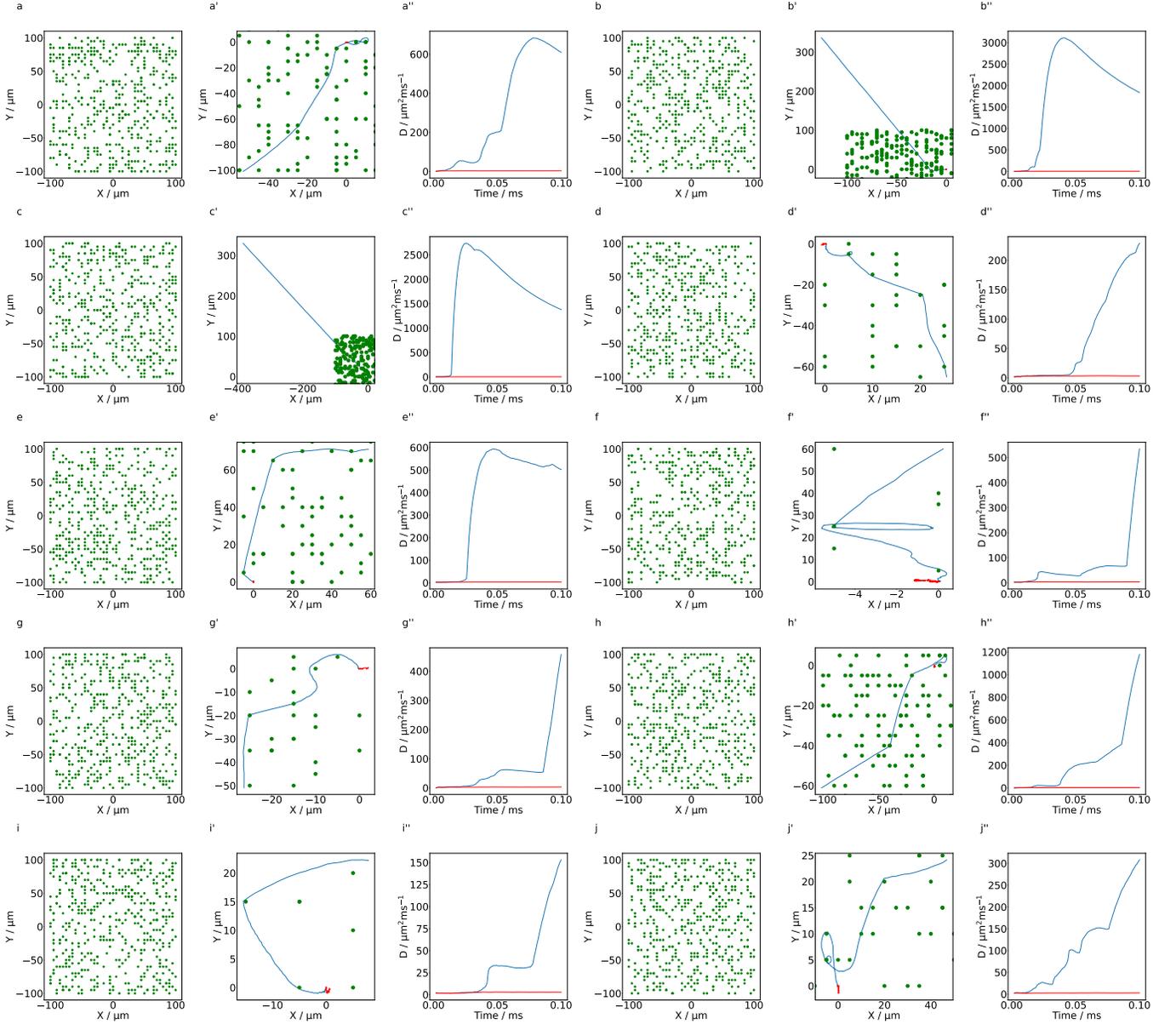}
    \caption{Simulated diffusion of a charged Atto-488-like single molecule under the influence of randomly positioned 500 surface charges (blue), compared to diffusion without surface charges (red), with details shown in insets. The particles always start at the origin (0 $\upmu$m, 0 $\upmu$m) and the minimum distance between two surface charges is 5 $\upmu$m.
    (a - j), surface charge distribution.
    (a' - j'), diffusion trajectories.
    (a" - j"), their respective diffusion coefficient plots.}
    \label{Figure:23}
\end{figure*}

\begin{figure*}[]
    \centering
    \includegraphics[width = 1\textwidth]{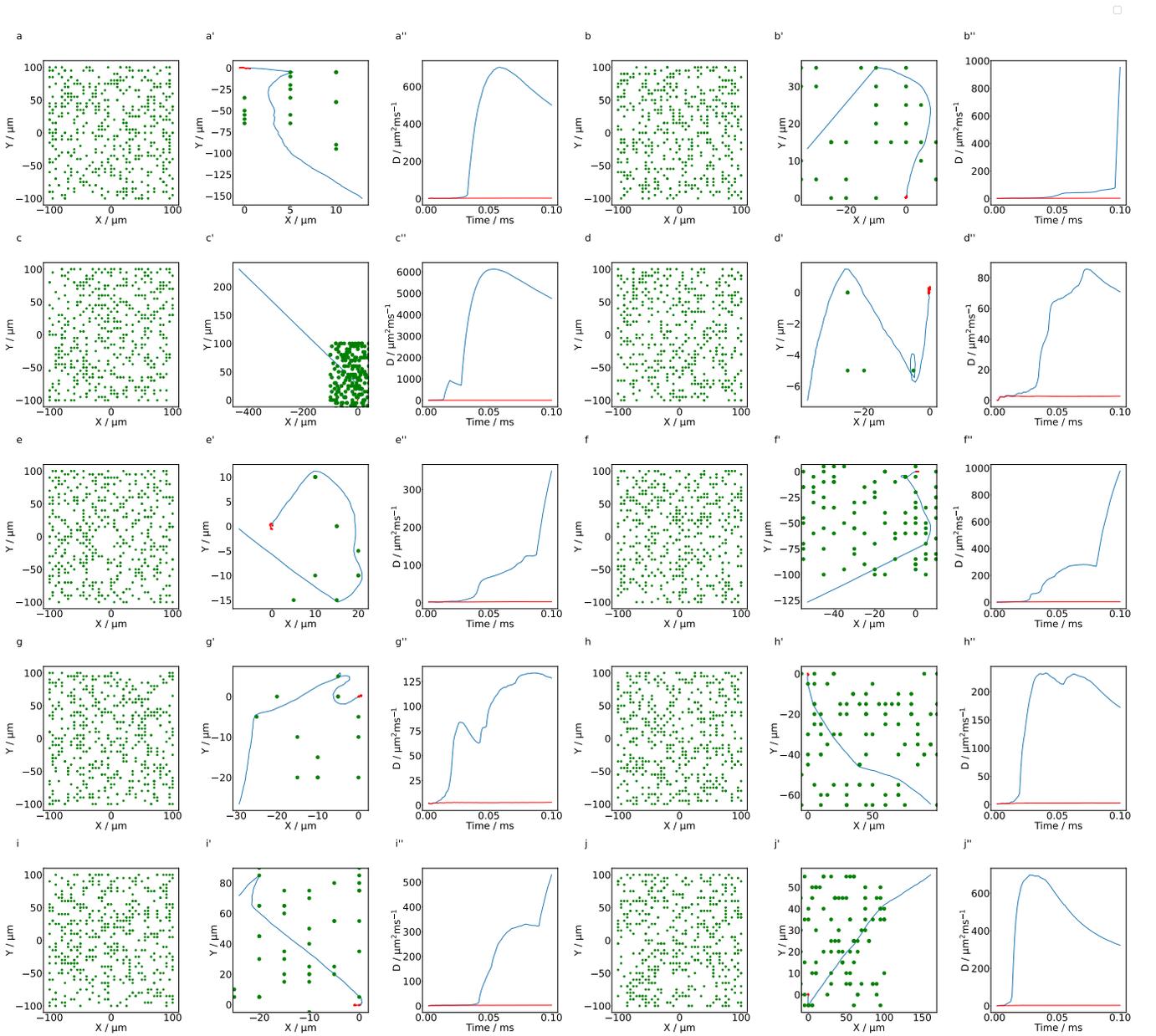}
    \caption{Simulated diffusion of a charged Atto-488-like single molecule under the influence of randomly positioned 500 surface charges (blue), compared to diffusion without surface charges (red), with details shown in insets. The particles always start at the origin (0 $\upmu$m, 0 $\upmu$m) and the minimum distance between two surface charges is 5 $\upmu$m.
    (a - j), surface charge distribution.
    (a' - j'), diffusion trajectories.
    (a" - j"), their respective diffusion coefficient plots.}
    \label{Figure:24}
\end{figure*}

\begin{figure*}[]
    \centering
    \includegraphics[width = 1\textwidth]{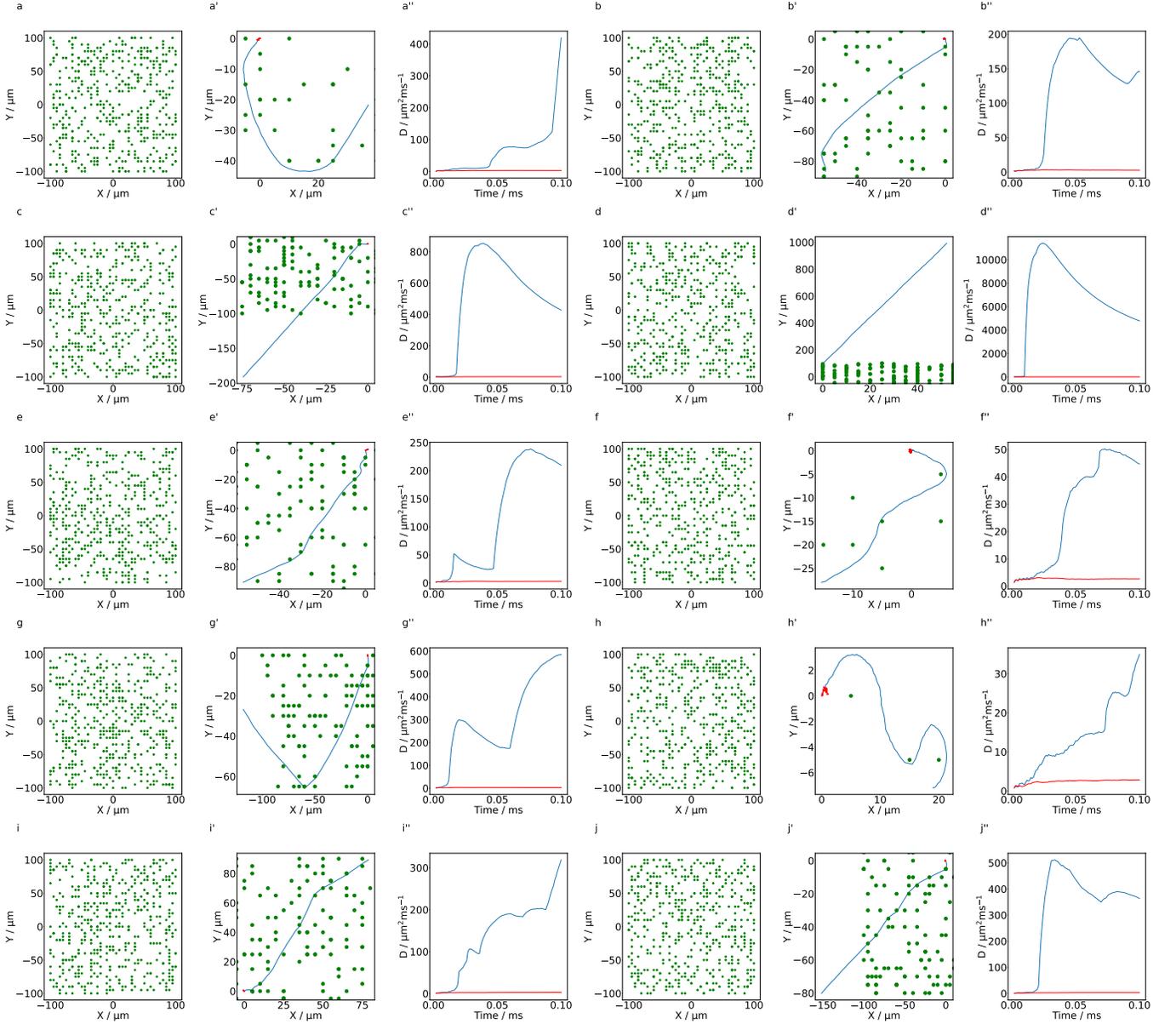}
    \caption{Simulated diffusion of a charged Atto-488-like single molecule under the influence of randomly positioned 500 surface charges (blue), compared to diffusion without surface charges (red), with details shown in insets. The particles always start at the origin (0 $\upmu$m, 0 $\upmu$m) and the minimum distance between two surface charges is 5 $\upmu$m.
    (a - j), surface charge distribution.
    (a' - j'), diffusion trajectories.
    (a" - j"), their respective diffusion coefficient plots.}
    \label{Figure:25}
\end{figure*}

\begin{figure*}[]
    \centering
    \includegraphics[width = 1\textwidth]{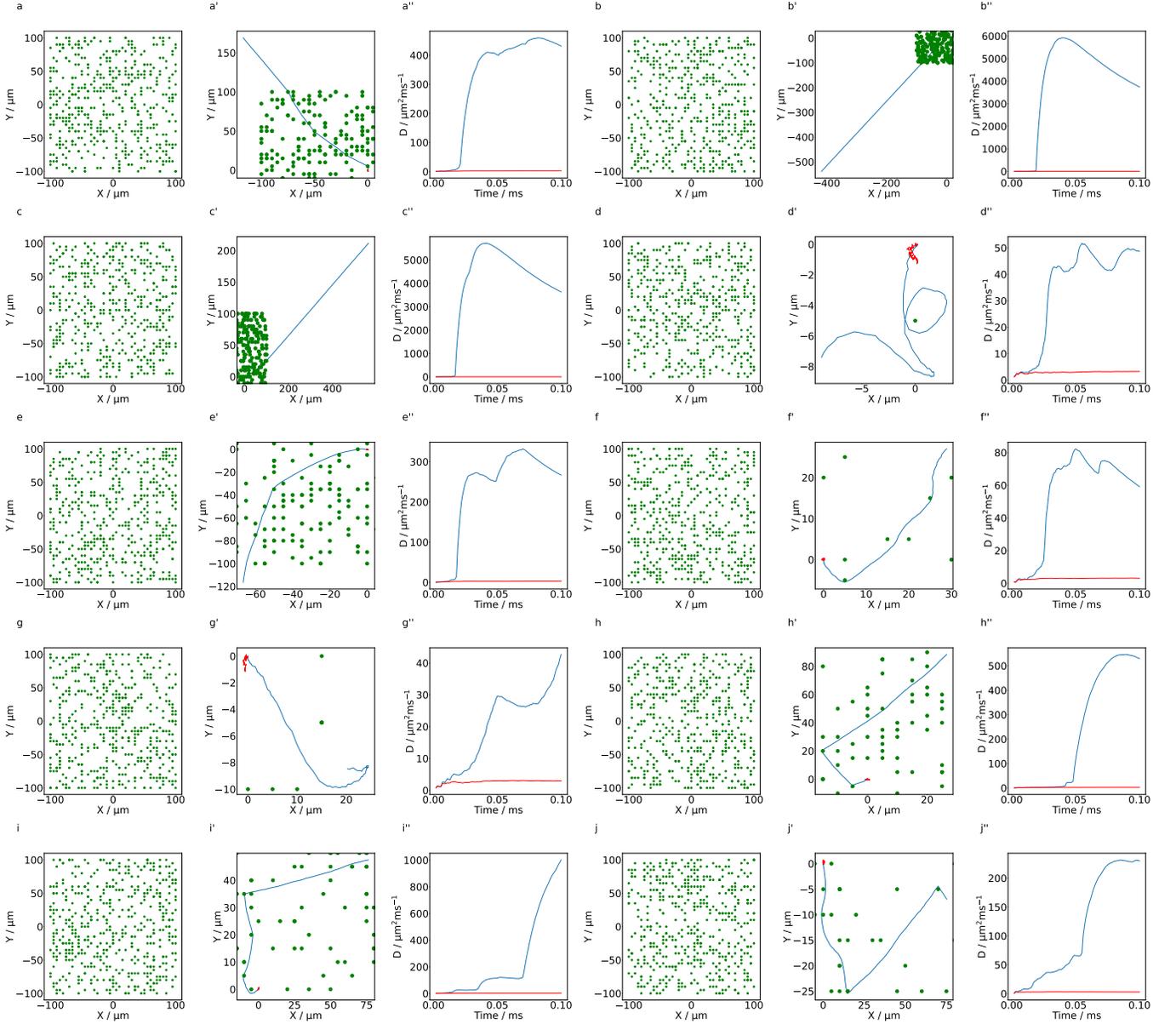}
    \caption{Simulated diffusion of a charged Atto-488-like single molecule under the influence of randomly positioned 500 surface charges (blue), compared to diffusion without surface charges (red), with details shown in insets. The particles always start at the origin (0 $\upmu$m, 0 $\upmu$m) and the minimum distance between two surface charges is 5 $\upmu$m.
    (a - j), surface charge distribution.
    (a' - j'), diffusion trajectories.
    (a" - j"), their respective diffusion coefficient plots.}
    \label{Figure:26}
\end{figure*}

\begin{figure*}[]
    \centering
    \includegraphics[width = 1\textwidth]{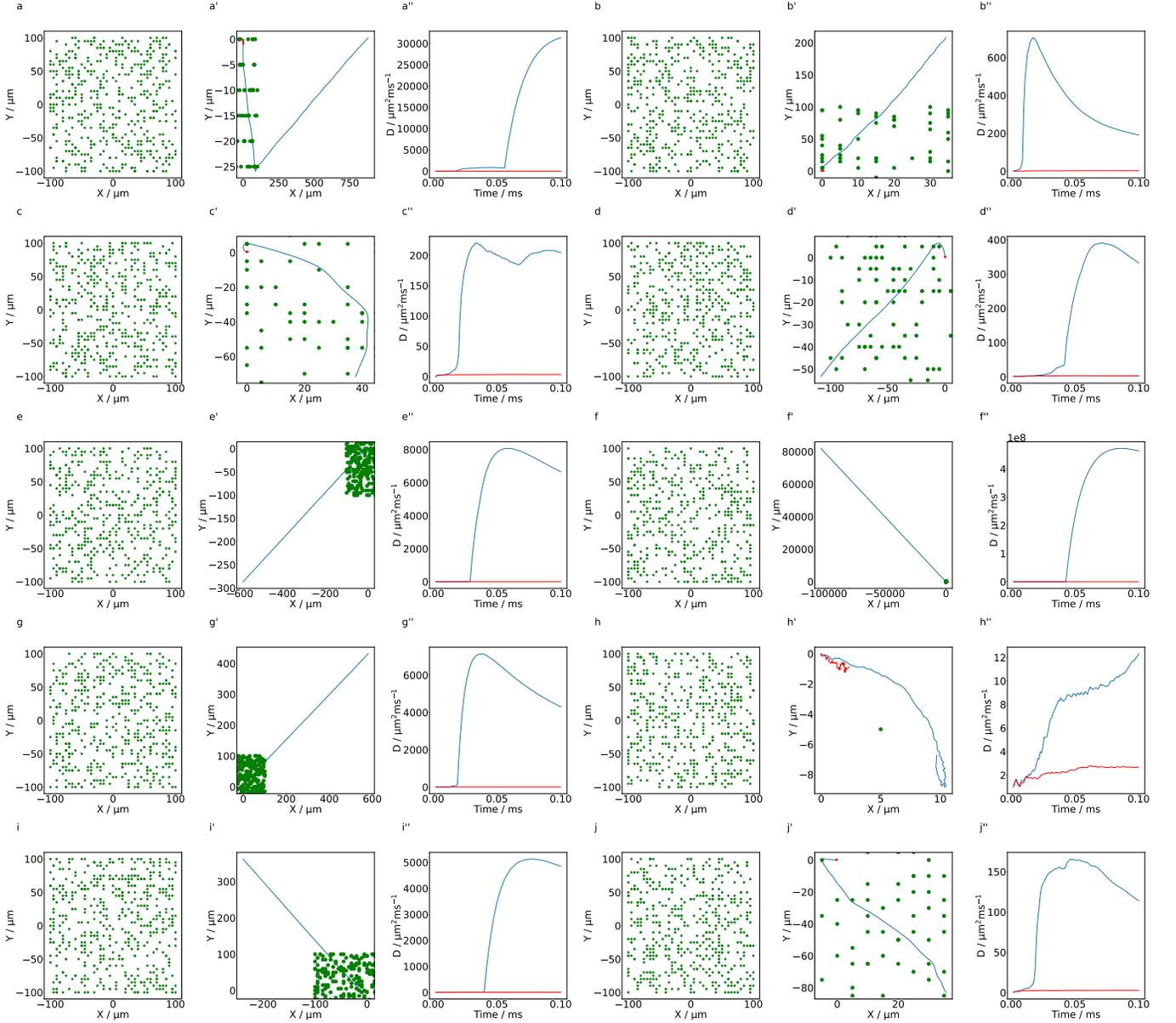}
    \caption{Simulated diffusion of a charged Atto-488-like single molecule under the influence of randomly positioned 500 surface charges (blue), compared to diffusion without surface charges (red), with details shown in insets. The particles always start at the origin (0 $\upmu$m, 0 $\upmu$m) and the minimum distance between two surface charges is 5 $\upmu$m.
    (a - j), surface charge distribution.
    (a' - j'), diffusion trajectories.
    (a" - j"), their respective diffusion coefficient plots.}
    \label{Figure:27}
\end{figure*}

\begin{figure*}[]
    \centering
    \includegraphics[width = 1\textwidth]{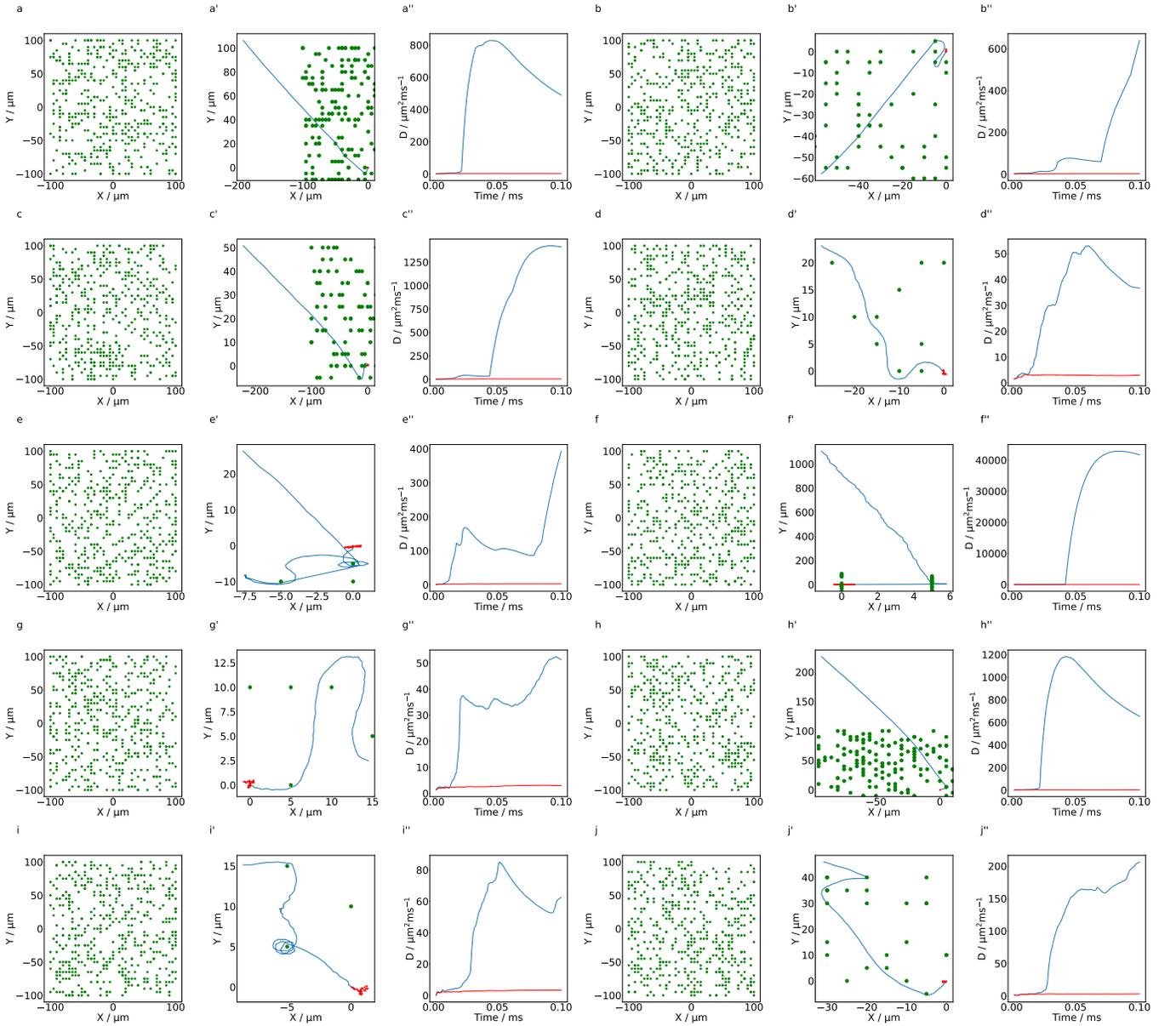}
    \caption{Simulated diffusion of a charged Atto-488-like single molecule under the influence of randomly positioned 500 surface charges (blue), compared to diffusion without surface charges (red), with details shown in insets. The particles always start at the origin (0 $\upmu$m, 0 $\upmu$m) and the minimum distance between two surface charges is 5 $\upmu$m.
    (a - j), surface charge distribution.
    (a' - j'), diffusion trajectories.
    (a" - j"), their respective diffusion coefficient plots.}
    \label{Figure:28}
\end{figure*}

\begin{figure*}[]
    \centering
    \includegraphics[width = 1\textwidth]{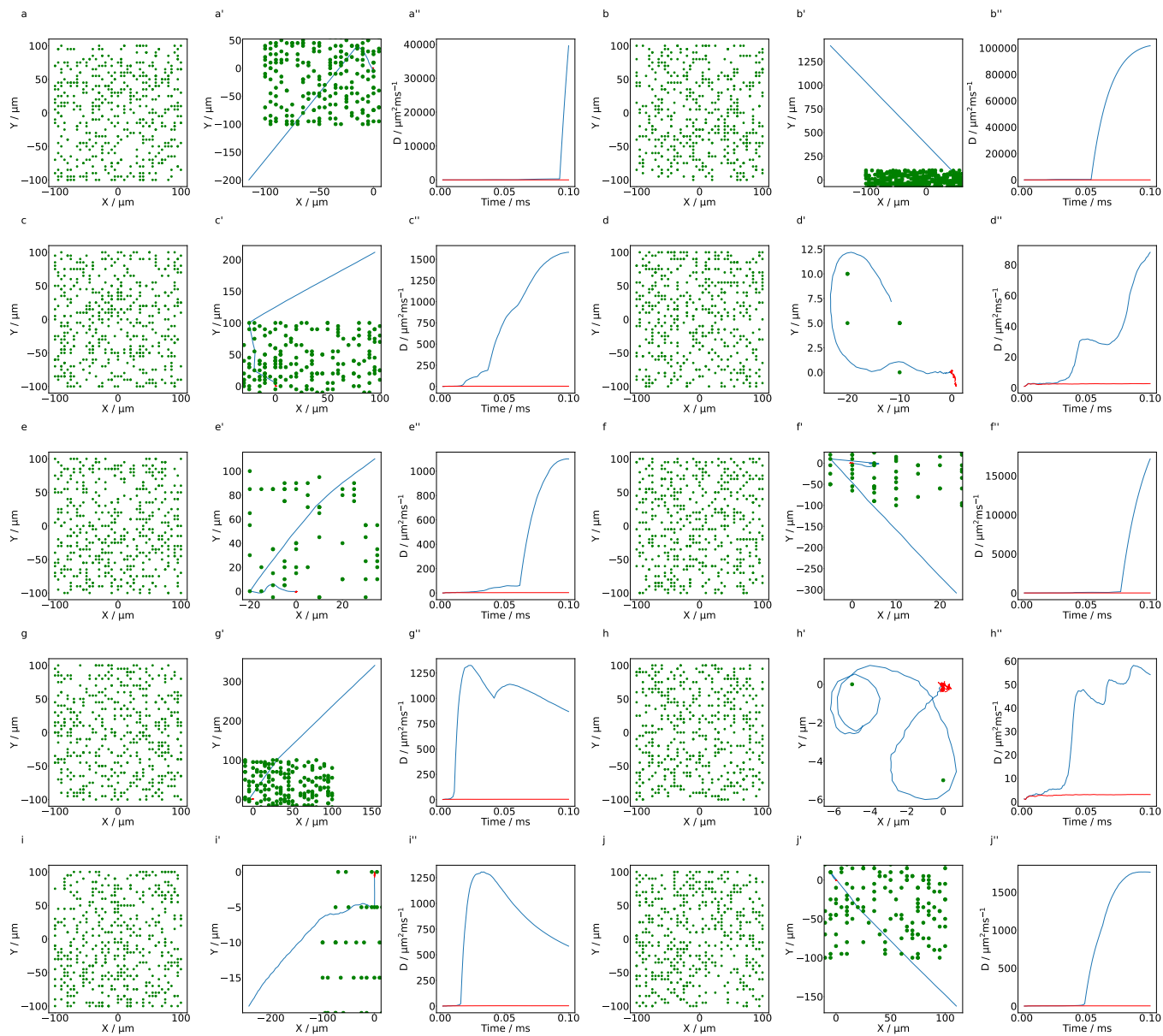}
    \caption{Simulated diffusion of a charged Atto-488-like single molecule under the influence of randomly positioned 500 surface charges (blue), compared to diffusion without surface charges (red), with details shown in insets. The particles always start at the origin (0 $\upmu$m, 0 $\upmu$m) and the minimum distance between two surface charges is 5 $\upmu$m.
    (a - j), surface charge distribution.
    (a' - j'), diffusion trajectories.
    (a" - j"), their respective diffusion coefficient plots.}
    \label{Figure:29}
\end{figure*}

\begin{figure*}[]
    \centering
    \includegraphics[width = 1\textwidth]{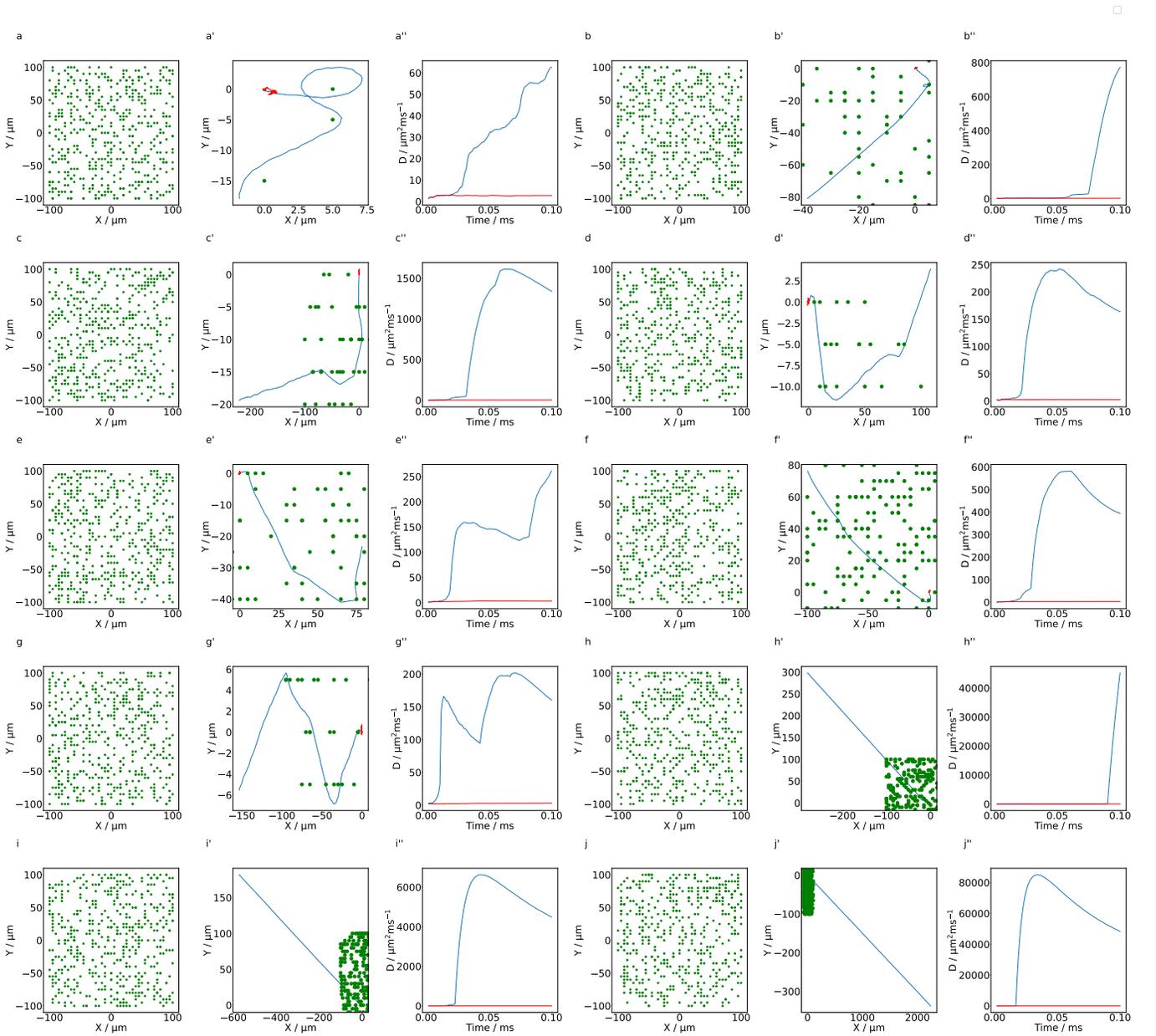}
    \caption{Simulated diffusion of a charged Atto-488-like single molecule under the influence of randomly positioned 500 surface charges (blue), compared to diffusion without surface charges (red), with details shown in insets. The particles always start at the origin (0 $\upmu$m, 0 $\upmu$m) and the minimum distance between two surface charges is 5 $\upmu$m.
    (a - j), surface charge distribution.
    (a' - j'), diffusion trajectories.
    (a" - j"), their respective diffusion coefficient plots.}
    \label{Figure:30}
\end{figure*}

\begin{figure*}[]
    \centering
    \includegraphics[width = 1\textwidth]{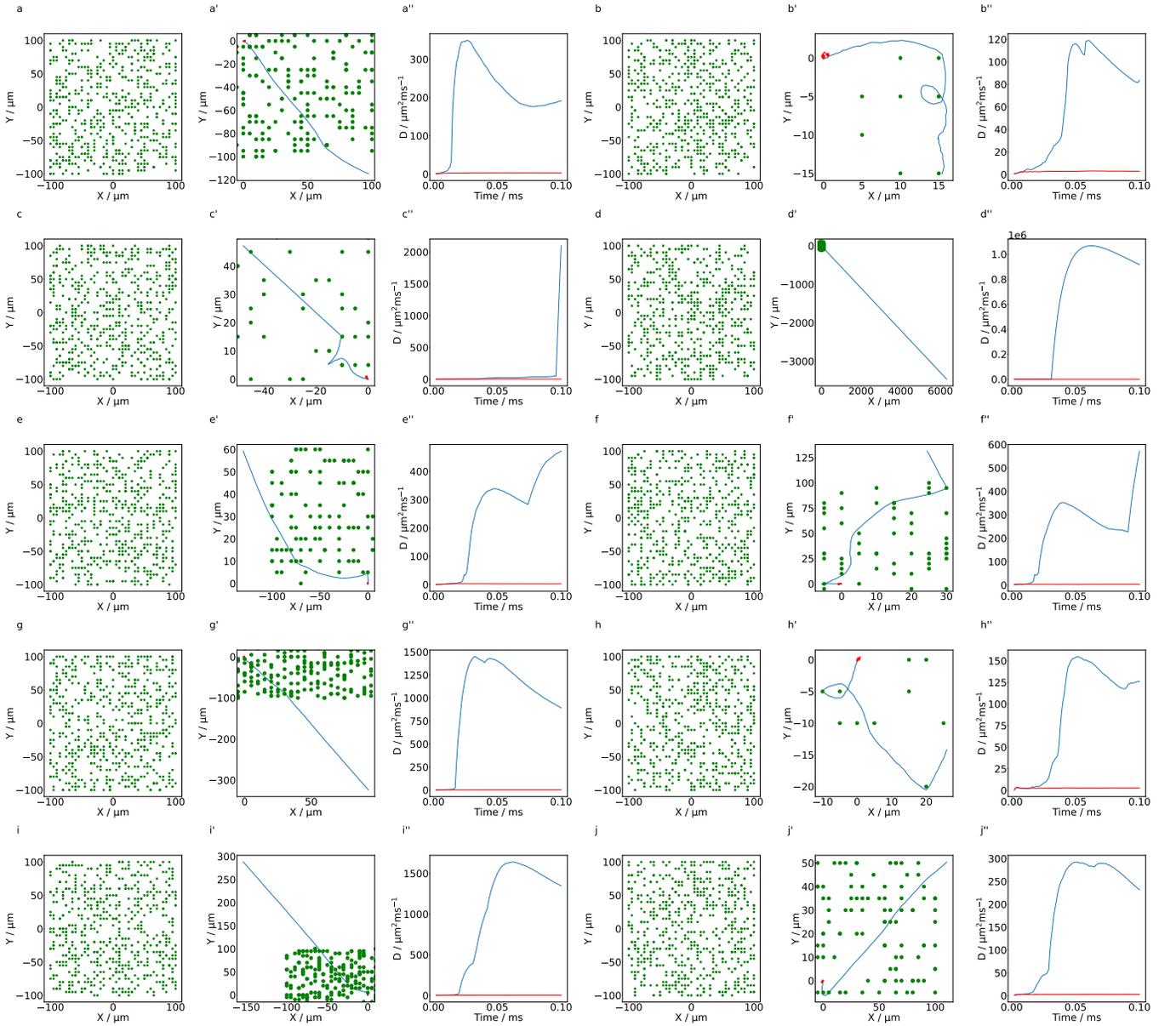}
    \caption{Simulated diffusion of a charged Atto-488-like single molecule under the influence of randomly positioned 600 surface charges (blue), compared to diffusion without surface charges (red), with details shown in insets. The particles always start at the origin (0 $\upmu$m, 0 $\upmu$m) and the minimum distance between two surface charges is 5 $\upmu$m.
    (a - j), surface charge distribution.
    (a' - j'), diffusion trajectories.
    (a" - j"), their respective diffusion coefficient plots.}
    \label{Figure:31}
\end{figure*}

\begin{figure*}[]
    \centering
    \includegraphics[width = 1\textwidth]{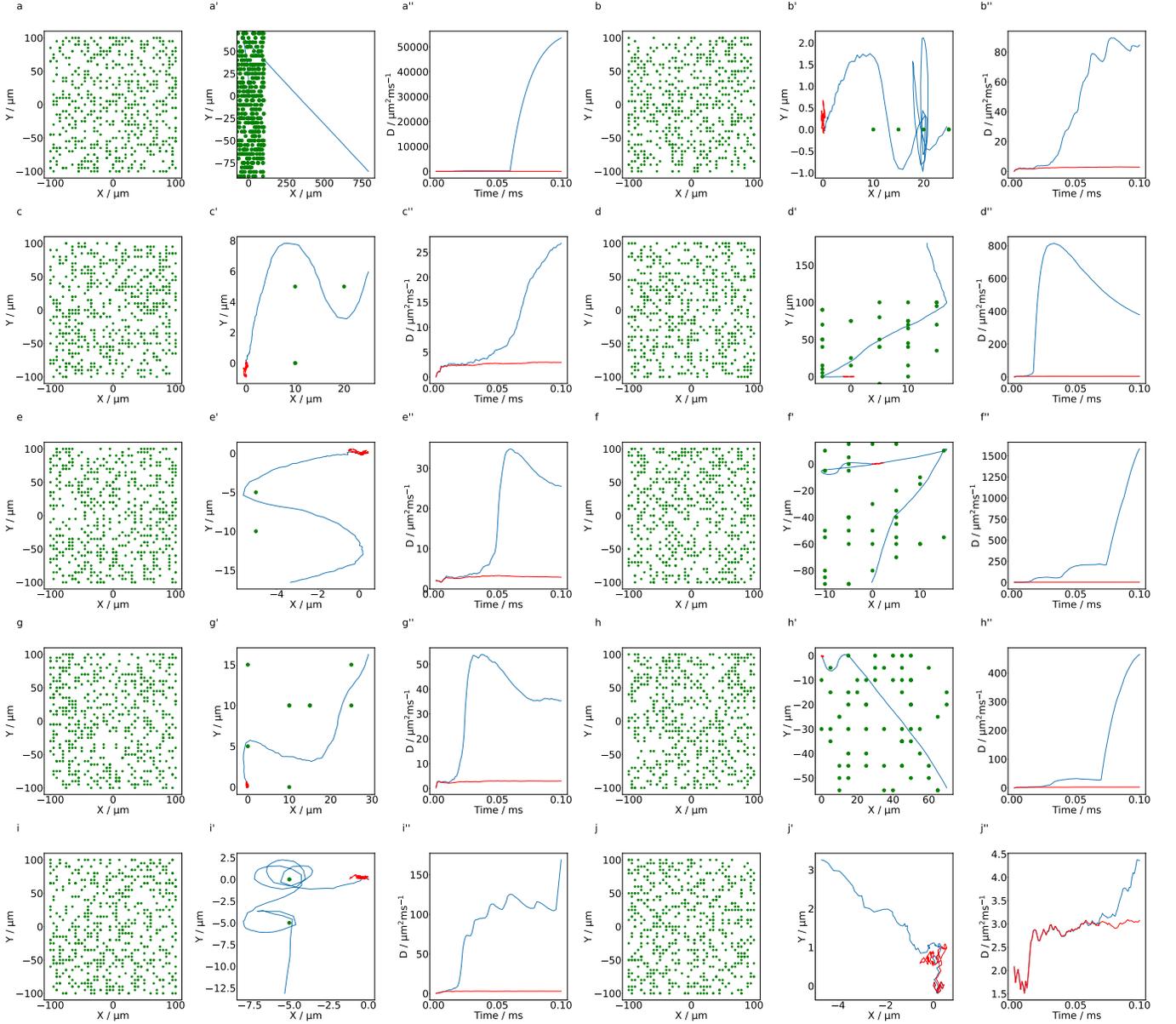}
    \caption{Simulated diffusion of a charged Atto-488-like single molecule under the influence of randomly positioned 600 surface charges (blue), compared to diffusion without surface charges (red), with details shown in insets. The particles always start at the origin (0 $\upmu$m, 0 $\upmu$m) and the minimum distance between two surface charges is 5 $\upmu$m.
    (a - j), surface charge distribution.
    (a' - j'), diffusion trajectories.
    (a" - j"), their respective diffusion coefficient plots.}
    \label{Figure:32}
\end{figure*}

\begin{figure*}[]
    \centering
    \includegraphics[width = 1\textwidth]{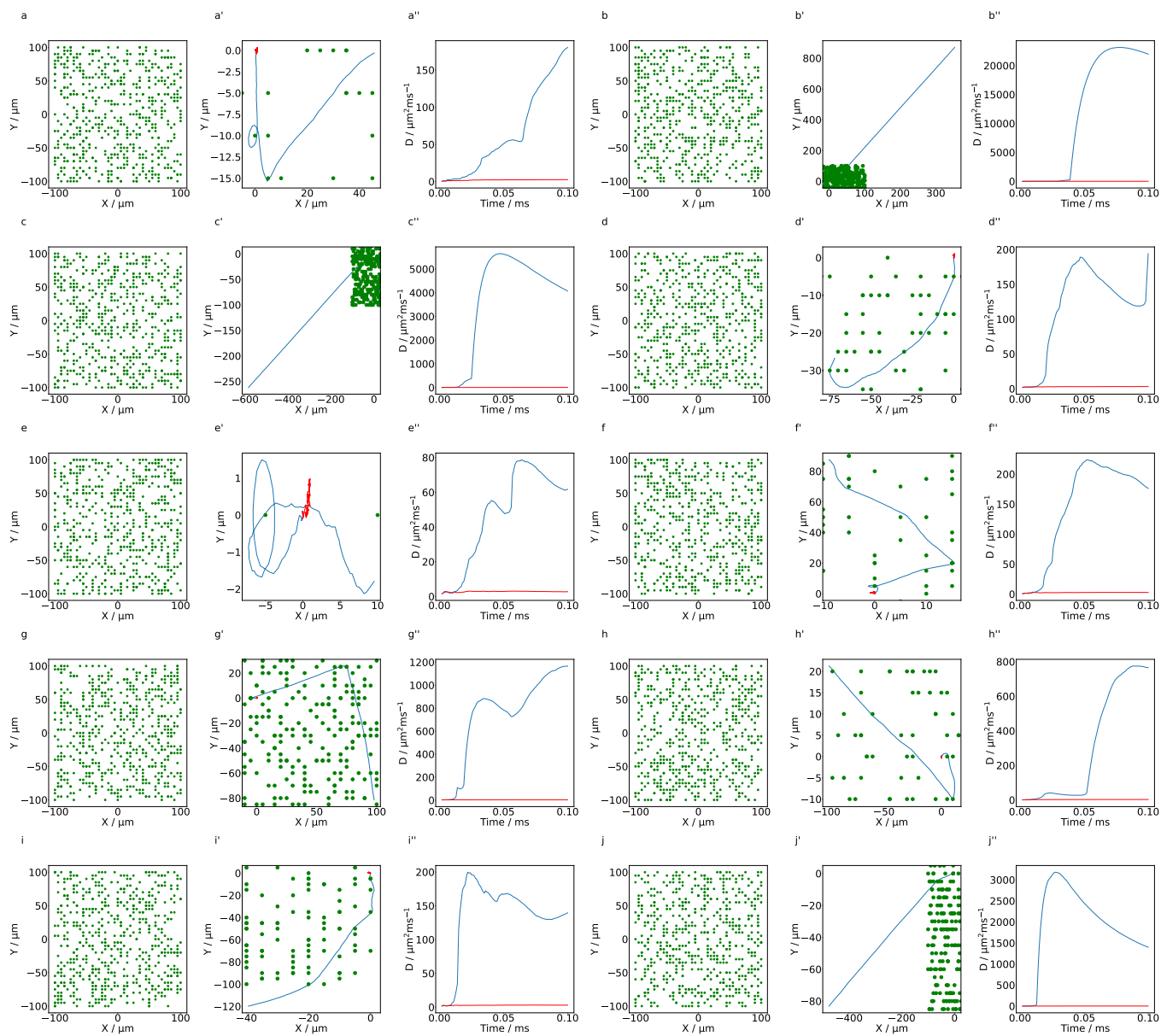}
    \caption{Simulated diffusion of a charged Atto-488-like single molecule under the influence of randomly positioned 600 surface charges (blue), compared to diffusion without surface charges (red), with details shown in insets. The particles always start at the origin (0 $\upmu$m, 0 $\upmu$m) and the minimum distance between two surface charges is 5 $\upmu$m.
    (a - j), surface charge distribution.
    (a' - j'), diffusion trajectories.
    (a" - j"), their respective diffusion coefficient plots.}
    \label{Figure:33}
\end{figure*}

\begin{figure*}[]
    \centering
    \includegraphics[width = 1\textwidth]{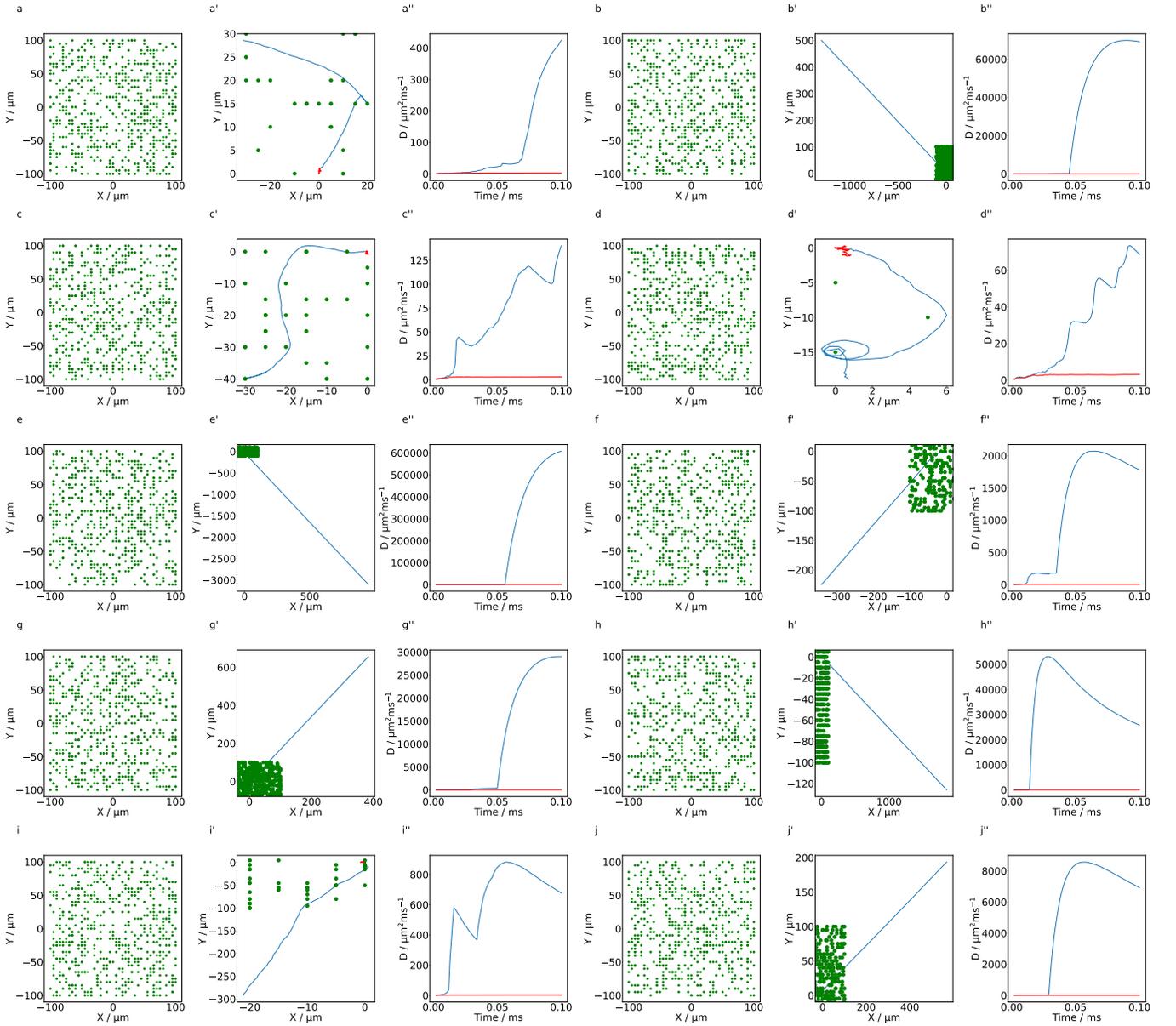}
    \caption{Simulated diffusion of a charged Atto-488-like single molecule under the influence of randomly positioned 600 surface charges (blue), compared to diffusion without surface charges (red), with details shown in insets. The particles always start at the origin (0 $\upmu$m, 0 $\upmu$m) and the minimum distance between two surface charges is 5 $\upmu$m.
    (a - j), surface charge distribution.
    (a' - j'), diffusion trajectories.
    (a" - j"), their respective diffusion coefficient plots.}
    \label{Figure:34}
\end{figure*}

\begin{figure*}[]
    \centering
    \includegraphics[width = 1\textwidth]{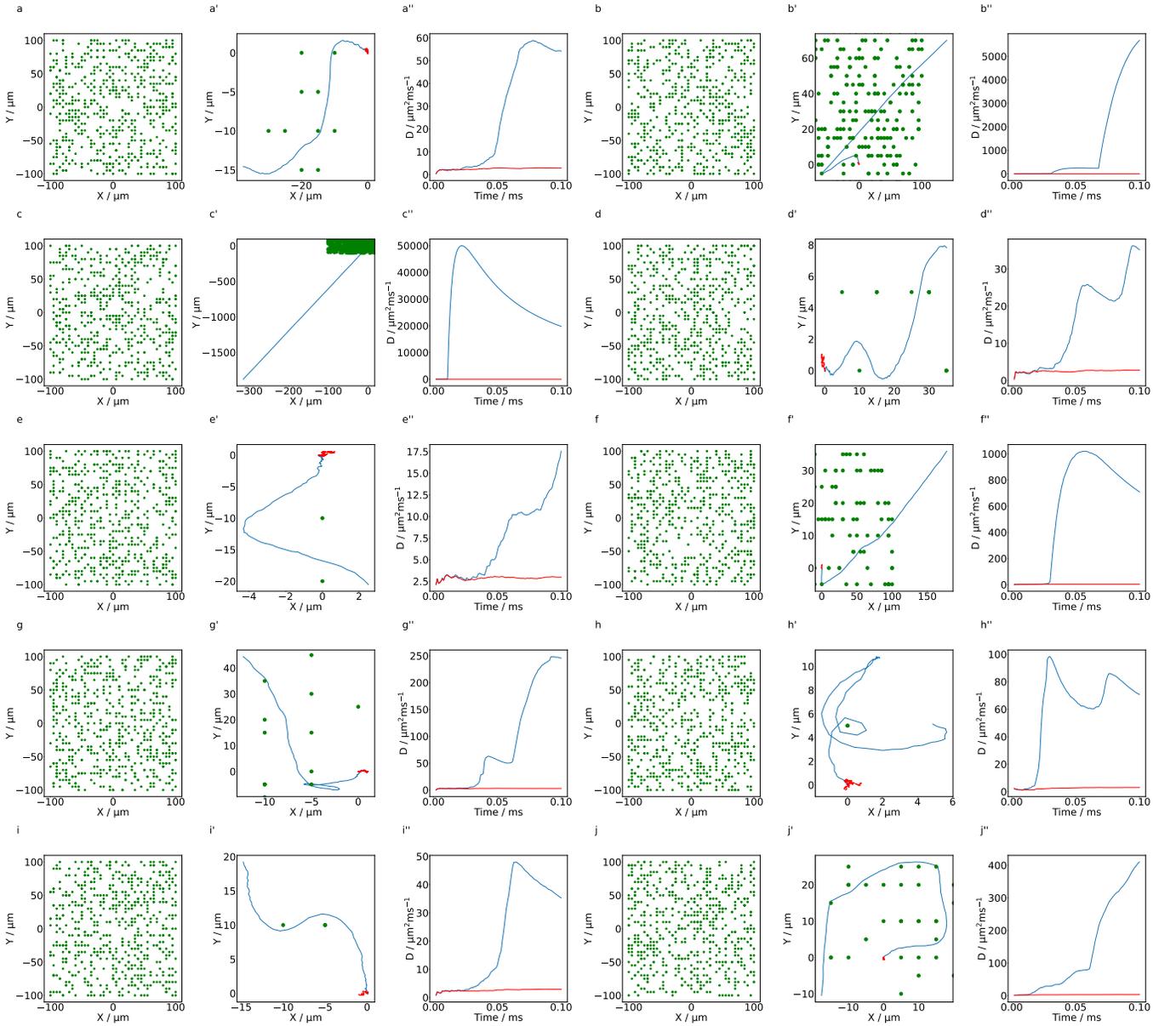}
    \caption{Simulated diffusion of a charged Atto-488-like single molecule under the influence of randomly positioned 600 surface charges (blue), compared to diffusion without surface charges (red), with details shown in insets. The particles always start at the origin (0 $\upmu$m, 0 $\upmu$m) and the minimum distance between two surface charges is 5 $\upmu$m.
    (a - j), surface charge distribution.
    (a' - j'), diffusion trajectories.
    (a" - j"), their respective diffusion coefficient plots.}
    \label{Figure:35}
\end{figure*}

\begin{figure*}[]
    \centering
    \includegraphics[width = 1\textwidth]{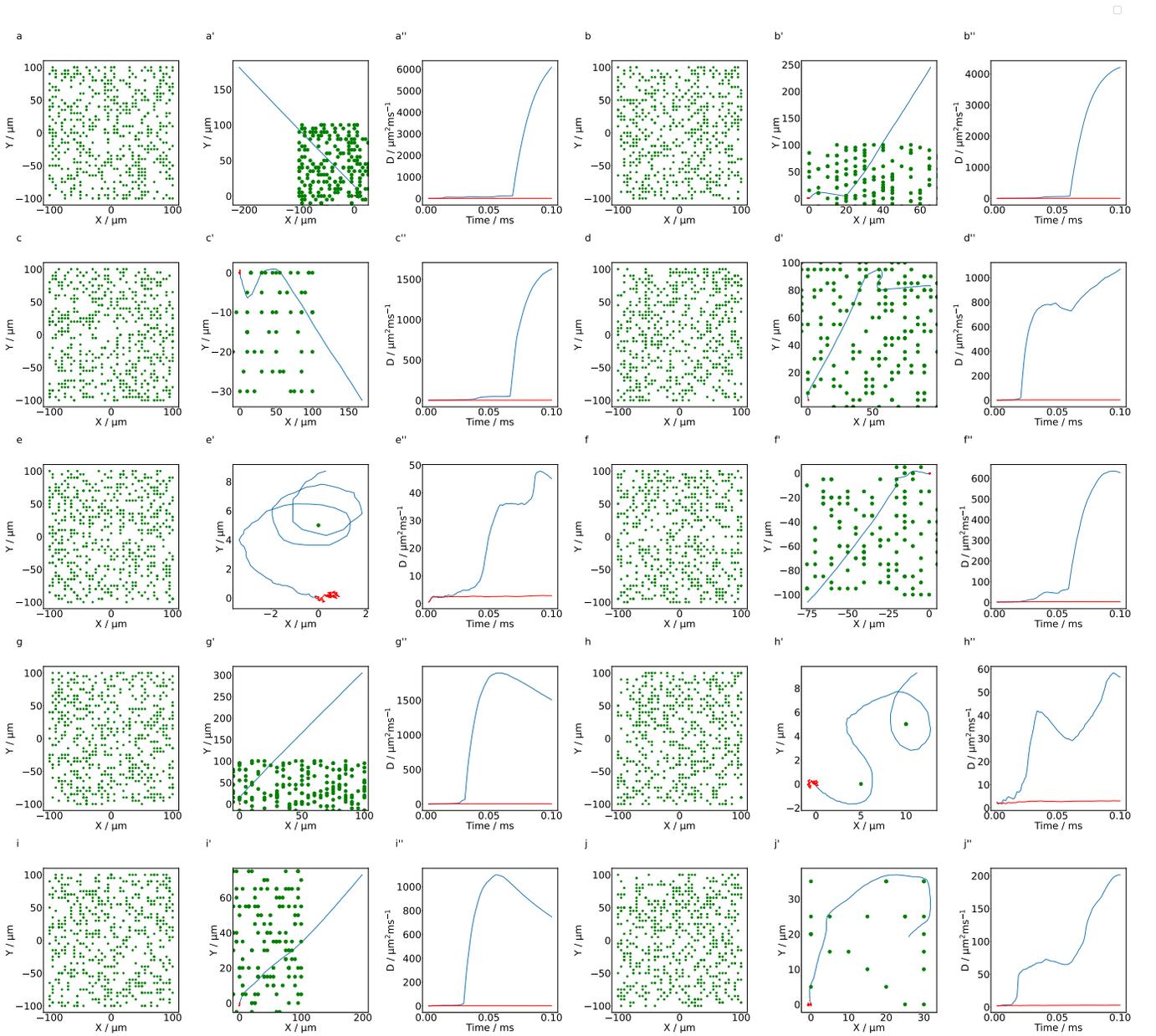}
    \caption{Simulated diffusion of a charged Atto-488-like single molecule under the influence of randomly positioned 600 surface charges (blue), compared to diffusion without surface charges (red), with details shown in insets. The particles always start at the origin (0 $\upmu$m, 0 $\upmu$m) and the minimum distance between two surface charges is 5 $\upmu$m.
    (a - j), surface charge distribution.
    (a' - j'), diffusion trajectories.
    (a" - j"), their respective diffusion coefficient plots.}
    \label{Figure:36}
\end{figure*}

\begin{figure*}[]
    \centering
    \includegraphics[width = 1\textwidth]{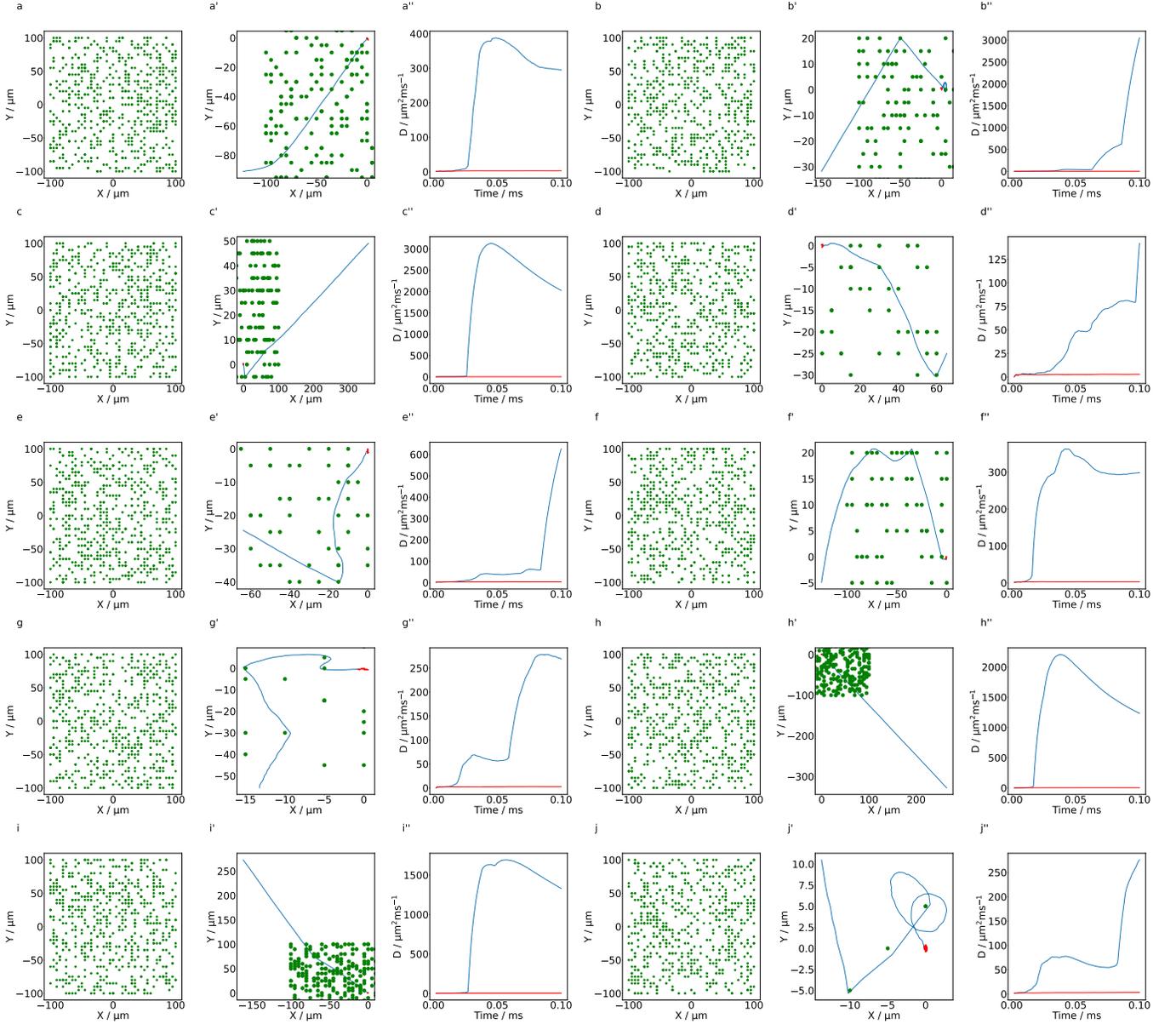}
    \caption{Simulated diffusion of a charged Atto-488-like single molecule under the influence of randomly positioned 600 surface charges (blue), compared to diffusion without surface charges (red), with details shown in insets. The particles always start at the origin (0 $\upmu$m, 0 $\upmu$m) and the minimum distance between two surface charges is 5 $\upmu$m.
    (a - j), surface charge distribution.
    (a' - j'), diffusion trajectories.
    (a" - j"), their respective diffusion coefficient plots.}
    \label{Figure:37}
\end{figure*}

\begin{figure*}[]
    \centering
    \includegraphics[width = 1\textwidth]{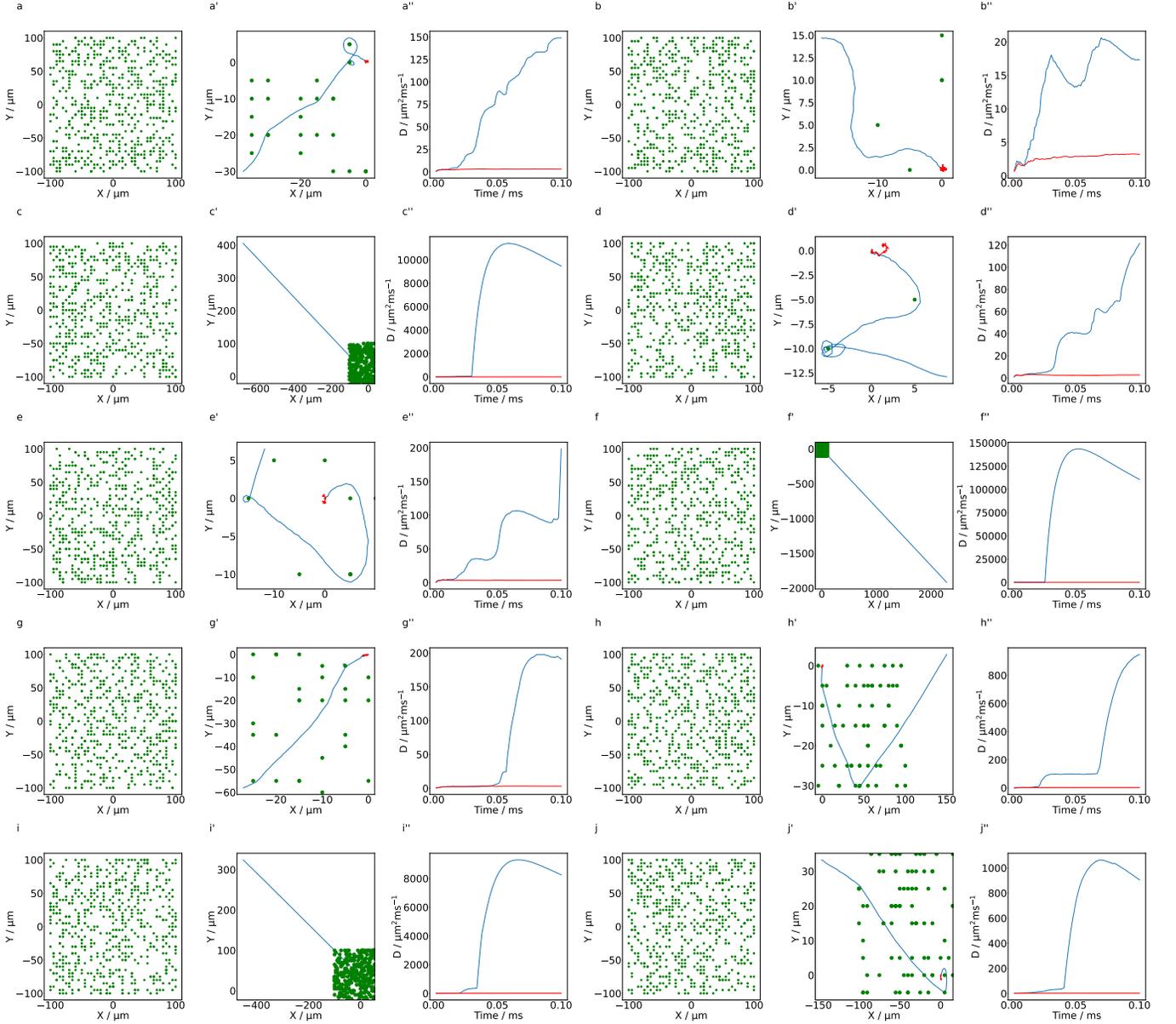}
    \caption{Simulated diffusion of a charged Atto-488-like single molecule under the influence of randomly positioned 600 surface charges (blue), compared to diffusion without surface charges (red), with details shown in insets. The particles always start at the origin (0 $\upmu$m, 0 $\upmu$m) and the minimum distance between two surface charges is 5 $\upmu$m.
    (a - j), surface charge distribution.
    (a' - j'), diffusion trajectories.
    (a" - j"), their respective diffusion coefficient plots.}
    \label{Figure:38}
\end{figure*}

\begin{figure*}[]
    \centering
    \includegraphics[width = 1\textwidth]{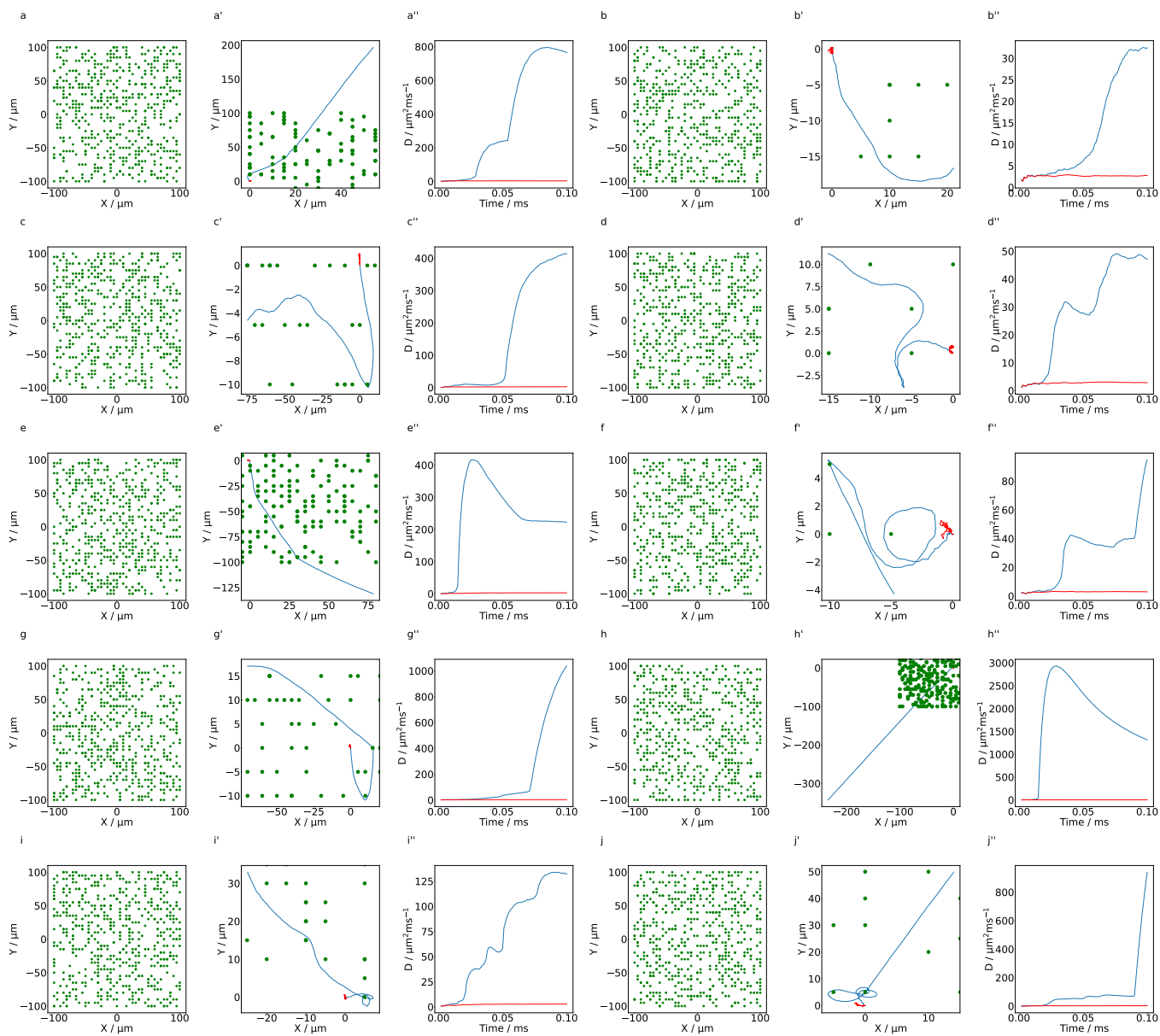}
    \caption{Simulated diffusion of a charged Atto-488-like single molecule under the influence of randomly positioned 600 surface charges (blue), compared to diffusion without surface charges (red), with details shown in insets. The particles always start at the origin (0 $\upmu$m, 0 $\upmu$m) and the minimum distance between two surface charges is 5 $\upmu$m.
    (a - j), surface charge distribution.
    (a' - j'), diffusion trajectories.
    (a" - j"), their respective diffusion coefficient plots.}
    \label{Figure:39}
\end{figure*}

\begin{figure*}[]
    \centering
    \includegraphics[width = 1\textwidth]{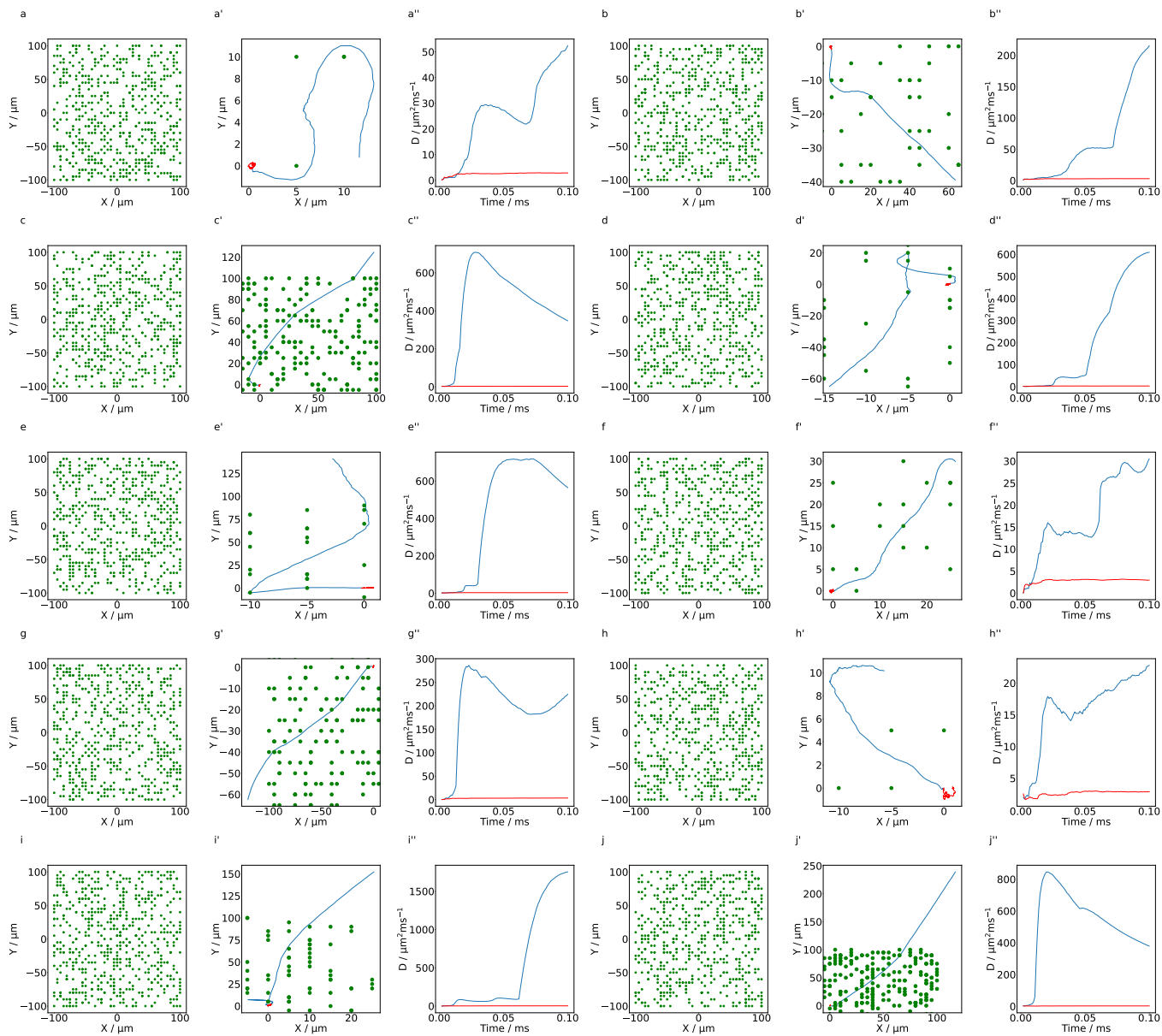}
    \caption{Simulated diffusion of a charged Atto-488-like single molecule under the influence of randomly positioned 600 surface charges (blue), compared to diffusion without surface charges (red), with details shown in insets. The particles always start at the origin (0 $\upmu$m, 0 $\upmu$m) and the minimum distance between two surface charges is 5 $\upmu$m.
    (a - j), surface charge distribution.
    (a' - j'), diffusion trajectories.
    (a" - j"), their respective diffusion coefficient plots.}
    \label{Figure:40}
\end{figure*}

\begin{figure*}[]
    \centering
    \includegraphics[width = 1\textwidth]{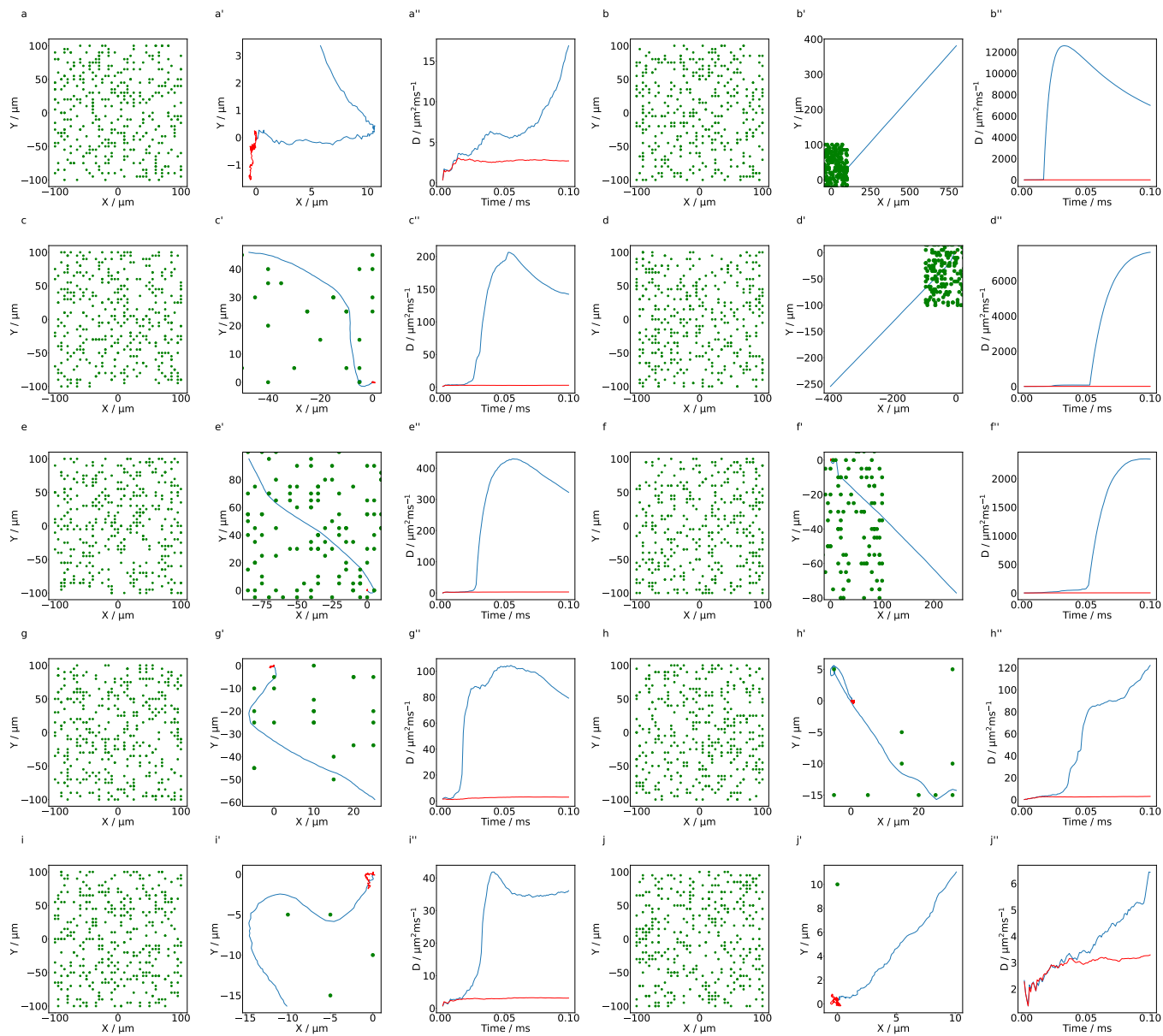}
    \caption{Simulated diffusion of a charged Atto-488-like single molecule under the influence of randomly positioned 400 surface charges (blue), compared to diffusion without surface charges (red), with details shown in insets. The particles always start at the origin (0 $\upmu$m, 0 $\upmu$m) and the minimum distance between two surface charges is 5 $\upmu$m.
    (a - j), surface charge distribution.
    (a' - j'), diffusion trajectories.
    (a" - j"), their respective diffusion coefficient plots.}
    \label{Figure:41}
\end{figure*}

\begin{figure*}[]
    \centering
    \includegraphics[width = 1\textwidth]{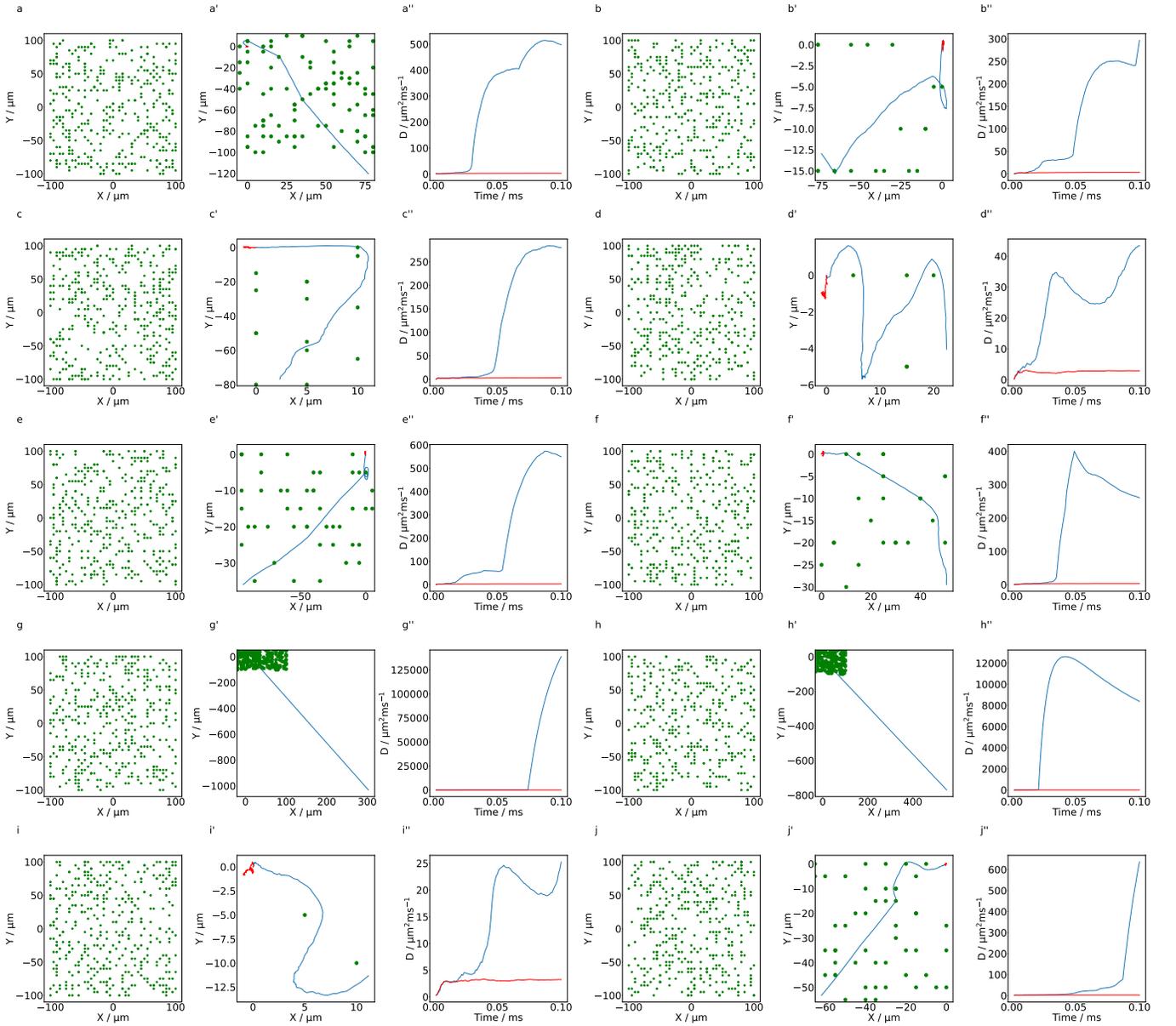}
    \caption{Simulated diffusion of a charged Atto-488-like single molecule under the influence of randomly positioned 400 surface charges (blue), compared to diffusion without surface charges (red), with details shown in insets. The particles always start at the origin (0 $\upmu$m, 0 $\upmu$m) and the minimum distance between two surface charges is 5 $\upmu$m.
    (a - j), surface charge distribution.
    (a' - j'), diffusion trajectories.
    (a" - j"), their respective diffusion coefficient plots.}
    \label{Figure:42}
\end{figure*}

\begin{figure*}[]
    \centering
    \includegraphics[width = 1\textwidth]{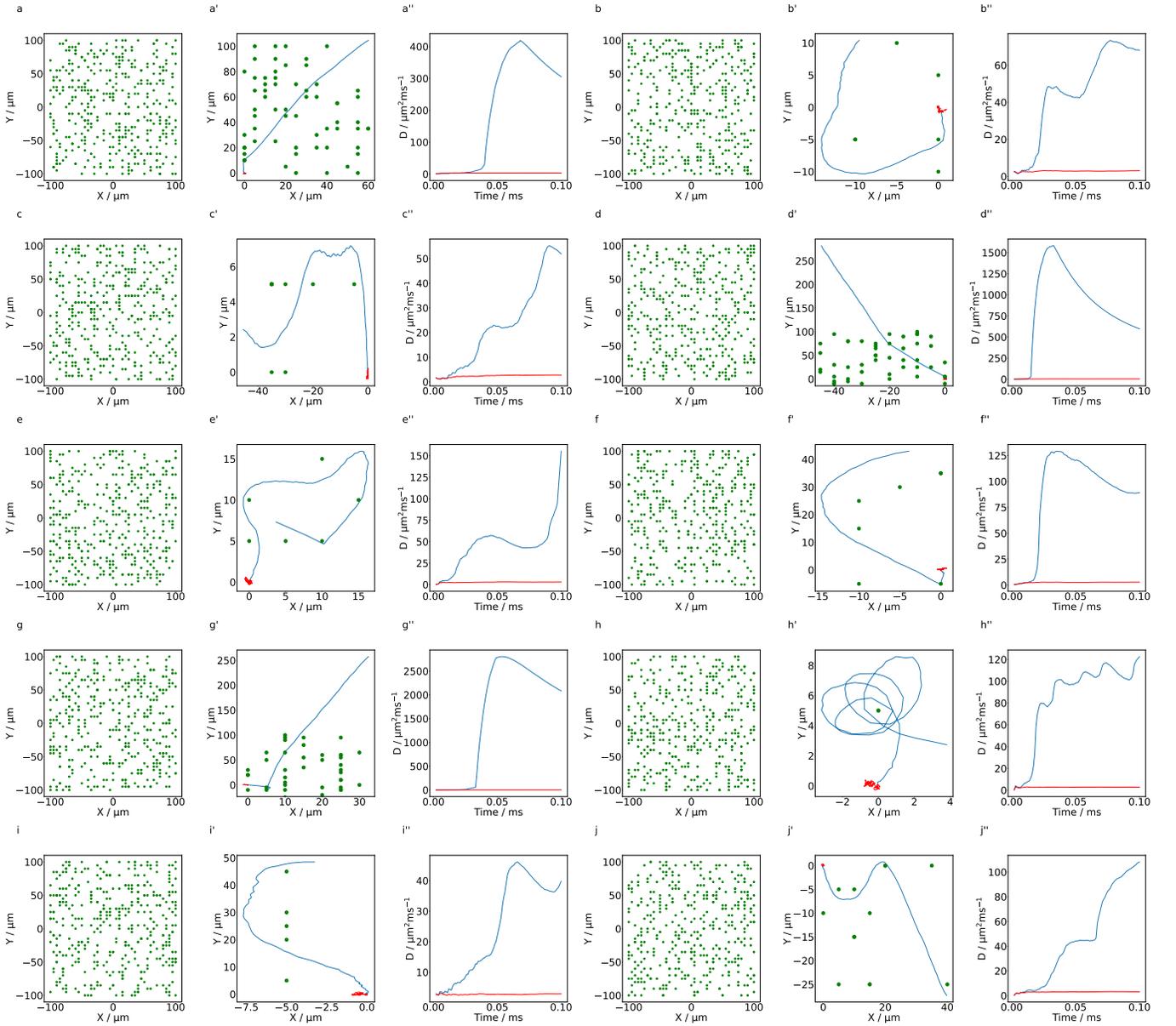}
    \caption{Simulated diffusion of a charged Atto-488-like single molecule under the influence of randomly positioned 400 surface charges (blue), compared to diffusion without surface charges (red), with details shown in insets. The particles always start at the origin (0 $\upmu$m, 0 $\upmu$m) and the minimum distance between two surface charges is 5 $\upmu$m.
    (a - j), surface charge distribution.
    (a' - j'), diffusion trajectories.
    (a" - j"), their respective diffusion coefficient plots.}
    \label{Figure:43}
\end{figure*}

\begin{figure*}[]
    \centering
    \includegraphics[width = 1\textwidth]{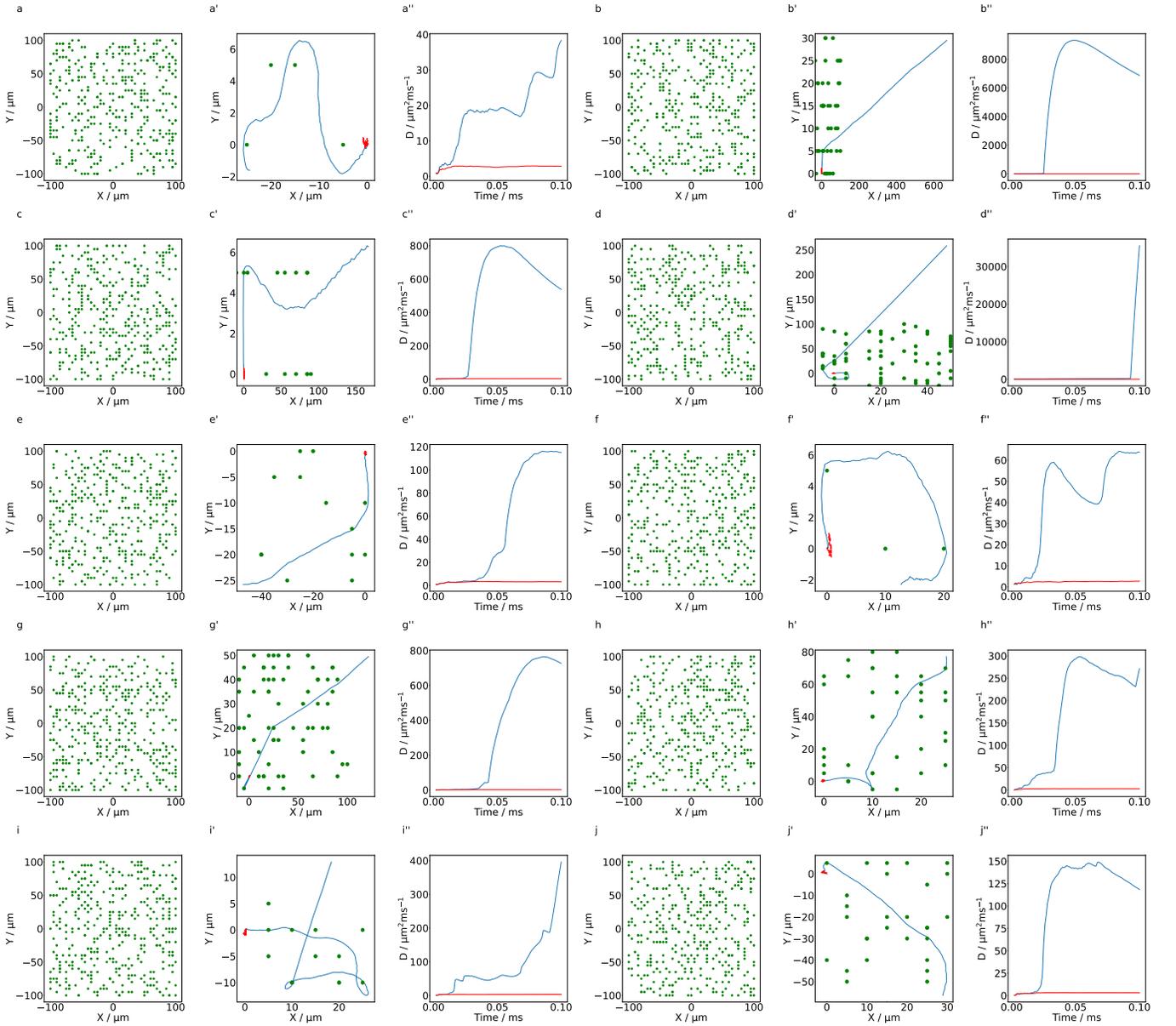}
    \caption{Simulated diffusion of a charged Atto-488-like single molecule under the influence of randomly positioned 400 surface charges (blue), compared to diffusion without surface charges (red), with details shown in insets. The particles always start at the origin (0 $\upmu$m, 0 $\upmu$m) and the minimum distance between two surface charges is 5 $\upmu$m.
    (a - j), surface charge distribution.
    (a' - j'), diffusion trajectories.
    (a" - j"), their respective diffusion coefficient plots.}
    \label{Figure:44}
\end{figure*}

\begin{figure*}[]
    \centering
    \includegraphics[width = 1\textwidth]{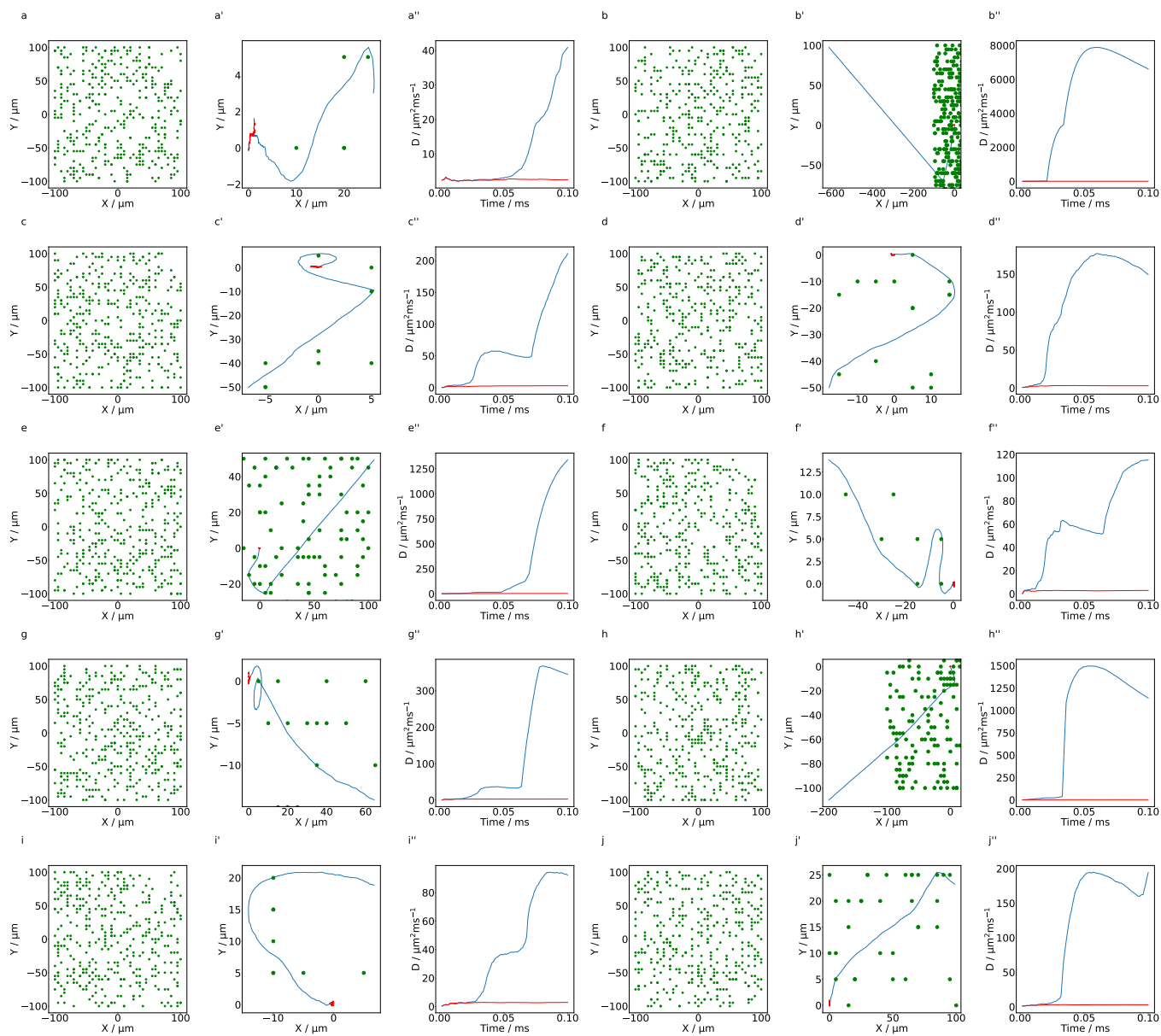}
    \caption{Simulated diffusion of a charged Atto-488-like single molecule under the influence of randomly positioned 400 surface charges (blue), compared to diffusion without surface charges (red), with details shown in insets. The particles always start at the origin (0 $\upmu$m, 0 $\upmu$m) and the minimum distance between two surface charges is 5 $\upmu$m.
    (a - j), surface charge distribution.
    (a' - j'), diffusion trajectories.
    (a" - j"), their respective diffusion coefficient plots.}
    \label{Figure:45}
\end{figure*}

\begin{figure*}[]
    \centering
    \includegraphics[width = 1\textwidth]{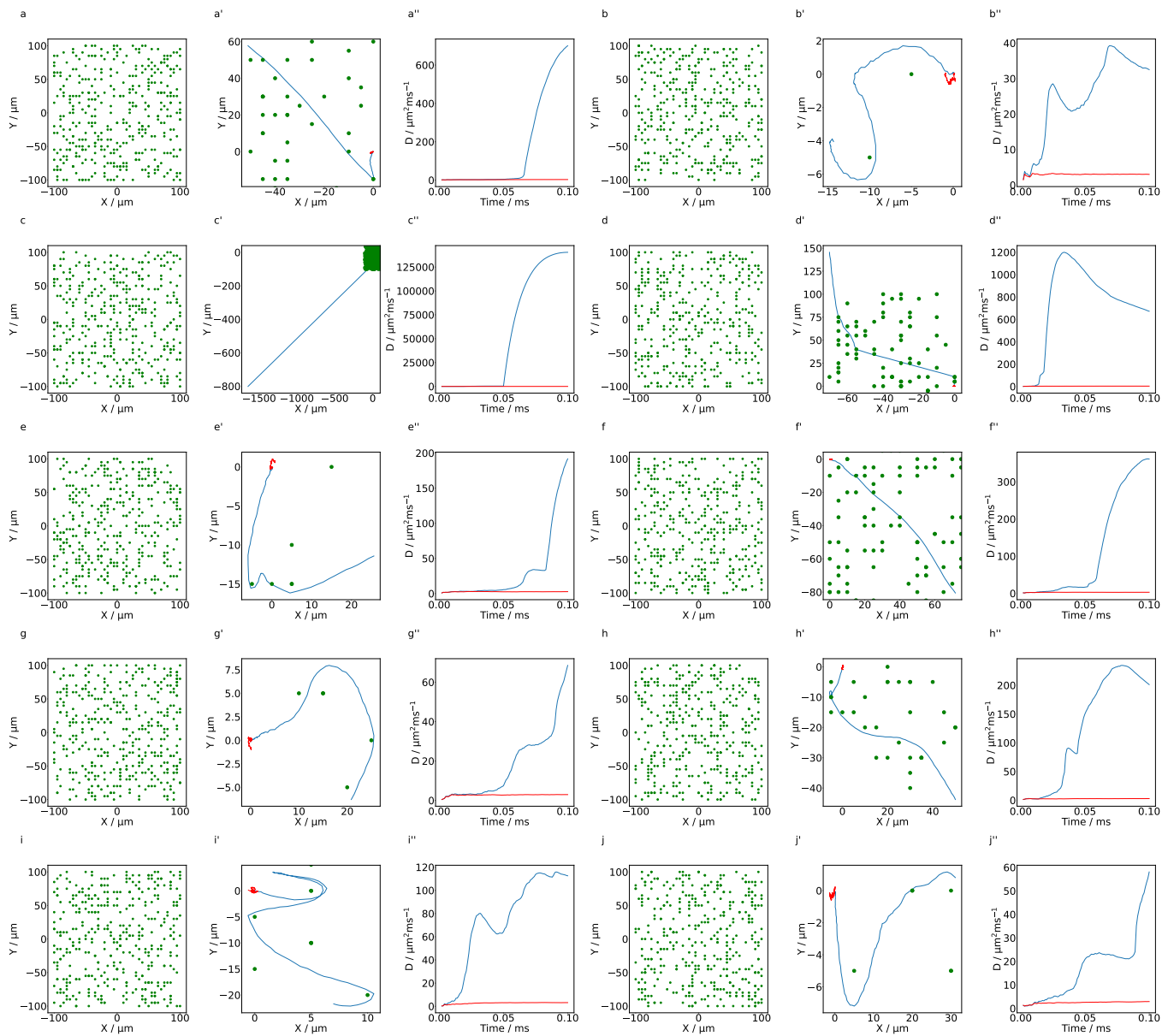}
    \caption{Simulated diffusion of a charged Atto-488-like single molecule under the influence of randomly positioned 400 surface charges (blue), compared to diffusion without surface charges (red), with details shown in insets. The particles always start at the origin (0 $\upmu$m, 0 $\upmu$m) and the minimum distance between two surface charges is 5 $\upmu$m.
    (a - j), surface charge distribution.
    (a' - j'), diffusion trajectories.
    (a" - j"), their respective diffusion coefficient plots.}
    \label{Figure:46}
\end{figure*}

\begin{figure*}[]
    \centering
    \includegraphics[width = 1\textwidth]{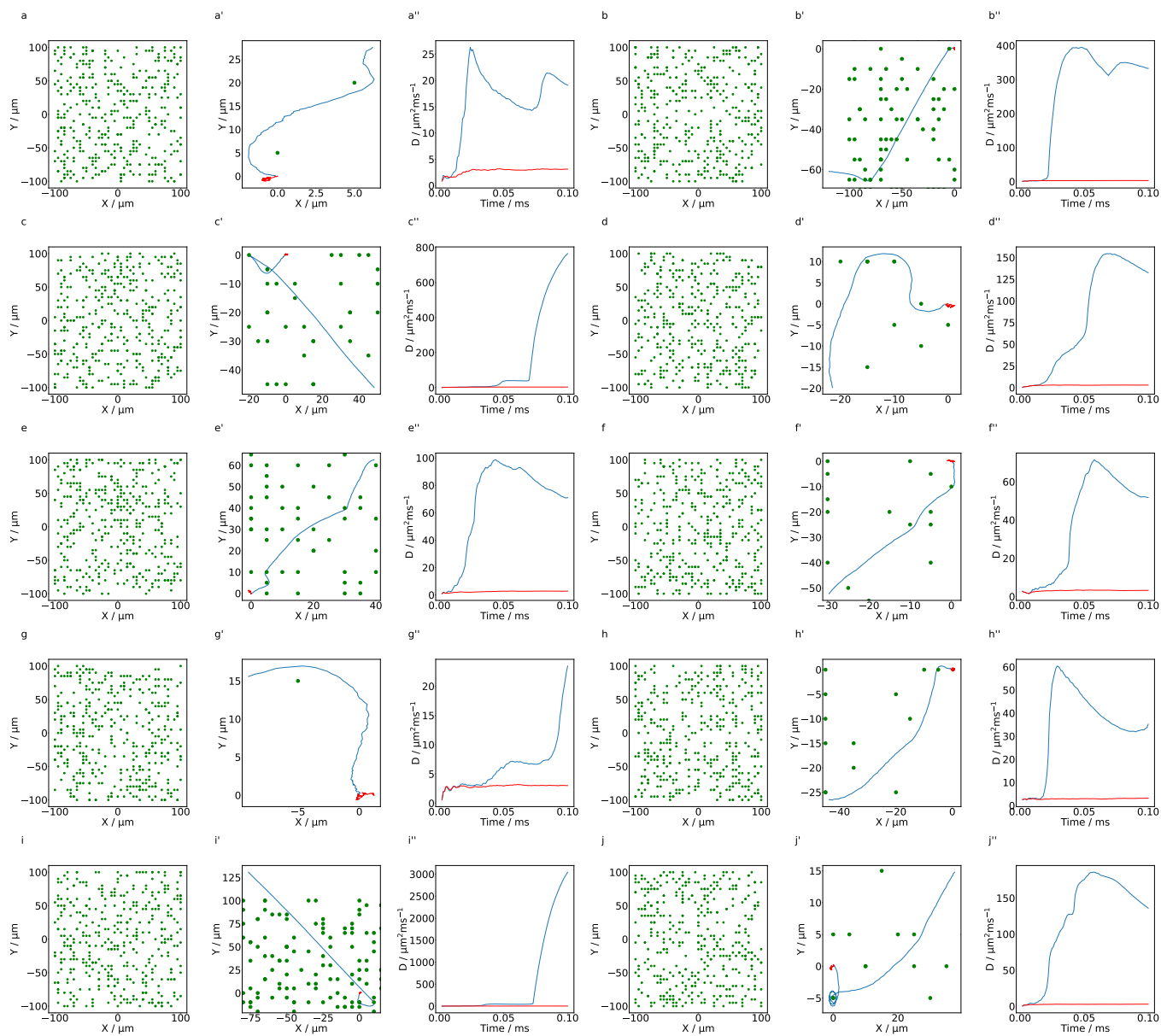}
    \caption{Simulated diffusion of a charged Atto-488-like single molecule under the influence of randomly positioned 400 surface charges (blue), compared to diffusion without surface charges (red), with details shown in insets. The particles always start at the origin (0 $\upmu$m, 0 $\upmu$m) and the minimum distance between two surface charges is 5 $\upmu$m.
    (a - j), surface charge distribution.
    (a' - j'), diffusion trajectories.
    (a" - j"), their respective diffusion coefficient plots.}
    \label{Figure:47}
\end{figure*}

\begin{figure*}[]
    \centering
    \includegraphics[width = 1\textwidth]{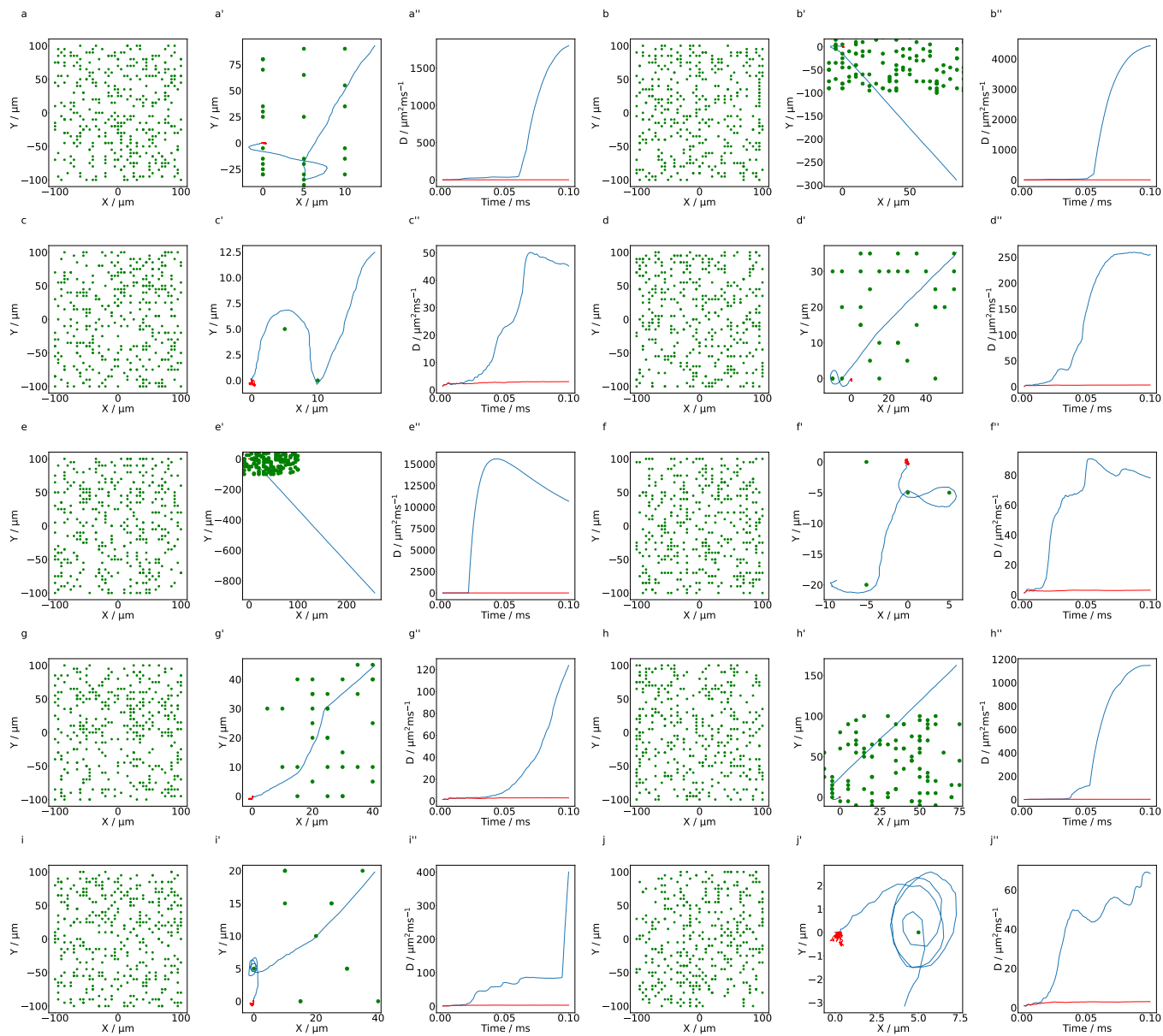}
    \caption{Simulated diffusion of a charged Atto-488-like single molecule under the influence of randomly positioned 400 surface charges (blue), compared to diffusion without surface charges (red), with details shown in insets. The particles always start at the origin (0 $\upmu$m, 0 $\upmu$m) and the minimum distance between two surface charges is 5 $\upmu$m.
    (a - j), surface charge distribution.
    (a' - j'), diffusion trajectories.
    (a" - j"), their respective diffusion coefficient plots.}
    \label{Figure:48}
\end{figure*}

\begin{figure*}[]
    \centering
    \includegraphics[width = 1\textwidth]{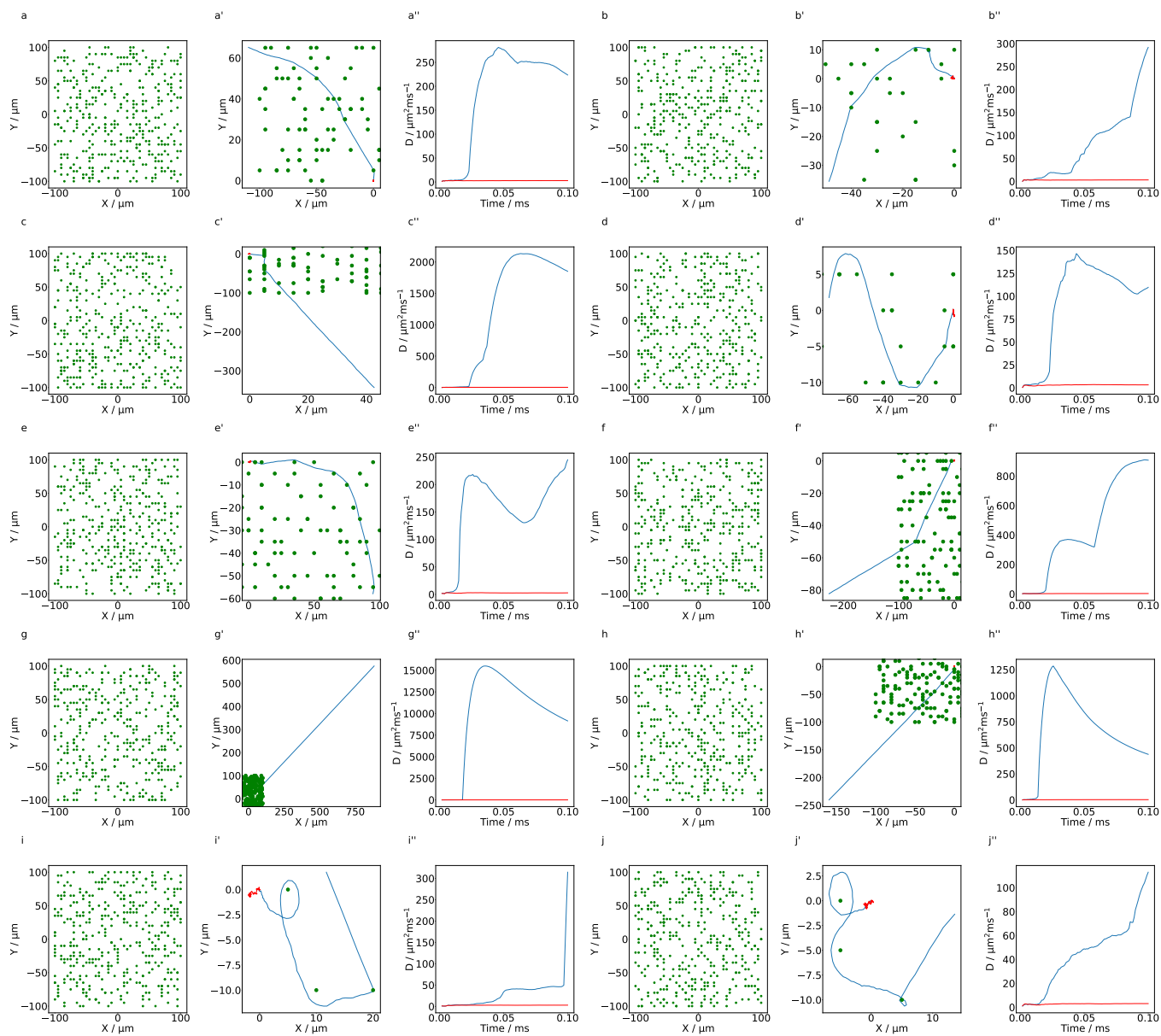}
    \caption{Simulated diffusion of a charged Atto-488-like single molecule under the influence of randomly positioned 400 surface charges (blue), compared to diffusion without surface charges (red), with details shown in insets. The particles always start at the origin (0 $\upmu$m, 0 $\upmu$m) and the minimum distance between two surface charges is 5 $\upmu$m.
    (a - j), surface charge distribution.
    (a' - j'), diffusion trajectories.
    (a" - j"), their respective diffusion coefficient plots.}
    \label{Figure:49}
\end{figure*}

\begin{figure*}[]
    \centering
    \includegraphics[width = 1\textwidth]{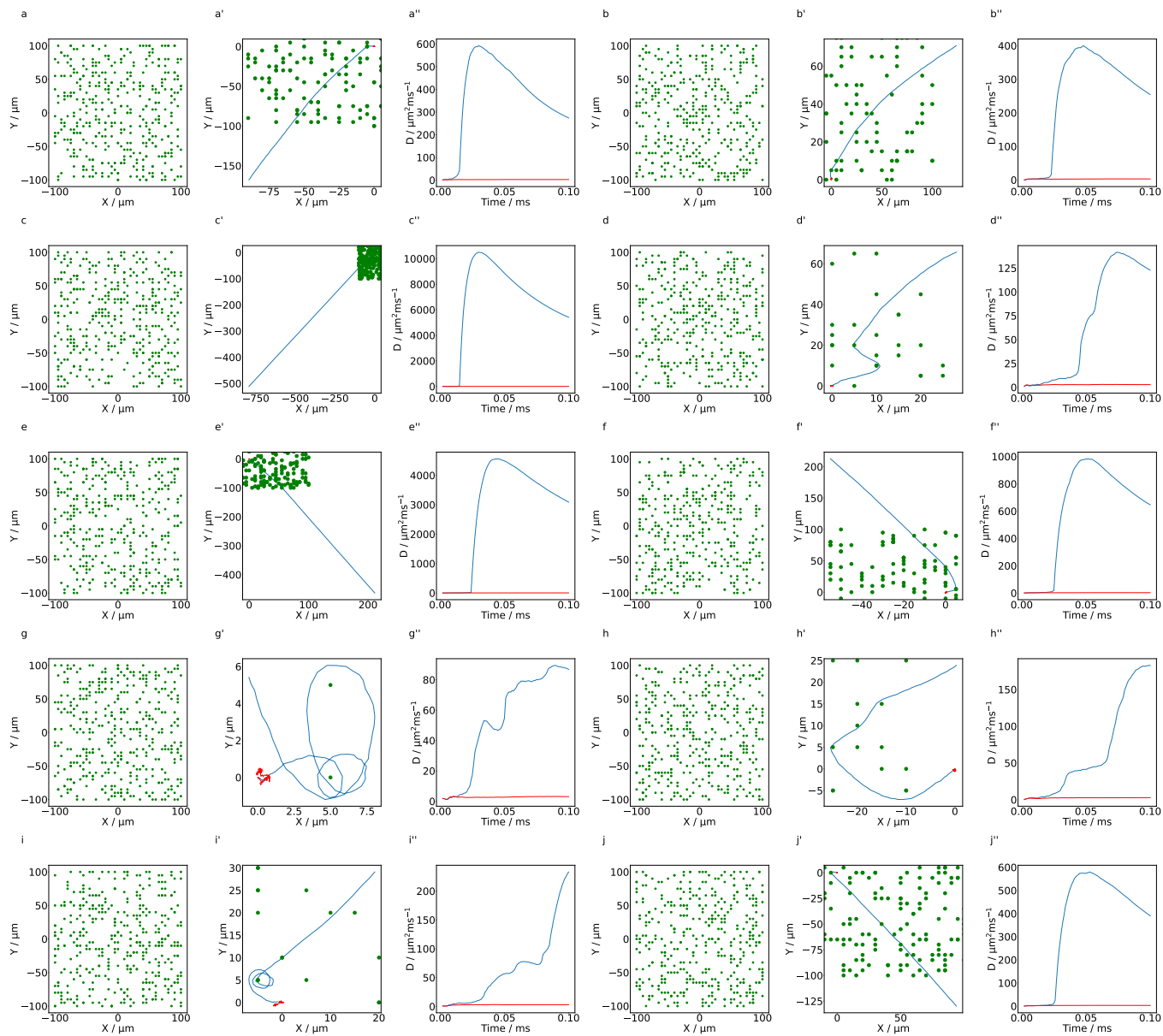}
    \caption{Simulated diffusion of a charged Atto-488-like single molecule under the influence of randomly positioned 400 surface charges (blue), compared to diffusion without surface charges (red), with details shown in insets. The particles always start at the origin (0 $\upmu$m, 0 $\upmu$m) and the minimum distance between two surface charges is 5 $\upmu$m.
    (a - j), surface charge distribution.
    (a' - j'), diffusion trajectories.
    (a" - j"), their respective diffusion coefficient plots.}
    \label{Figure:50}
\end{figure*}

\begin{figure*}[]
    \centering
    \includegraphics[width = 1\textwidth]{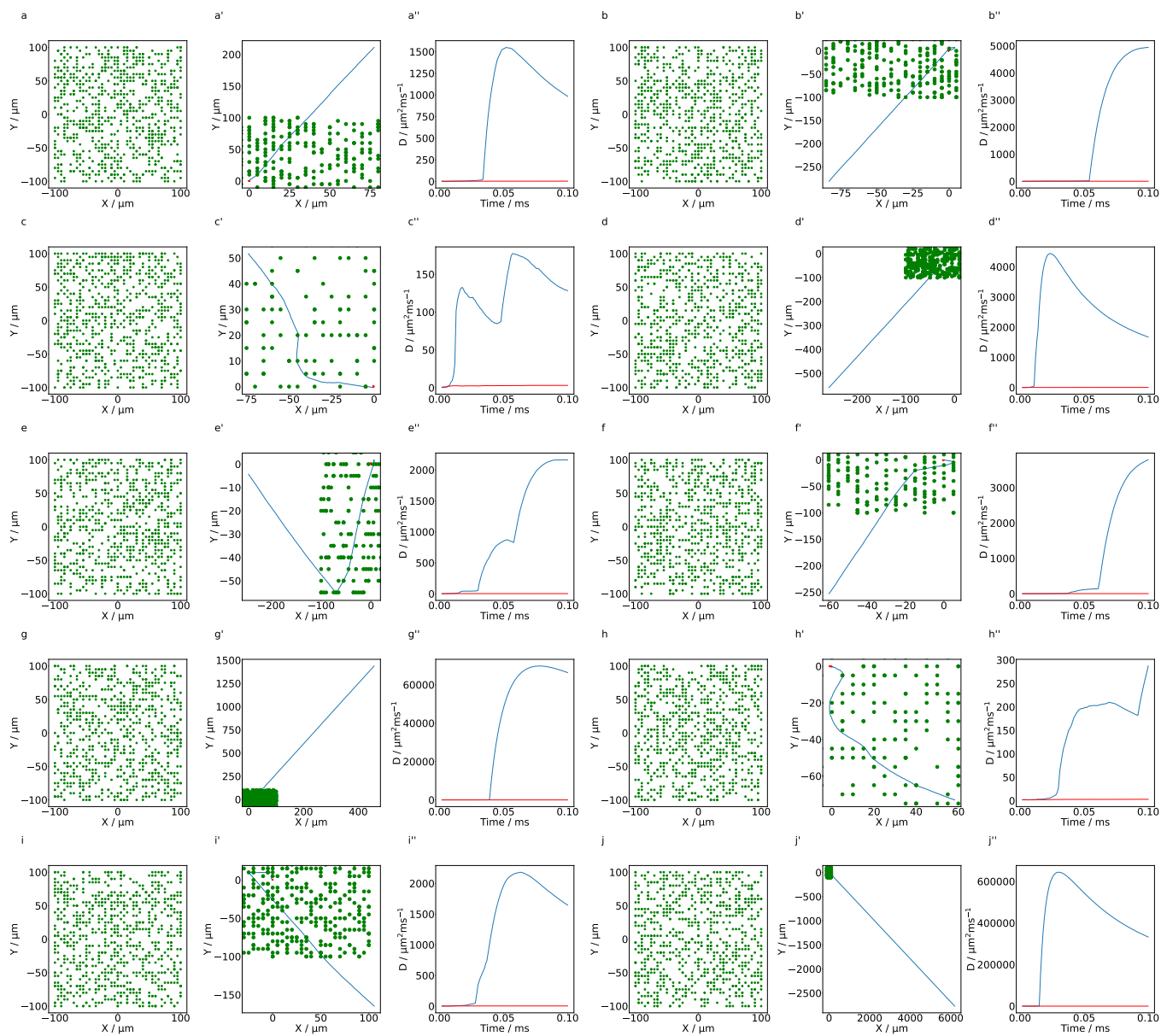}
    \caption{Simulated diffusion of a charged Atto-488-like single molecule under the influence of randomly positioned 800 surface charges (blue), compared to diffusion without surface charges (red), with details shown in insets. The particles always start at the origin (0 $\upmu$m, 0 $\upmu$m) and the minimum distance between two surface charges is 5 $\upmu$m.
    (a - j), surface charge distribution.
    (a' - j'), diffusion trajectories.
    (a" - j"), their respective diffusion coefficient plots.}
    \label{Figure:51}
\end{figure*}

\begin{figure*}[]
    \centering
    \includegraphics[width = 1\textwidth]{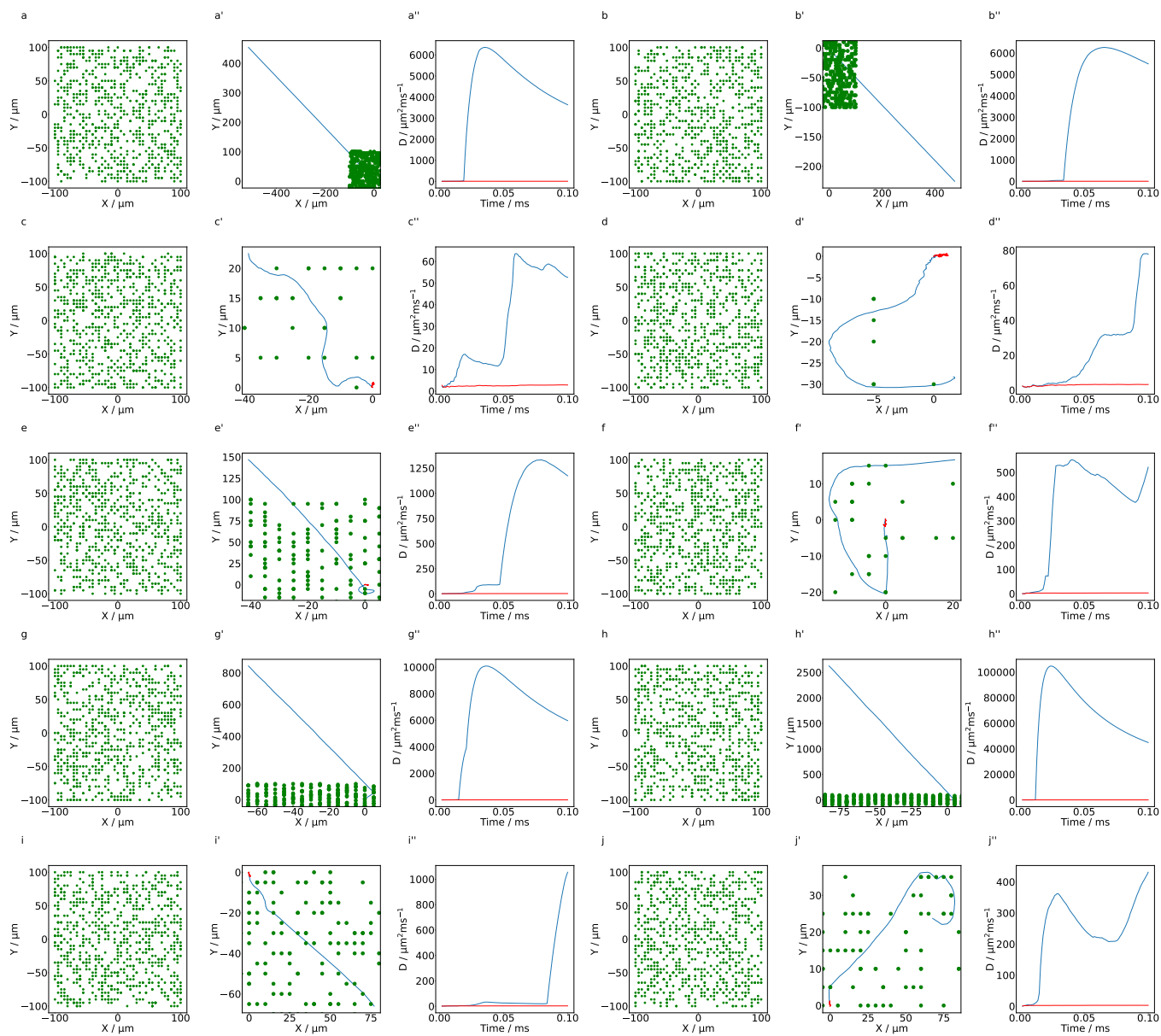}
    \caption{Simulated diffusion of a charged Atto-488-like single molecule under the influence of randomly positioned 800 surface charges (blue), compared to diffusion without surface charges (red), with details shown in insets. The particles always start at the origin (0 $\upmu$m, 0 $\upmu$m) and the minimum distance between two surface charges is 5 $\upmu$m.
    (a - j), surface charge distribution.
    (a' - j'), diffusion trajectories.
    (a" - j"), their respective diffusion coefficient plots.}
    \label{Figure:52}
\end{figure*}

\begin{figure*}[]
    \centering
    \includegraphics[width = 1\textwidth]{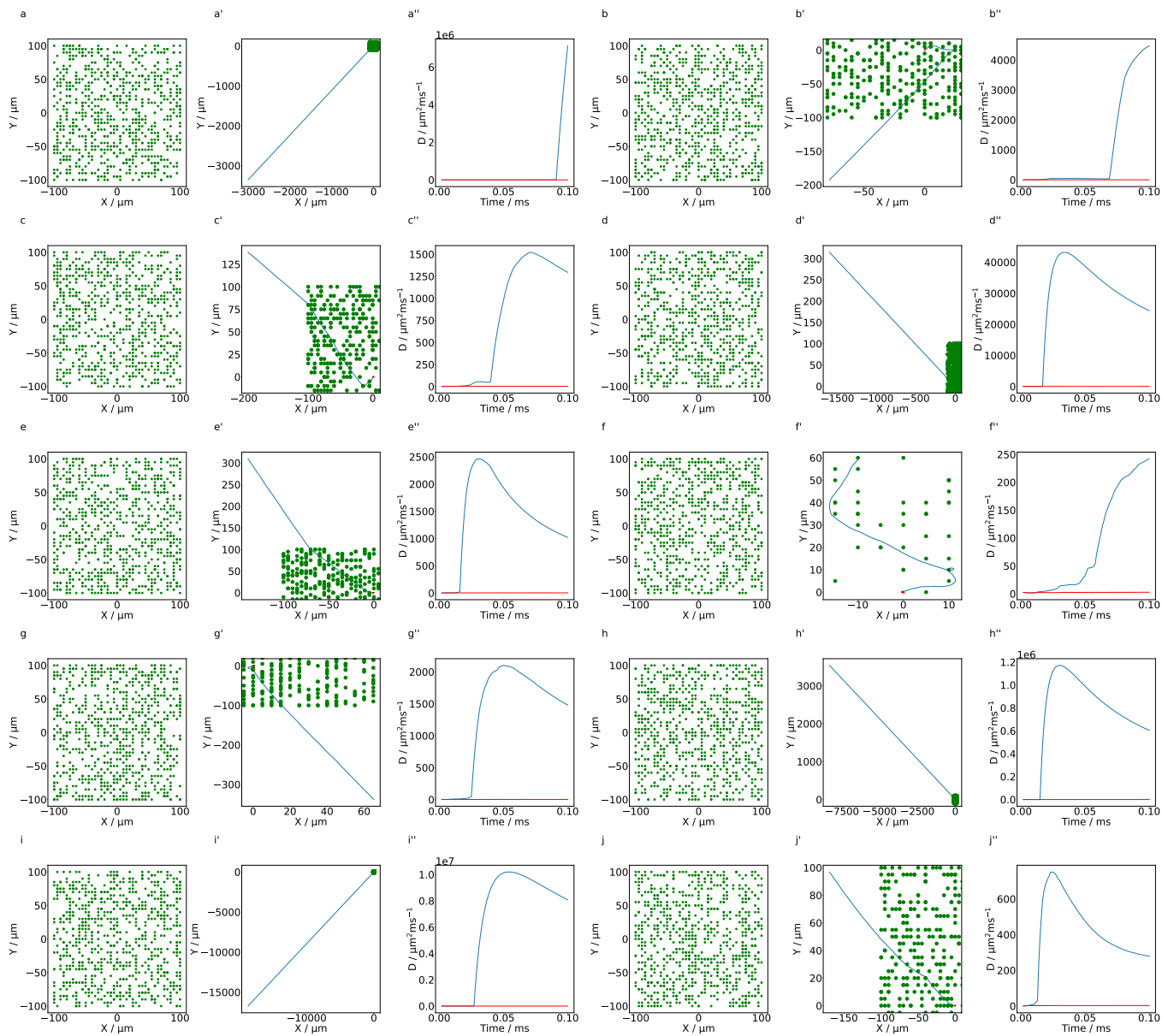}
    \caption{Simulated diffusion of a charged Atto-488-like single molecule under the influence of randomly positioned 800 surface charges (blue), compared to diffusion without surface charges (red), with details shown in insets. The particles always start at the origin (0 $\upmu$m, 0 $\upmu$m) and the minimum distance between two surface charges is 5 $\upmu$m.
    (a - j), surface charge distribution.
    (a' - j'), diffusion trajectories.
    (a" - j"), their respective diffusion coefficient plots.}
    \label{Figure:53}
\end{figure*}

\begin{figure*}[]
    \centering
    \includegraphics[width = 1\textwidth]{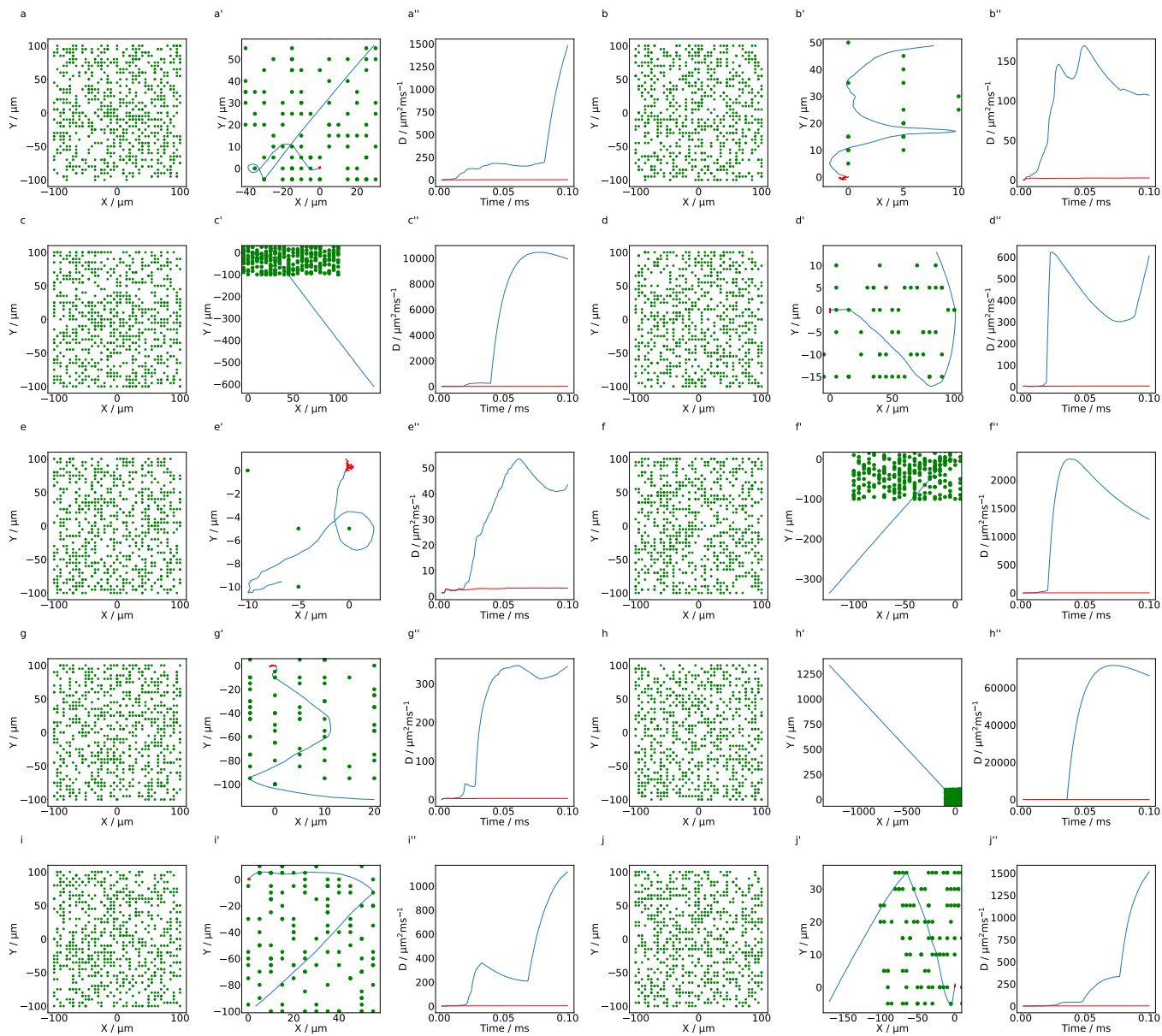}
    \caption{Simulated diffusion of a charged Atto-488-like single molecule under the influence of randomly positioned 800 surface charges (blue), compared to diffusion without surface charges (red), with details shown in insets. The particles always start at the origin (0 $\upmu$m, 0 $\upmu$m) and the minimum distance between two surface charges is 5 $\upmu$m.
    (a - j), surface charge distribution.
    (a' - j'), diffusion trajectories.
    (a" - j"), their respective diffusion coefficient plots.}
    \label{Figure:54}
\end{figure*}

\begin{figure*}[]
    \centering
    \includegraphics[width = 1\textwidth]{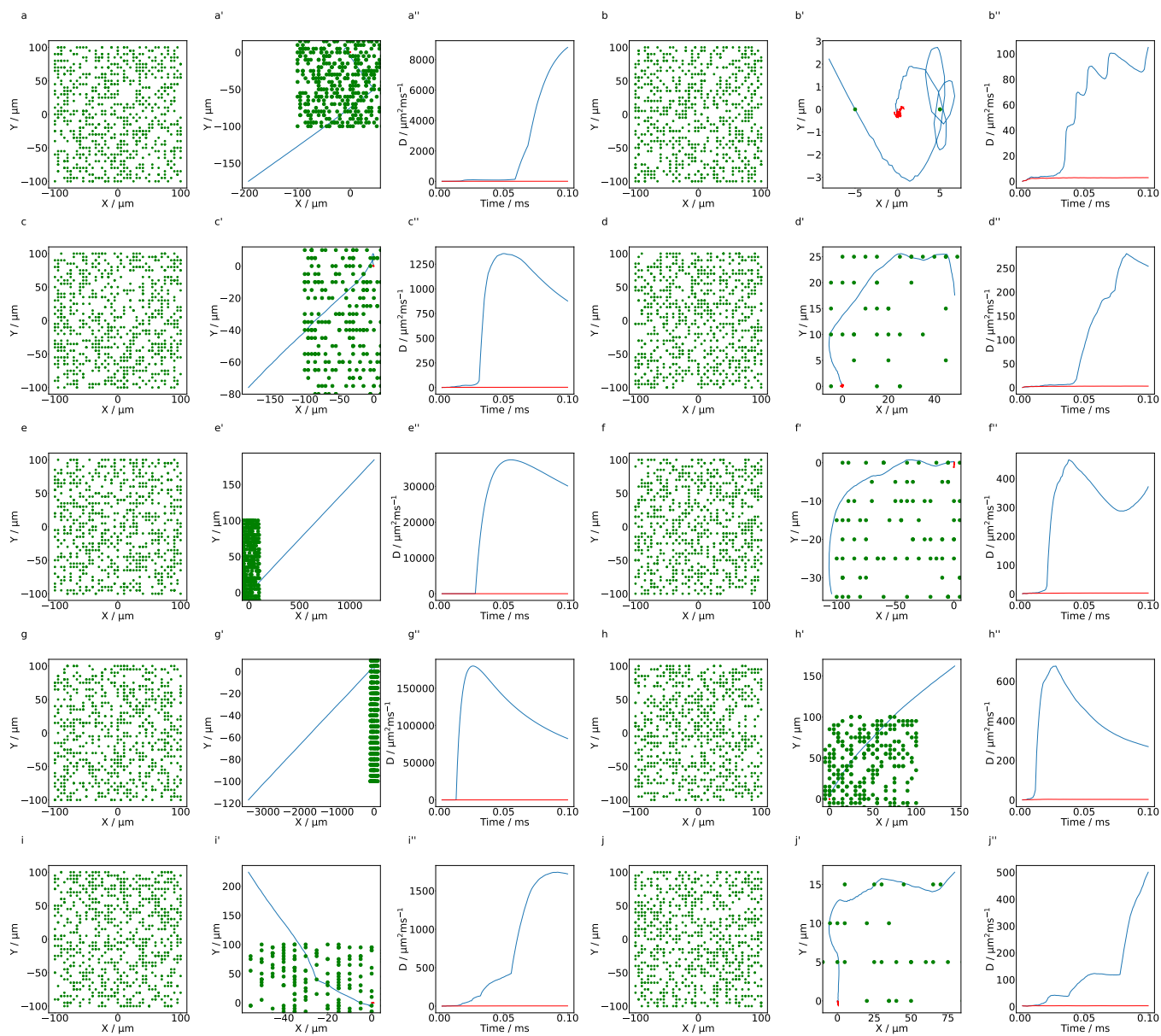}
    \caption{Simulated diffusion of a charged Atto-488-like single molecule under the influence of randomly positioned 800 surface charges (blue), compared to diffusion without surface charges (red), with details shown in insets. The particles always start at the origin (0 $\upmu$m, 0 $\upmu$m) and the minimum distance between two surface charges is 5 $\upmu$m.
    (a - j), surface charge distribution.
    (a' - j'), diffusion trajectories.
    (a" - j"), their respective diffusion coefficient plots.}
    \label{Figure:55}
\end{figure*}

\begin{figure*}[]
    \centering
    \includegraphics[width = 1\textwidth]{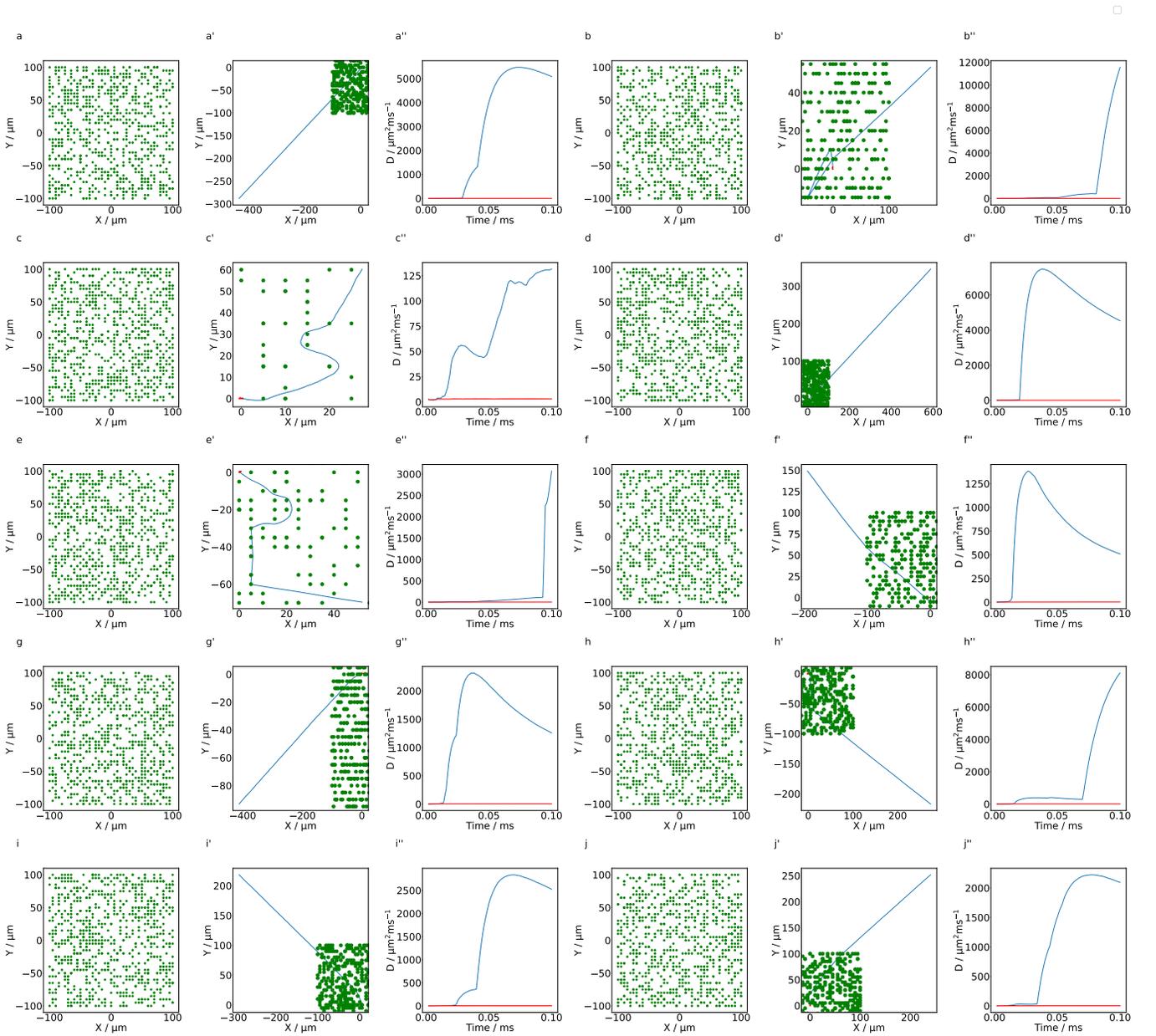}
    \caption{Simulated diffusion of a charged Atto-488-like single molecule under the influence of randomly positioned 800 surface charges (blue), compared to diffusion without surface charges (red), with details shown in insets. The particles always start at the origin (0 $\upmu$m, 0 $\upmu$m) and the minimum distance between two surface charges is 5 $\upmu$m.
    (a - j), surface charge distribution.
    (a' - j'), diffusion trajectories.
    (a" - j"), their respective diffusion coefficient plots.}
    \label{Figure:56}
\end{figure*}

\begin{figure*}[]
    \centering
    \includegraphics[width = 1\textwidth]{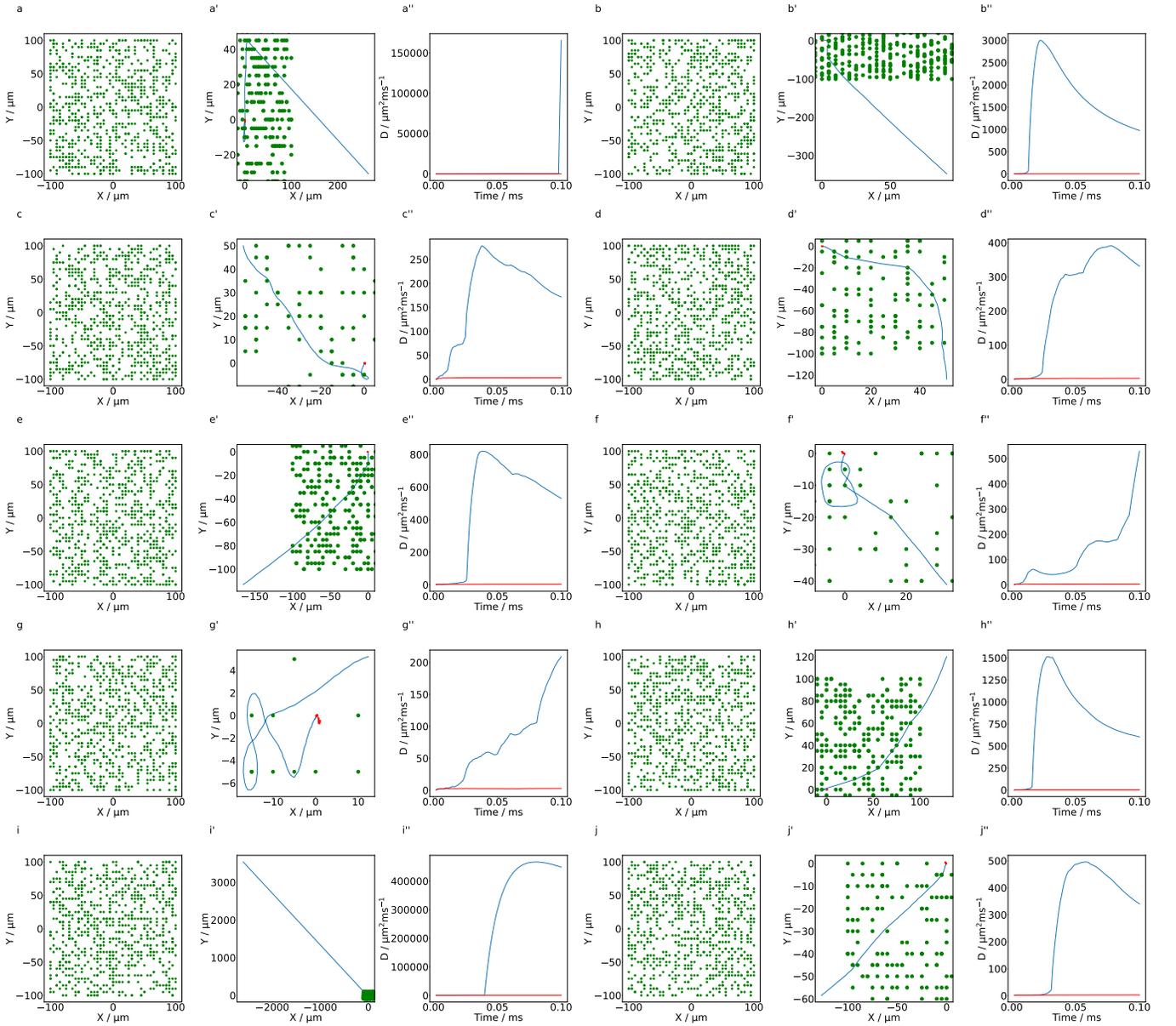}
    \caption{Simulated diffusion of a charged Atto-488-like single molecule under the influence of randomly positioned 800 surface charges (blue), compared to diffusion without surface charges (red), with details shown in insets. The particles always start at the origin (0 $\upmu$m, 0 $\upmu$m) and the minimum distance between two surface charges is 5 $\upmu$m.
    (a - j), surface charge distribution.
    (a' - j'), diffusion trajectories.
    (a" - j"), their respective diffusion coefficient plots.}
    \label{Figure:57}
\end{figure*}

\begin{figure*}[]
    \centering
    \includegraphics[width = 1\textwidth]{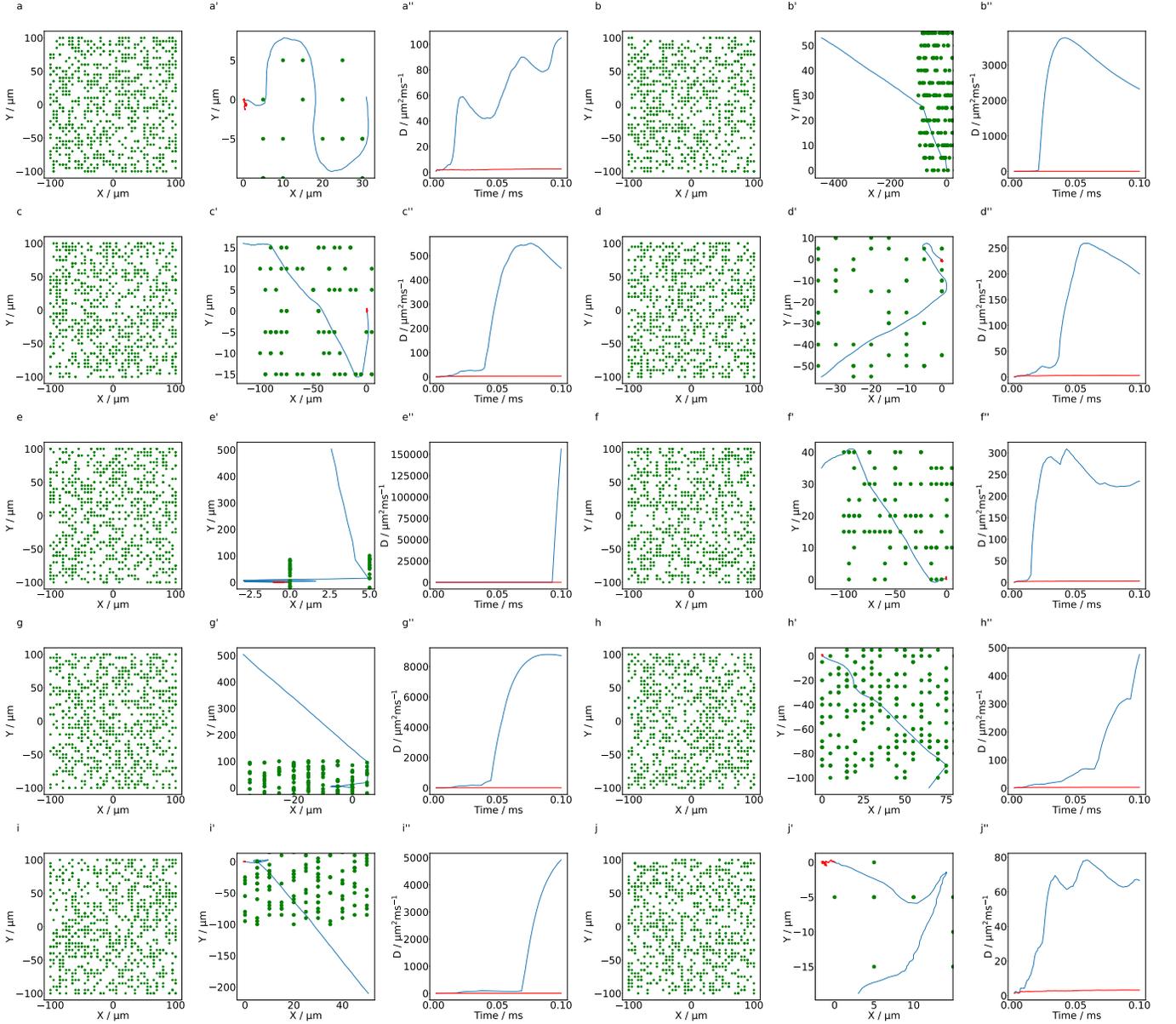}
    \caption{Simulated diffusion of a charged Atto-488-like single molecule under the influence of randomly positioned 800 surface charges (blue), compared to diffusion without surface charges (red), with details shown in insets. The particles always start at the origin (0 $\upmu$m, 0 $\upmu$m) and the minimum distance between two surface charges is 5 $\upmu$m.
    (a - j), surface charge distribution.
    (a' - j'), diffusion trajectories.
    (a" - j"), their respective diffusion coefficient plots.}
    \label{Figure:58}
\end{figure*}

\begin{figure*}[]
    \centering
    \includegraphics[width = 1\textwidth]{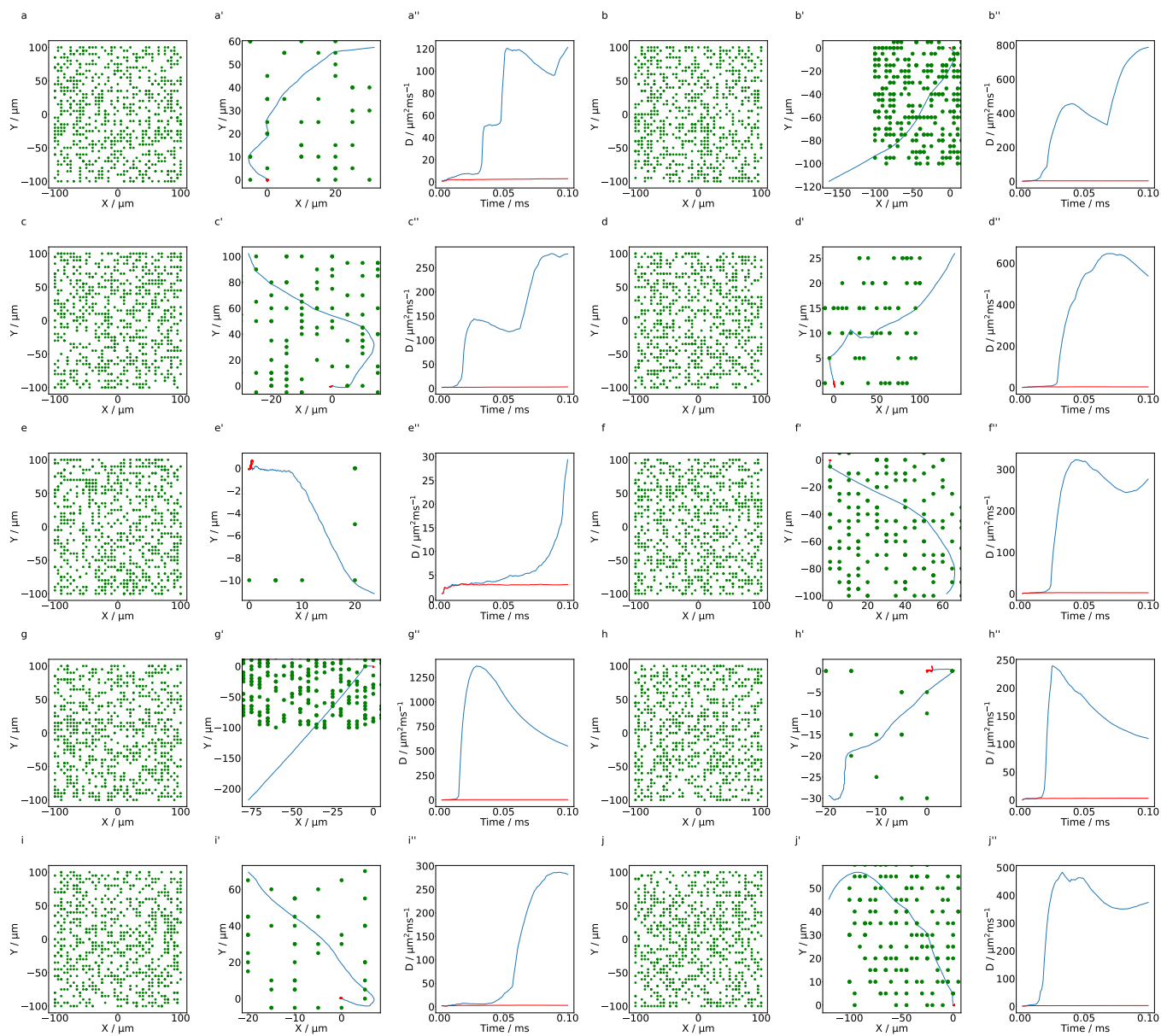}
    \caption{Simulated diffusion of a charged Atto-488-like single molecule under the influence of randomly positioned 800 surface charges (blue), compared to diffusion without surface charges (red), with details shown in insets. The particles always start at the origin (0 $\upmu$m, 0 $\upmu$m) and the minimum distance between two surface charges is 5 $\upmu$m.
    (a - j), surface charge distribution.
    (a' - j'), diffusion trajectories.
    (a" - j"), their respective diffusion coefficient plots.}
    \label{Figure:59}
\end{figure*}

\begin{figure*}[]
    \centering
    \includegraphics[width = 1\textwidth]{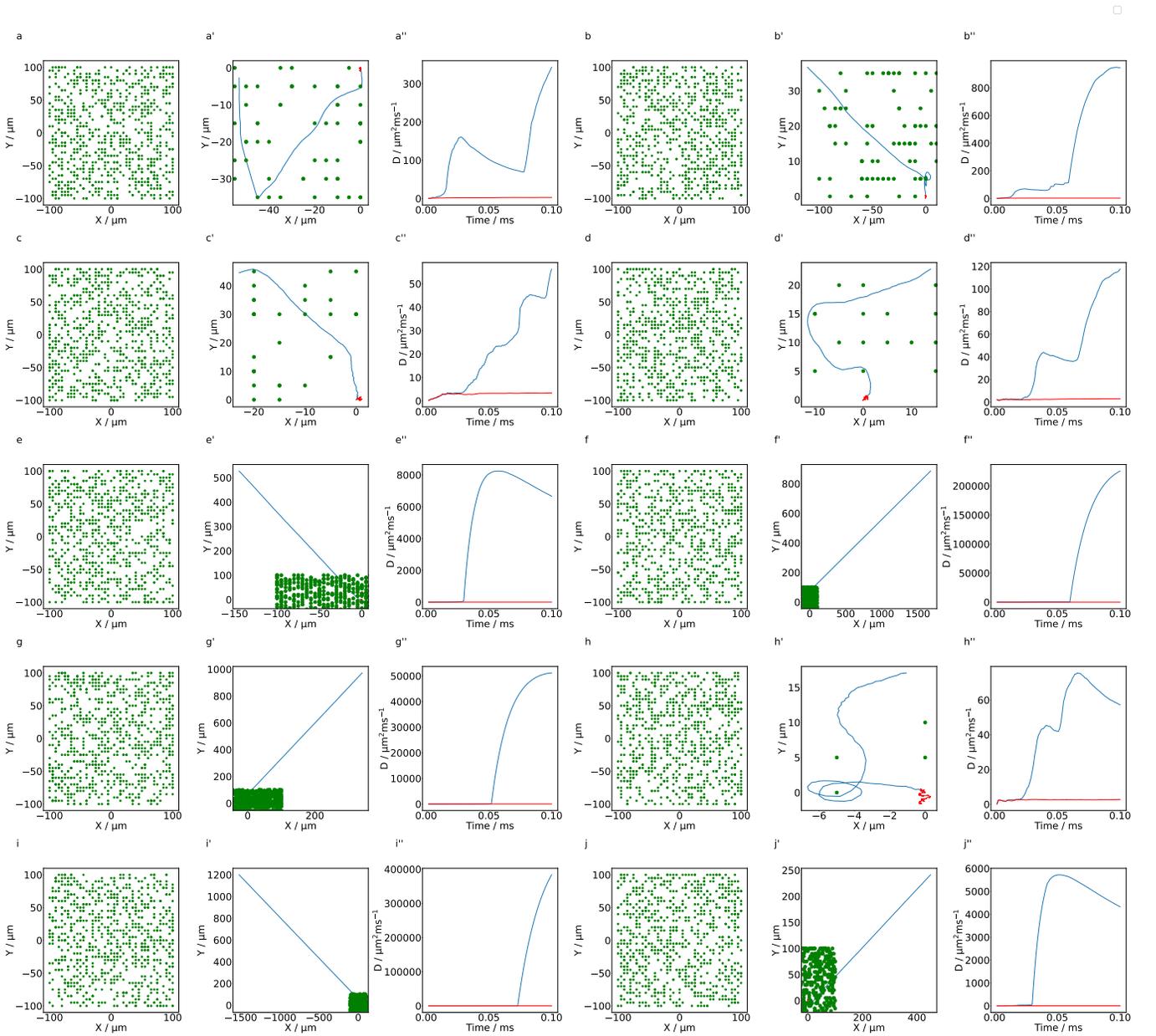}
    \caption{Simulated diffusion of a charged Atto-488-like single molecule under the influence of randomly positioned 800 surface charges (blue), compared to diffusion without surface charges (red), with details shown in insets. The particles always start at the origin (0 $\upmu$m, 0 $\upmu$m) and the minimum distance between two surface charges is 5 $\upmu$m.
    (a - j), surface charge distribution.
    (a' - j'), diffusion trajectories.
    (a" - j"), their respective diffusion coefficient plots.}
    \label{Figure:60}
\end{figure*}

\begin{figure*}[]
    \centering
    \includegraphics[width = 1\textwidth]{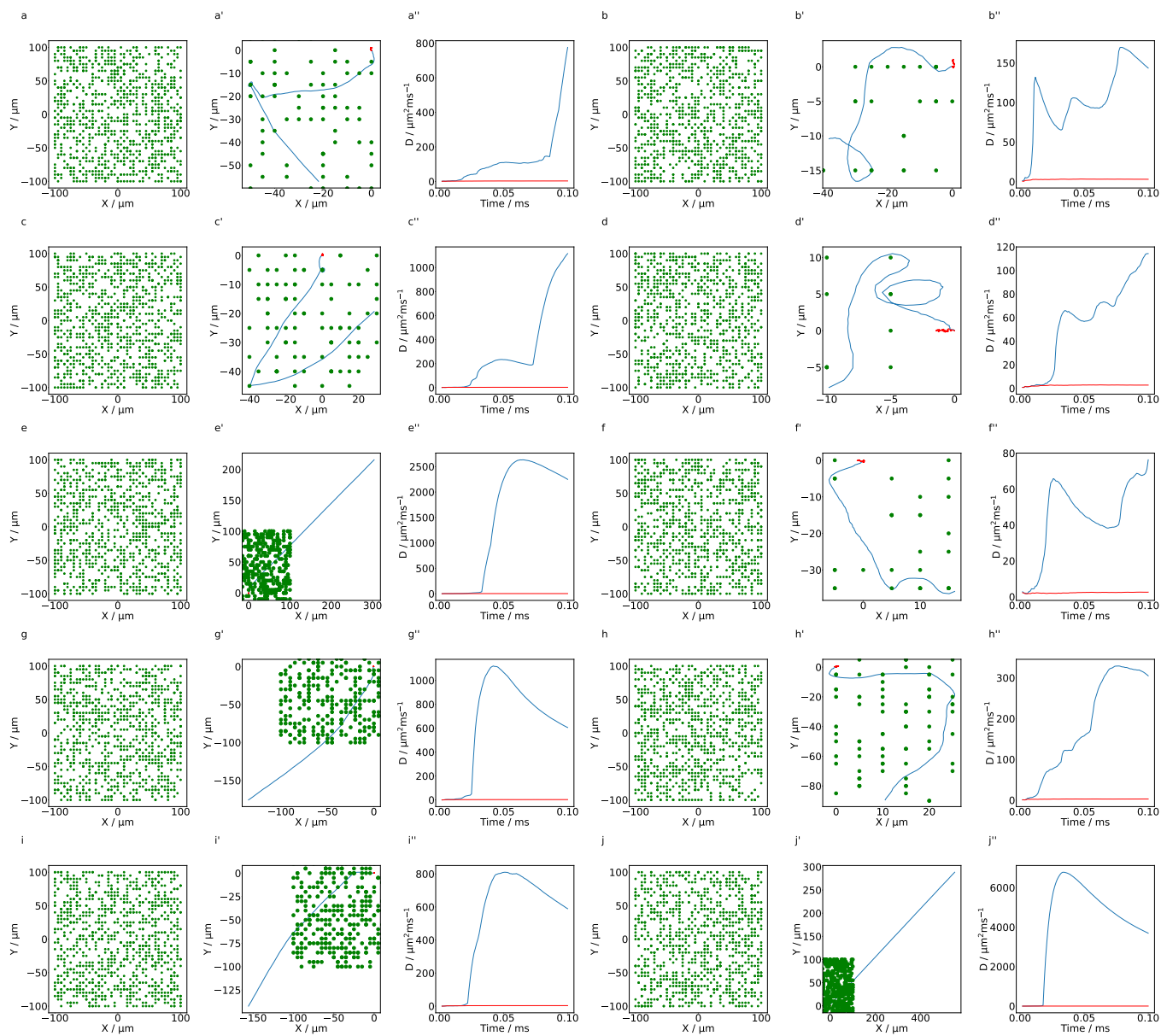}
    \caption{Simulated diffusion of a charged Atto-488-like single molecule under the influence of randomly positioned 1000 surface charges (blue), compared to diffusion without surface charges (red), with details shown in insets. The particles always start at the origin (0 $\upmu$m, 0 $\upmu$m) and the minimum distance between two surface charges is 5 $\upmu$m.
    (a - j), surface charge distribution.
    (a' - j'), diffusion trajectories.
    (a" - j"), their respective diffusion coefficient plots.}
    \label{Figure:61}
\end{figure*}

\begin{figure*}[]
    \centering
    \includegraphics[width = 1\textwidth]{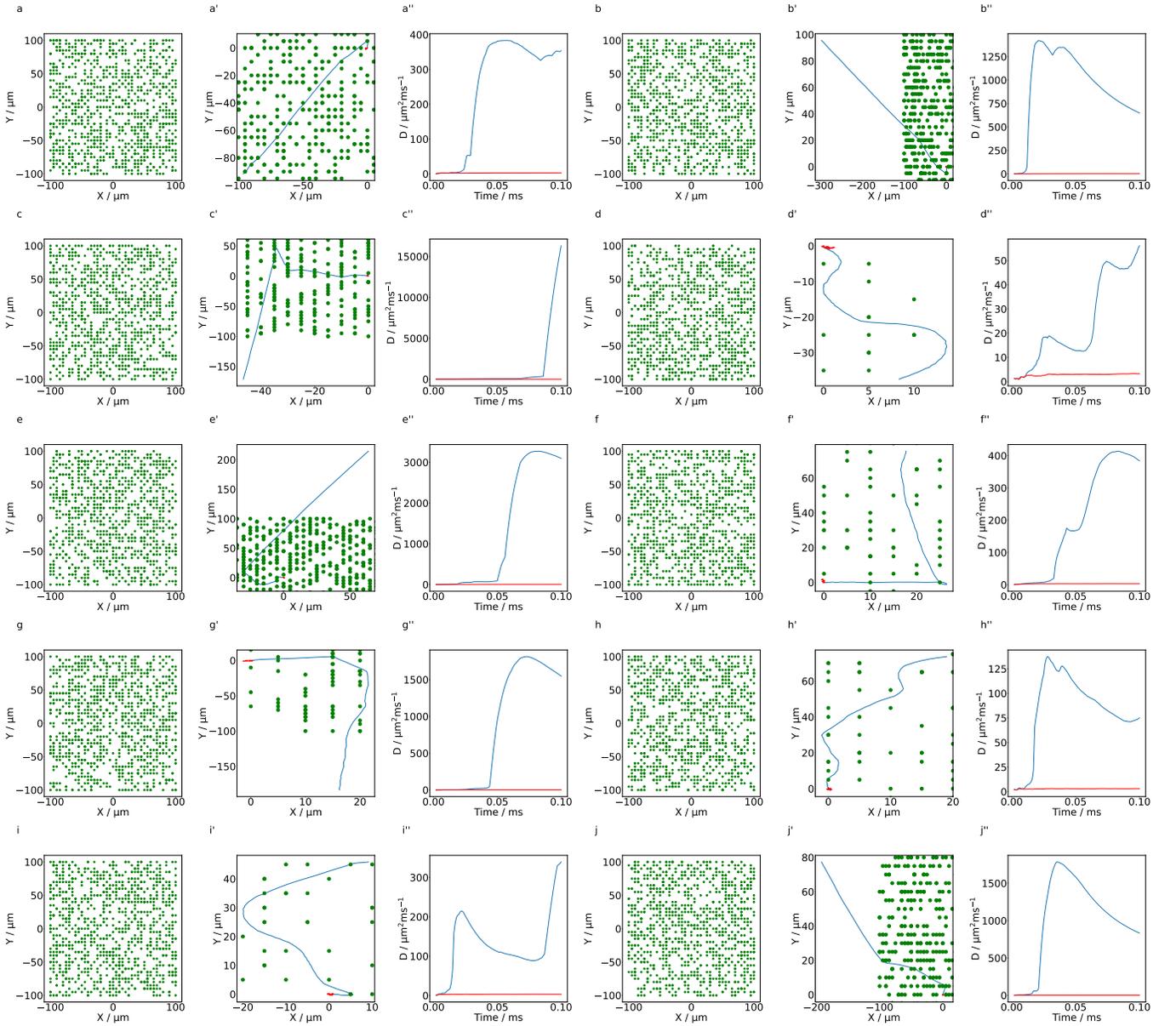}
    \caption{Simulated diffusion of a charged Atto-488-like single molecule under the influence of randomly positioned 1000 surface charges (blue), compared to diffusion without surface charges (red), with details shown in insets. The particles always start at the origin (0 $\upmu$m, 0 $\upmu$m) and the minimum distance between two surface charges is 5 $\upmu$m.
    (a - j), surface charge distribution.
    (a' - j'), diffusion trajectories.
    (a" - j"), their respective diffusion coefficient plots.}
    \label{Figure:62}
\end{figure*}

\begin{figure*}[]
    \centering
    \includegraphics[width = 1\textwidth]{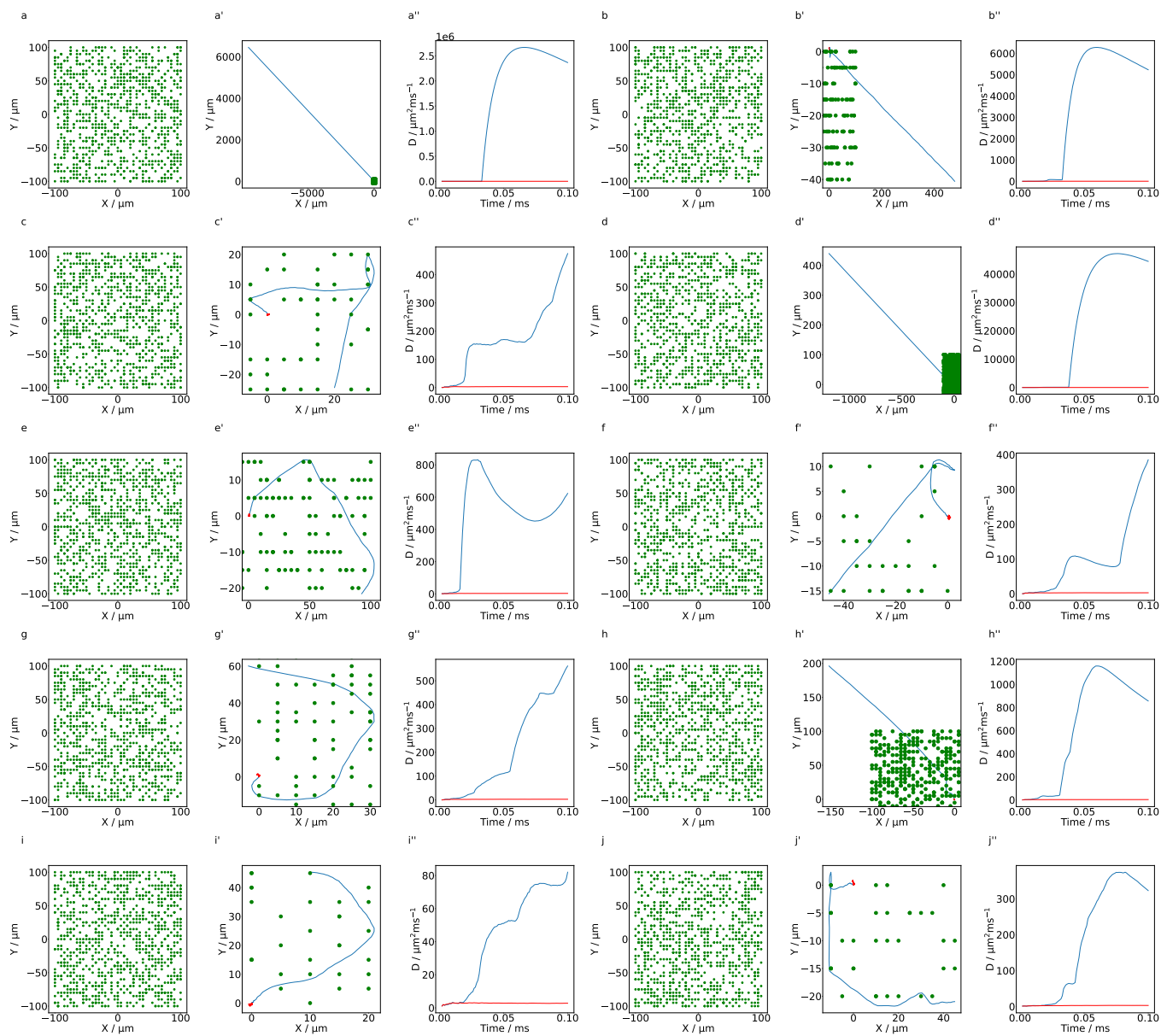}
    \caption{Simulated diffusion of a charged Atto-488-like single molecule under the influence of randomly positioned 1000 surface charges (blue), compared to diffusion without surface charges (red), with details shown in insets. The particles always start at the origin (0 $\upmu$m, 0 $\upmu$m) and the minimum distance between two surface charges is 5 $\upmu$m.
    (a - j), surface charge distribution.
    (a' - j'), diffusion trajectories.
    (a" - j"), their respective diffusion coefficient plots.}
    \label{Figure:63}
\end{figure*}

\begin{figure*}[]
    \centering
    \includegraphics[width = 1\textwidth]{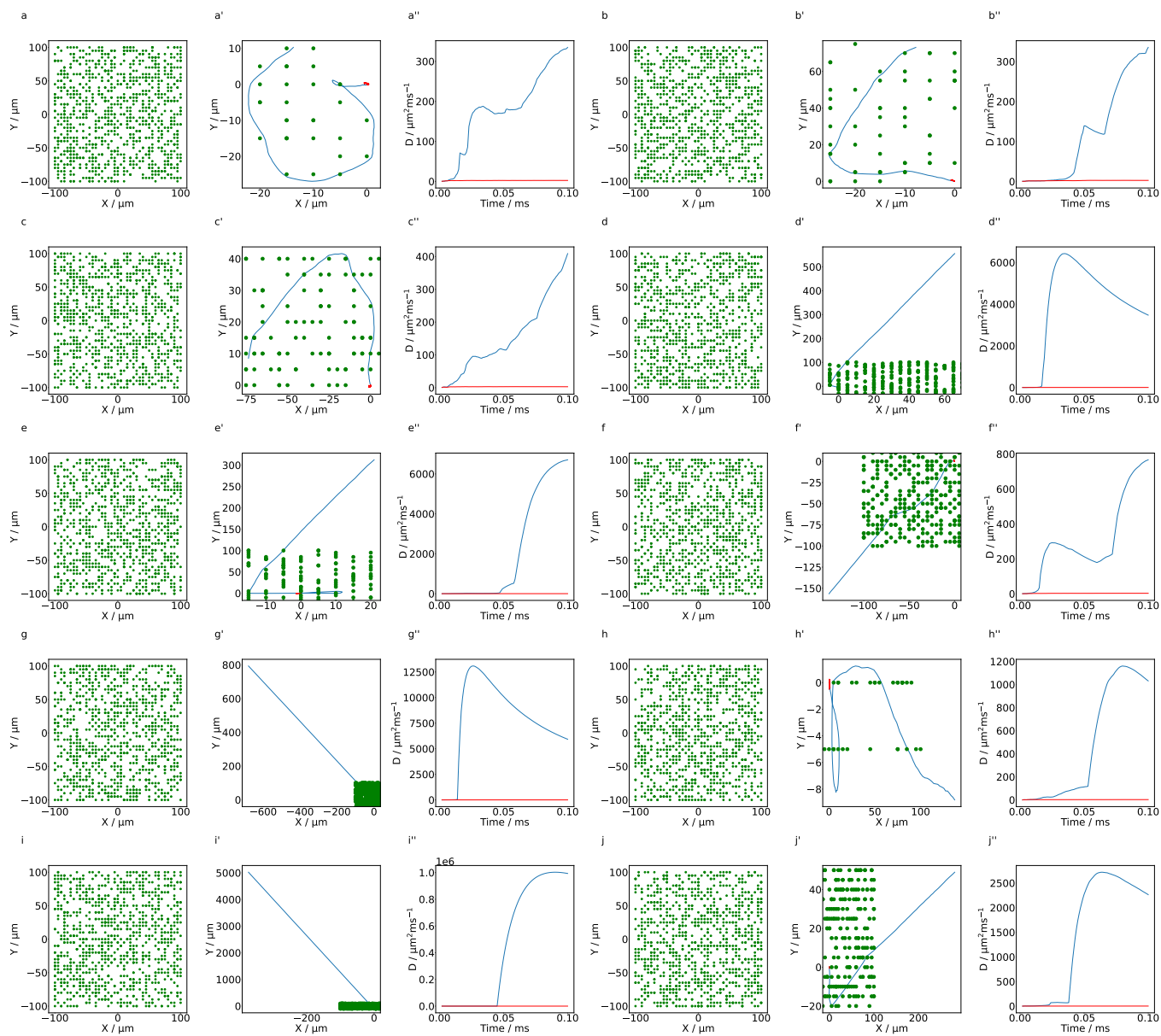}
    \caption{Simulated diffusion of a charged Atto-488-like single molecule under the influence of randomly positioned 1000 surface charges (blue), compared to diffusion without surface charges (red), with details shown in insets. The particles always start at the origin (0 $\upmu$m, 0 $\upmu$m) and the minimum distance between two surface charges is 5 $\upmu$m.
    (a - j), surface charge distribution.
    (a' - j'), diffusion trajectories.
    (a" - j"), their respective diffusion coefficient plots.}
    \label{Figure:64}
\end{figure*}

\begin{figure*}[]
    \centering
    \includegraphics[width = 1\textwidth]{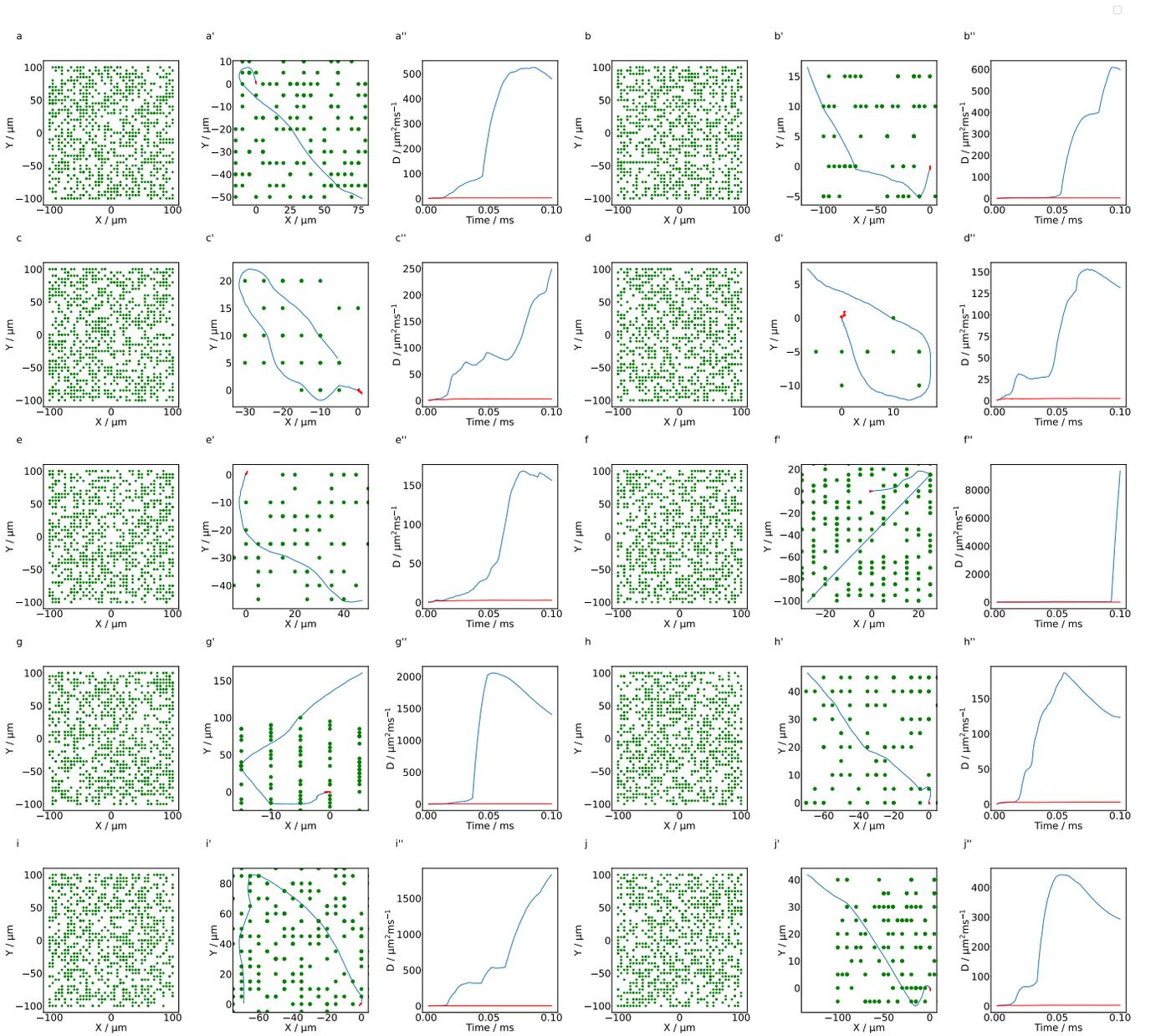}
    \caption{Simulated diffusion of a charged Atto-488-like single molecule under the influence of randomly positioned 1000 surface charges (blue), compared to diffusion without surface charges (red), with details shown in insets. The particles always start at the origin (0 $\upmu$m, 0 $\upmu$m) and the minimum distance between two surface charges is 5 $\upmu$m.
    (a - j), surface charge distribution.
    (a' - j'), diffusion trajectories.
    (a" - j"), their respective diffusion coefficient plots.}
    \label{Figure:65}
\end{figure*}

\begin{figure*}[]
    \centering
    \includegraphics[width = 1\textwidth]{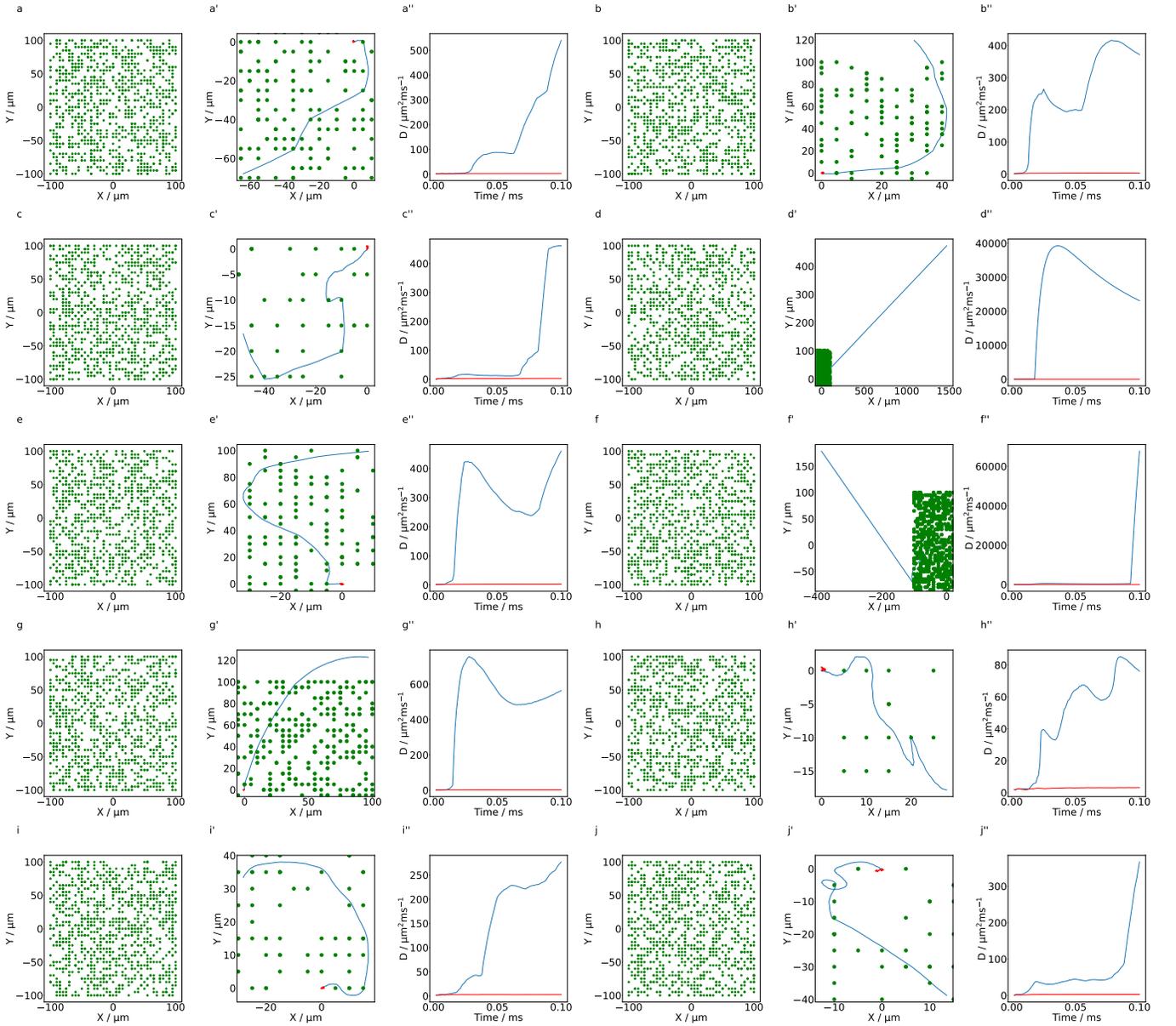}
    \caption{Simulated diffusion of a charged Atto-488-like single molecule under the influence of randomly positioned 1000 surface charges (blue), compared to diffusion without surface charges (red), with details shown in insets. The particles always start at the origin (0 $\upmu$m, 0 $\upmu$m) and the minimum distance between two surface charges is 5 $\upmu$m.
    (a - j), surface charge distribution.
    (a' - j'), diffusion trajectories.
    (a" - j"), their respective diffusion coefficient plots.}
    \label{Figure:66}
\end{figure*}

\begin{figure*}[]
    \centering
    \includegraphics[width = 1\textwidth]{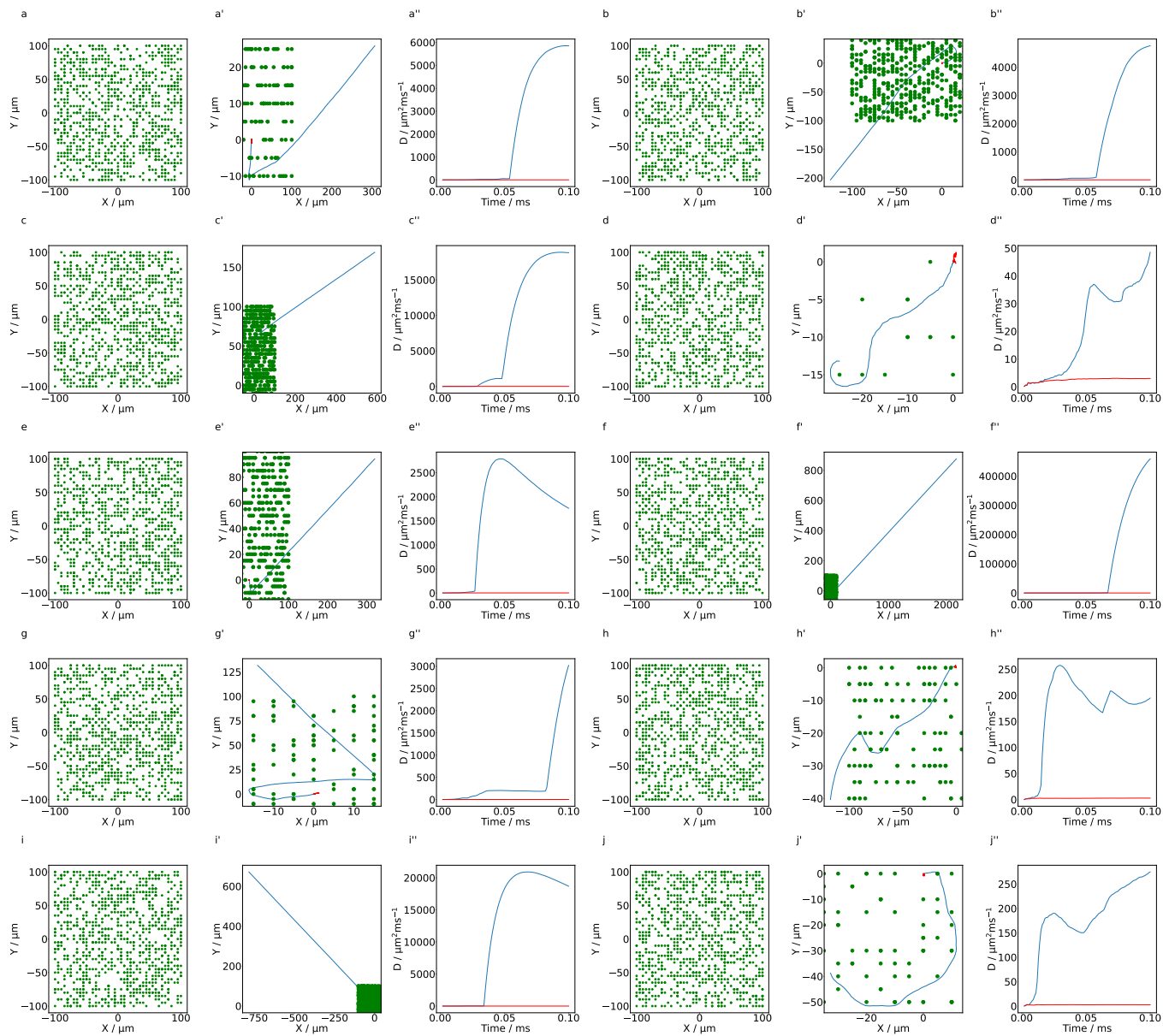}
    \caption{Simulated diffusion of a charged Atto-488-like single molecule under the influence of randomly positioned 1000 surface charges (blue), compared to diffusion without surface charges (red), with details shown in insets. The particles always start at the origin (0 $\upmu$m, 0 $\upmu$m) and the minimum distance between two surface charges is 5 $\upmu$m.
    (a - j), surface charge distribution.
    (a' - j'), diffusion trajectories.
    (a" - j"), their respective diffusion coefficient plots.}
    \label{Figure:67}
\end{figure*}

\begin{figure*}[]
    \centering
    \includegraphics[width = 1\textwidth]{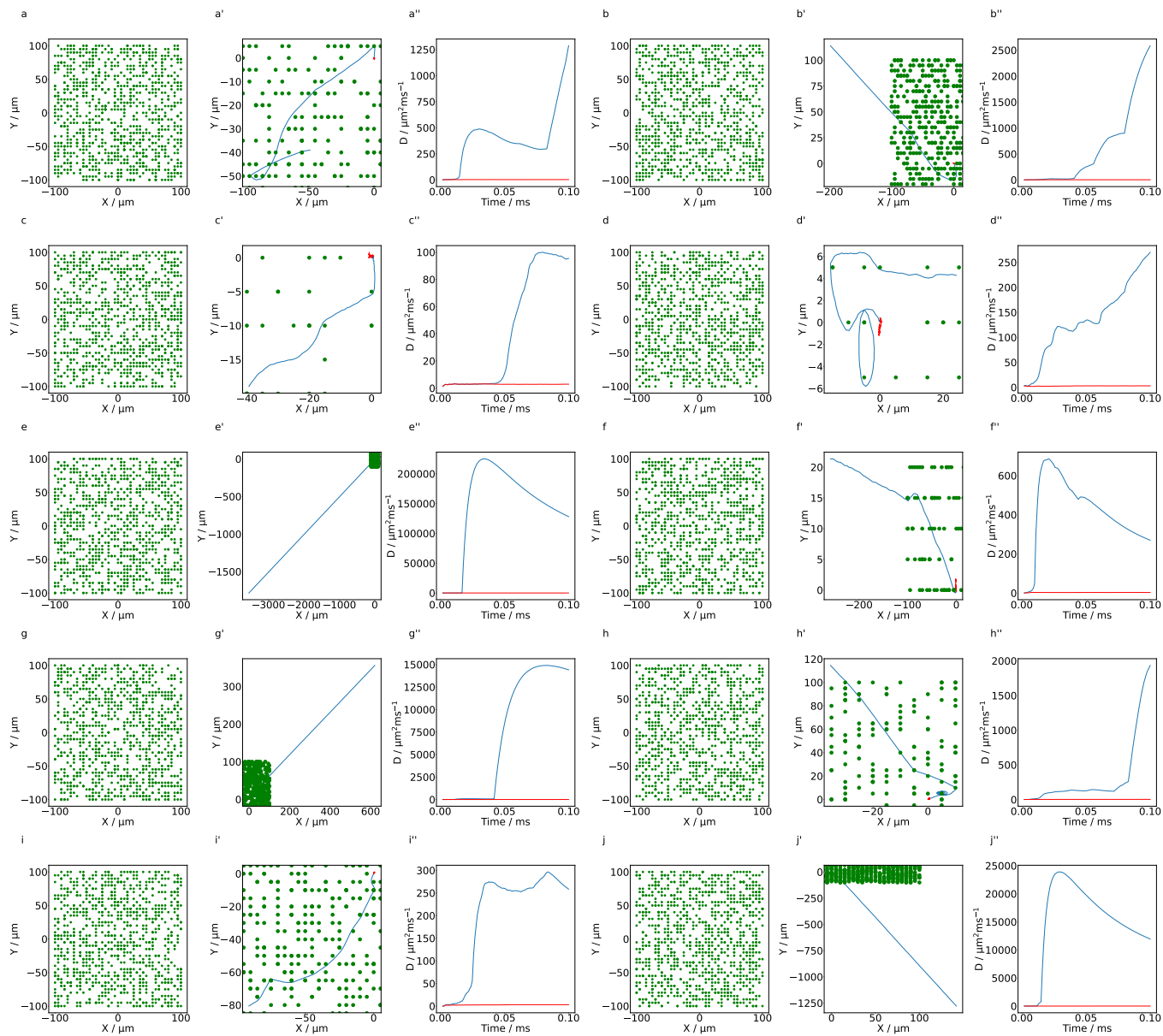}
    \caption{Simulated diffusion of a charged Atto-488-like single molecule under the influence of randomly positioned 1000 surface charges (blue), compared to diffusion without surface charges (red), with details shown in insets. The particles always start at the origin (0 $\upmu$m, 0 $\upmu$m) and the minimum distance between two surface charges is 5 $\upmu$m.
    (a - j), surface charge distribution.
    (a' - j'), diffusion trajectories.
    (a" - j"), their respective diffusion coefficient plots.}
    \label{Figure:68}
\end{figure*}

\begin{figure*}[]
    \centering
    \includegraphics[width = 1\textwidth]{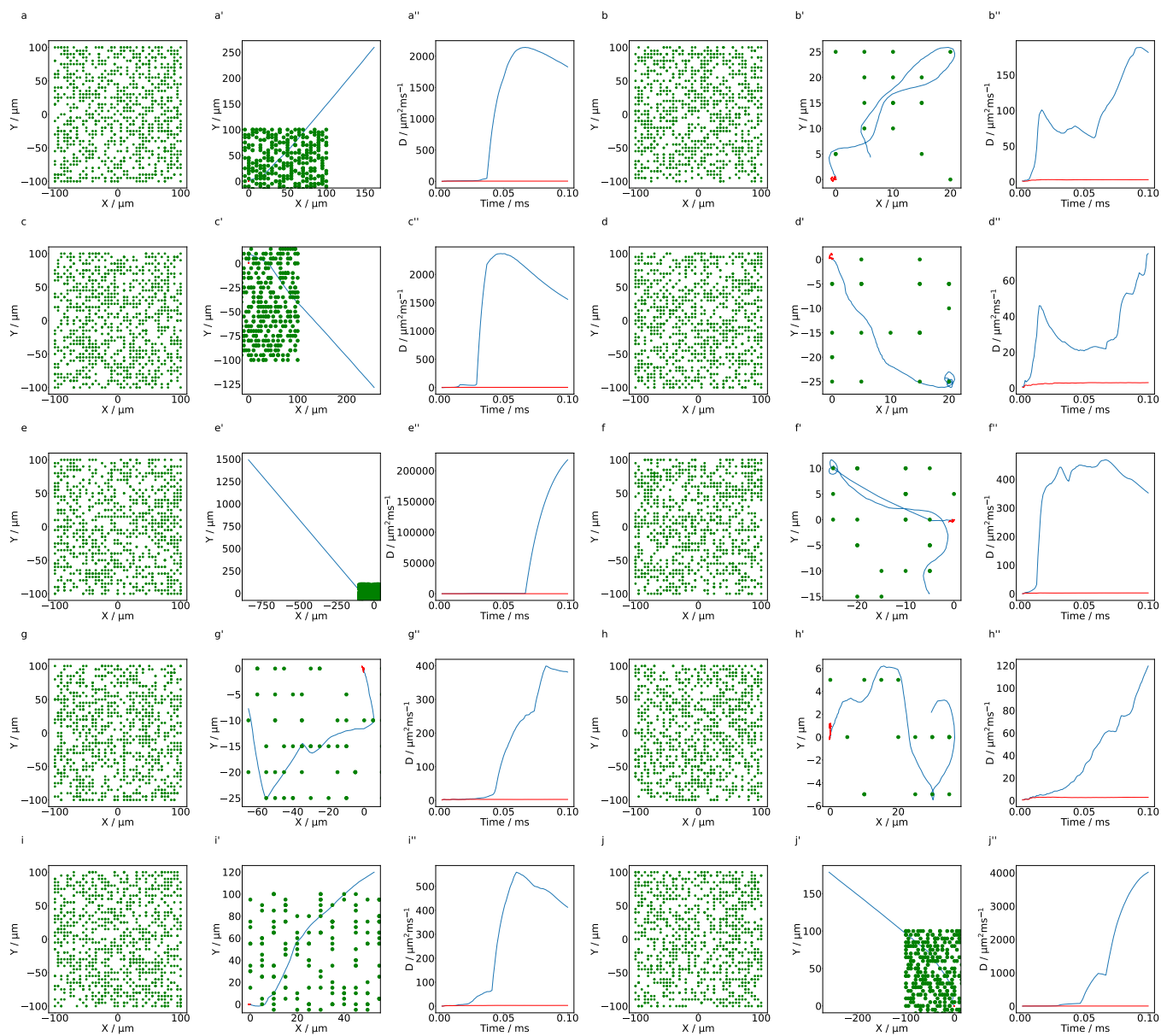}
    \caption{Simulated diffusion of a charged Atto-488-like single molecule under the influence of randomly positioned 1000 surface charges (blue), compared to diffusion without surface charges (red), with details shown in insets. The particles always start at the origin (0 $\upmu$m, 0 $\upmu$m) and the minimum distance between two surface charges is 5 $\upmu$m.
    (a - j), surface charge distribution.
    (a' - j'), diffusion trajectories.
    (a" - j"), their respective diffusion coefficient plots.}
    \label{Figure:69}
\end{figure*}

\begin{figure*}[]
    \centering
    \includegraphics[width = 1\textwidth]{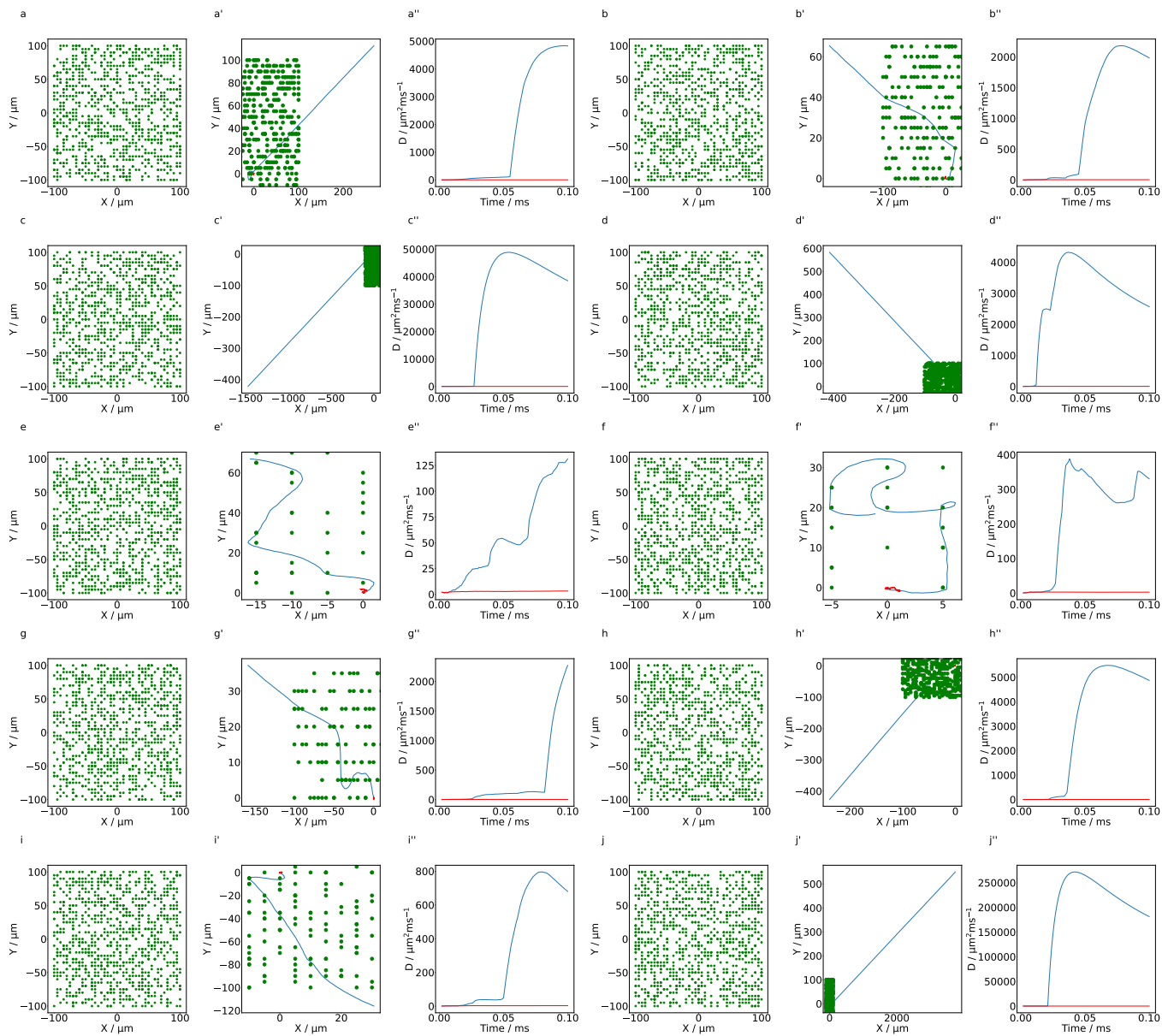}
    \caption{Simulated diffusion of a charged Atto-488-like single molecule under the influence of randomly positioned 1000 surface charges (blue), compared to diffusion without surface charges (red), with details shown in insets. The particles always start at the origin (0 $\upmu$m, 0 $\upmu$m) and the minimum distance between two surface charges is 5 $\upmu$m.
    (a - j), surface charge distribution.
    (a' - j'), diffusion trajectories.
    (a" - j"), their respective diffusion coefficient plots.}
    \label{Figure:70}
\end{figure*}

\end{document}